\documentclass[12pt]{article}

\usepackage{newtxtext,newtxmath}

\usepackage{graphicx}

\usepackage[letterpaper,margin=1in]{geometry}

\renewenvironment{abstract}
	{\quotation}
	{\endquotation}

\date{}

\makeatletter
\renewcommand{\fnum@figure}{\textbf{Figure \thefigure}}
\renewcommand{\fnum@table}{\textbf{Table \thetable}}
\makeatother

\usepackage{scicite}

\usepackage{url}

\def\scititle{
	Gas-depleted planet formation occurred in the four-planet system around the red dwarf LHS\,1903
}
\title{\bfseries \boldmath \scititle}

\author{
{\small Thomas G. Wilson$^{1,2,\ast}$, 
Anna M. Simpson$^{3}$, 
Andrew Collier Cameron$^{1}$, 
Ryan Cloutier$^{4,5}$,}\and 
{\small Vardan Adibekyan$^{6,7}$, 
Ancy Anna John$^{1}$, 
Yann Alibert$^{8,9}$, 
Manu Stalport$^{10,11}$, 
Jo Ann Egger$^{8}$,}\and 
{\small Andrea Bonfanti$^{12}$
% ...
}\and 
\scriptsize{$^{1}$ Centre for Exoplanet Science, School of Physics and Astronomy, University of St Andrews, St Andrews, UK}\and
\scriptsize{$^{2}$ Department of Physics, University of Warwick, Coventry, UK}\and
\scriptsize{$^{3}$ Department of Astronomy and Department of Physics, University of Michigan, Ann Arbor, USA}\and
\scriptsize{$^{4}$ Department of Physics \& Astronomy, McMaster University, Hamilton, Canada}\and
\scriptsize{$^{5}$ Center for Astrophysics, Harvard \& Smithsonian, Cambridge, USA}\and
\scriptsize{$^{6}$ Instituto de Astrofisica e Ciencias do Espaco, Universidade do Porto, Porto, Portugal}\and
\scriptsize{$^{7}$ Departamento de Fisica e Astronomia, Faculdade de Ciencias, Universidade do Porto, Porto, Portugal}\and
\scriptsize{$^{8}$ Physikalisches Institut, University of Bern, Bern, Switzerland}\and
\scriptsize{$^{9}$ Center for Space and Habitability, University of Bern, Bern, Switzerland}\and
\scriptsize{$^{10}$ Space sciences, Technologies and Astrophysics Research Institute, Université de Liège, Liège, Belgium}\and
\scriptsize{$^{11}$ Observatoire Astronomique de l'Université de Genève, Versoix, Switzerland}\and
\scriptsize{$^{12}$ Space Research Institute, Austrian Academy of Sciences, Graz, Austria}\and
\scriptsize{$^{13}$ Laboratoire d'Astrophysique de Marseille, Marseille, France}\and
\scriptsize{$^{14}$ Division Technique Institute National Des Sciences De L'univers, La Seyne-sur-Mer, France}\and
\scriptsize{$^{15}$ Astrophysics Group, Keele University, UK}\and
\scriptsize{$^{16}$ Leiden Observatory, University of Leiden, Leiden, The Netherlands}\and
\scriptsize{$^{17}$ Department of Space, Earth and Environment, Chalmers University of Technology, Onsala Space Observatory, Onsala, Sweden}\and
\scriptsize{$^{18}$ Lund Observatory, Division of Astrophysics, Department of Physics, Lund University, Lund, Sweden}\and
\scriptsize{$^{19}$ Lund Observatory, Department of Astronomy \& Theoretical Physics, Lund University, Lund, Sweden}\and
\scriptsize{$^{20}$ Department of Physics and Kavli Institute for Astrophysics and Space Research, Massachusetts Institute of Technology, Cambridge, USA}\and
\scriptsize{$^{21}$ Dipartimento di Fisica e Astronomia "Galileo Galilei", Universita degli Studi di Padova, Padova, Italy}\and
\scriptsize{$^{22}$ Istituto Nazionale di Astrofisica, Osservatorio Astronomico di Padova, Padova, Italy}\and
\scriptsize{$^{23}$ Astrophysics Group, Cavendish Laboratory, University of Cambridge, Cambridge, UK}\and
\scriptsize{$^{24}$ Istituto Nazionale di Astrofisica, Osservatorio Astronomico di Palermo, Palermo, Italy}\and
\scriptsize{$^{25}$ Sub-department of Astrophysics, University of Oxford, Oxford, UK}\and
\scriptsize{$^{26}$ Instituto de Astrof\'isica de Canarias, La Laguna, Spain}\and
\scriptsize{$^{27}$ Departamento de Astrof\'isica, Universidad de La Laguna, La Laguna, Spain}\and
\scriptsize{$^{28}$ Institut de Ciencies de l'Espai, Consejo Superior de Investigaciones Científicas, Bellaterra, Spain}\and
\scriptsize{$^{29}$ Institut d'Estudis Espacials de Catalunya, Barcelona, Spain}\and
\scriptsize{$^{30}$ Admatis, Miskolc, Hungary}\and
\scriptsize{$^{31}$ Departamento de Astrofisica, Centro de Astrobiologia, Consejo Superior de Investigaciones Científicas, Instituto Nacional de Técnica Aeroespacial, Villanueva de la Cañada, Spain}\and
\scriptsize{$^{32}$ Max Planck Institut für Extraterrestrische Physik, Garching bei München, Germany}\and
\scriptsize{$^{33}$ Université Grenoble Alpes, L'Institut de Planétologie et d'Astrophysique de Grenoble, Grenoble, France}\and
\scriptsize{$^{34}$ Department of Astronomy, Stockholm University, AlbaNova University Center, Stockholm, Sweden}\and
\scriptsize{$^{35}$ Danmarks Tekniske Universitet Space, National Space Institute, Technical University of Denmark, Kongens Lyngby, Denmark}\and
\scriptsize{$^{36}$ Institute of Optical Sensor Systems, German Aerospace Center, Berlin, Germany}\and
\scriptsize{$^{37}$ Institute of Planetary Research, German Aerospace Center, Berlin, Germany}\and
\scriptsize{$^{38}$ Universidad Nacional Aut\'onoma de M\'exico, Instituto de Astronom\'ia, Ciudad de México, M\'exico}\and
\scriptsize{$^{39}$ Department of Physics and Astronomy, Georgia State University, Atlanta, USA}\and
\scriptsize{$^{40}$ Research Consortium On Nearby Stars Institute, Chambersburg, USA}\and
\scriptsize{$^{41}$ Université de Paris Cité, Institut de physique du globe de Paris, Centre National de la Recherche Scientifique, Paris, France}\and
\scriptsize{$^{42}$ NASA Exoplanet Science Institute, Caltech/Infrared Processing and Analysis Center, Pasadena, USA}\and
\scriptsize{$^{43}$ George Mason University, Fairfax, USA}\and
\scriptsize{$^{44}$ Fundacion Galileo Galilei, Istituto Nazionale di Astrofisica, Breña Baja, Spain}\and
\scriptsize{$^{45}$ Istituto Nazionale di Astrofisica, Osservatorio Astrofisico di Catania, Catania, Italy}\and
\scriptsize{$^{46}$ Istituto Nazionale di Astrofisica, Osservatorio Astrofisico di Torino, Pino Torinese, Italy}\and
\scriptsize{$^{47}$ Department of Astrophysics, University of Exeter, Exeter, UK}\and
\scriptsize{$^{48}$ Astronomy Department, University of Washington, Seattle, USA}\and
\scriptsize{$^{49}$ Centre for Mathematical Sciences, Lund University, Lund, Sweden}\and
\scriptsize{$^{50}$ Astrobiology Research Unit, Université de Liège, Liège, Belgium}\and
\scriptsize{$^{51}$ Department of Physics and Astronomy, University of New Mexico, Albuquerque, USA}\and
\scriptsize{$^{52}$ Department of Astronomy, University of California Berkeley, Berkeley, USA}\and
\scriptsize{$^{53}$ Département d’astronomie de l'Université de Genève, Versoix, Switzerland}\and
\scriptsize{$^{54}$ Centre Vie dans l’Univers, Faculté des sciences, Université de Genève, Genève, Switzerland}\and
\scriptsize{$^{55}$ Department of Multi-Disciplinary Sciences, Graduate School of Arts and Sciences, The University of Tokyo, Tokyo, Japan}\and
\scriptsize{$^{56}$ Komaba Institute for Science, The University of Tokyo, Tokyo, Japan}\and
\scriptsize{$^{57}$ Dipartimento di Fisica, Universita degli Studi di Torino, Torino, Italy}\and
\scriptsize{$^{58}$ Department of Astrophysics, University of Vienna, Türkenschanzstrasse 17, 1180 Vienna, Austria}\and
\scriptsize{$^{59}$ Observatori Astronòmic Albanyà, Girona, Spain}\and
\scriptsize{$^{60}$ European Space Research and Technology Centre, European Space Agency, Noordwijk, The Netherlands}\and
\scriptsize{$^{61}$ Department of Physics \& Astronomy, University of California Irvine, Irvine, USA}\and
\scriptsize{$^{62}$ Centre for Astrophysics, University of Southern Queensland, Toowoomba, Australia}\and
\scriptsize{$^{63}$ Division of Science, National Astronomical Observatory of Japan, Tokyo, Japan}\and
\scriptsize{$^{64}$ School of Earth and Space Exploration, Arizona State University, Tempe, USA}\and
\scriptsize{$^{65}$ NASA Ames Research Center, Moffett Field, USA}\and
\scriptsize{$^{66}$ Department of Physics \& Astronomy, Swarthmore College, Swarthmore, USA}\and
\scriptsize{$^{67}$ Department of Physics, University of the Pacific, Stockton, USA}\and
\scriptsize{$^{68}$ Konkoly Observatory, Research Centre for Astronomy and Earth Sciences, Budapest, Hungary}\and
\scriptsize{$^{69}$ E\"otv\"os Lor\'and University, Institute of Physics, Budapest, Hungary}\and
\scriptsize{$^{70}$ Institute for Celestial Mechanics and Computation of Ephemerides, Observatoire de Paris, Paris, France}\and
\scriptsize{$^{71}$ Institut d'astrophysique de Paris, Université Pierre \& Marie Curie, Paris, France}\and
\scriptsize{$^{72}$ Department of Astronomy, Graduate School of Science, The University of Tokyo, Tokyo, Japan}\and
\scriptsize{$^{73}$ Department of Astronomy \& Astrophysics, University of Chicago, Chicago, USA}\and
\scriptsize{$^{74}$ School of Physics \& Astronomy, University of Birmingham, Birmingham, UK}\and
\scriptsize{$^{75}$ Astrobiology Center, Tokyo, Japan}\and
\scriptsize{$^{76}$ Institute for Astronomy, University of Edinburgh, Edinburgh, UK}\and
\scriptsize{$^{77}$ Centre for Exoplanet Science, University of Edinburgh, Edinburgh, UK}\and
\scriptsize{$^{78}$ Istituto Nazionale di Astrofisica, Osservatorio Astronomico di Brera, Merate, Italy}\and
\scriptsize{$^{79}$ Department of Physics, Eidgenössische Technische Hochschule Zurich, Zurich, Switzerland}\and
\scriptsize{$^{80}$ Research Institute for Advanced Computer Science, Universities Space Research Association, Washington DC, USA}\and
\scriptsize{$^{81}$ Zentrum für Astronomie und Astrophysik, Technische Universität Berlin, Berlin, Germany}\and
\scriptsize{$^{82}$ Institut fuer Geologische Wissenschaften, Freie Universitaet Berlin, Berlin, Germany}\and
\scriptsize{$^{83}$ Universidad Nacional Aut\'onoma de M\'exico, Instituto de Astronom\'ia, Ensenada, M\'exico}\and
\scriptsize{$^{84}$ Department of Astronomy, University of Maryland, College Park, USA}\and
\scriptsize{$^{85}$ NASA Goddard Space Flight Center, Greenbelt, USA}\and
\scriptsize{$^{86}$ Department of Earth, Atmospheric and Planetary Sciences, Massachusetts Institute of Technology, Cambridge, USA}\and
\scriptsize{$^{87}$ Department of Aeronautics and Astronautics, Massachusetts Institute of Technology, Cambridge, USA}\and
\scriptsize{$^{88}$ South African Astronomical Observatory, Cape Town, South Africa}\and
\scriptsize{$^{89}$ E\"otv\"os Lor\'and University, Gothard Astrophysical Observatory, Szombathely, Hungary}\and
\scriptsize{$^{90}$ Magyar Tudományos Akadémia-E\"otv\"os Lor\'and University Exoplanet Research Group, Szombathely, Hungary}\and
\scriptsize{$^{91}$ Space Science Data Center, Roma, Italy}\and
\scriptsize{$^{92}$ Istituto Nazionale di Astrofisica, Osservatorio Astronomico di Roma, Monte Porzio Catone, Italy}\and
\scriptsize{$^{93}$ Institute of Astronomy, University of Cambridge, Cambridge, UK}\and
\scriptsize{$^{94}$ Astrophysics Research Centre, School of Mathematics and Physics, Queen’s University Belfast, Belfast, UK}\and
\scriptsize{$^{95}$ Department of Astrophysical Sciences, Princeton University, Princeton, USA}\and
\scriptsize{$^{96}$ Società Astronomica Lunae, Castelnuovo Magra, Italy}\and
\scriptsize{$^{97}$ Department of Physics, Engineering and Astronomy, Stephen F. Austin State University, Nacogdoches, USA}\and
	\small$^\ast$Corresponding author. Email: thomas.g.wilson@warwick.ac.uk\and
}

\begin{document} 

\maketitle

\begin{abstract} \bfseries \boldmath
  Small exoplanet radii show two populations, referred to as super-Earths and sub-Neptunes, separated by a gap known as the radius valley. This may be produced by the removal of atmospheres due to stellar or internal heating, or lack of an initial envelope. We use transit photometry and radial velocity measurements to detect and characterize four planets orbiting LHS~1903, a red dwarf (M-dwarf) star in the Milky Way's thick disk. The planets have orbital periods between 2.2 and 29.3~days, and span the radius valley within a single planetary system. The derived densities indicate that LHS~1903~b is rocky, while LHS~1903~c and LHS~1903~d have extended atmospheres. Although the most distant planet from the host star, LHS~1903~e, has no gaseous envelope, indicating it formed from gas-depleted material.
\end{abstract}

\noindent
Planets form in protoplanetary disk environments and evolve under irradiation from their host stars or via internal mechanisms. Theoretical models of these processes can be built from first principles or constrained by comparing them to the demographics of observed exoplanets. Systems containing multiple planets provide further information because all planets formed together in the same disk and have been exposed to the same star throughout their lifetimes, eliminating factors (i.e. mass, temperature, metallicity) which vary between systems and complicate the interpretation of larger samples.

One such demographic trend is the radius valley\cite{Fulton2017,VanEylen2018,Cloutier2020a}, a dearth of planets with radii between 1.6 to 1.8\,Earth radius (R$_\oplus$), is thought to separate smaller, rocky (iron- and silicate-rich) planets from larger bodies that have water or gas layers\cite{Rogers2015,Zeng2019,Luque2022}. Planet formation\cite{Lopez2018,Venturini2020,Izidoro2021,Venturini2024} and evolution\cite{Owen2017,Ginzburg2018,Burn2024} theories have reproduced the radius valley for planets orbiting Sun-like stars\cite{VanEylen2018}. However, for red dwarf (M-dwarf) host stars, studies disagree on the properties or existence of a radius valley \cite{Hsu2020,VanEylen2021,Petigura2022,Bonfanti2024}. 

Proposed explanations of the radius valley include thermally-driven mass loss (T-DML) or gas-depleted formation (G-DF)\cite{Owen2017,Lopez2018,Ginzburg2018}. T-DML models predict that the radius valley is evolution-based caused by planet atmospheric escape due to stellar irradiation (via photo-evaporation)\cite{Owen2017} or due to internal planet heating (core-powered mass loss)\cite{Ginzburg2018}. G-DF theory explains the radius valley by a lack of material within the protoplanetary disk during formation\cite{Lopez2018}. Those mechanisms would produce different trends in planet composition (rocky, water, or gaseous) as a function of distance from the host star\cite{Lopez2018}. However, there are few known multi-planet M-dwarf systems with sufficiently long-period planets to test these theories. Most previously identified multi-planet systems orbiting M-dwarf stars do not have sufficiently precise radius and mass measurements\cite{Cloutier2020a,Otegi2020} to draw strong conclusions.

\subsubsection*{Identification of the LHS\,1903 planetary system}

LHS\,1903 (TOI-1730, Gaia\,DR3\,978086481343568128) is a M-dwarf star\cite{note:MM} that was photometrically observed by the Transiting Exoplanet Survey Satellite (TESS)\cite{Ricker2015} between 2019 and 2023 to obtain $\sim$69.8\,d of science data. Calibrated TESS images were corrected for non-astrophysical signals and processed into light curves that can be used to find planetary systems. Orbiting objects that pass between their host stars and our telescopes produce photometric dips called transits. We searched the TESS data for such signals and found three sets of periodic features that passed diagnostic tests\cite{note:MM}, as seen in a plot of flux as a function of time (Fig.~S1). 

These were ascribed to three transiting planets with planet radius ($R_{\rm p}$)\,$\sim$1.5, 2.0, and 2.5\,R$_\oplus$, with orbital periods of 2.16, 6.23, and 12.57\,day (d), respectively\cite{note:MM}. With these properties, these objects span the M-dwarf radius valley and thus offer a unique opportunity to test theoretical formation and evolution models. To determine the masses of the planets and characterize the system, we initiated observations of the LHS\,1903 planets using multiple facilities.

\subsubsection*{Follow-up observations of LHS\,1903}

We obtained space- and ground-based photometry from the Characterising Exoplanets Satellite (CHEOPS)\cite{Benz2021}, Las Cumbres Observatory Global Telescope (LCOGT)\cite{Brown2013}, Multicolor Simultaneous Camera for Studying Atmospheres of Transiting Exoplanets 2 (MuSCAT2)\cite{Narita2019} and 3 (MuSCAT3)\cite{Narita2020}, and Search and Characterisation of Transiting Exoplanets (SAINT-EX)\cite{Demory2020}. We also obtained high-resolution spectroscopy of the host star using the High Accuracy Radial Velocity Planet Searcher-North (HARPS-N)\cite{Cosentino2012} spectrograph. High angular resolution images were obtained using the adaptive optics instruments Palomar High Angular Resolution Observer (PHARO)\cite{Hayward2001} and Shane Adaptive Optics Infrared Camera-Spectrograph (ShARCS)\cite{Savel2020}. A summary of the observational datasets is provided in Tables~S1 and~S2, and Figs.~S2-S8. 

We corrected, calibrated, and processed the PHARO and ShARCS near-infrared time series data into stacked images for each instrument. By performing injection and recovery analyses, we converted these images into sensitivity curves that allowed us to rule out the presence of main sequence stellar companions 8.9\,au from LHS\,1903\cite{note:MM}. The lack of excess noise in Gaia astrometry rejects Jupiter mass planets and brown dwarfs out to 1 and 10\,au, respectively\cite{note:MM}. Therefore, by refuting binarity for LHS\,1903, we confirmed that the TESS transits were caused by orbiting planets (hereafter LHS\,1903\,b,\,c, and\,d) and that the host star is a single, old M-dwarf. 
We conducted spectral analysis on the HARPS-N data to determine the stellar effective temperature ($T_{\rm eff}$)\,=\,3664$\pm$70\,kelvin (K), stellar surface gravity ($\log{g}$)\,=\,4.75$\pm$0.12\,centimetre per square second (cm~s$^{-2}$), and stellar iron abundance ([Fe/H])\,=\,-0.11$\pm$0.09 values\cite{note:MM}. From broadband photometry and stellar atmospheric and evolutionary models, we executed infrared flux method and isochronal analyses to determine the stellar radius ($R_\star$)\,=\,0.539$\pm$0.014\,Solar radius (R$_\odot$), stellar mass ($M_\star$)\,=\,0.538$^{+0.039}_{-0.030}$\,Solar mass (M$_\odot$), stellar density ($\rho_\star$)\,=\,3.44$\pm$0.35\,Solar density ($\rho_\odot$) (4850$\pm$490\,kilogram per cubic metre (${\rm kg\,m^{-3}}$)), and stellar age ($t_\star$)\,=\,7.08$^{+2.87}_{-1.98}$\,gigayear (Gyr)\cite{note:MM}. The uncertainties listed here and hereafter are 1$\sigma$, unless stated otherwise. The measured properties of the host star are listed in Tables~1 and~S3. The kinematic properties of the host star were obtained by computing the Galactic velocities from Gaia astrometry and HARPS-N radial velocities (RV) and via comparison to known stellar populations, we found that LHS\,1903 is more likely than not (63$\%$ to 37$\%$) to be part of the thick disk compared to the thin disk\cite{note:MM} (Fig.~S9). The stellar position and velocity highlights that LHS\,1903 is in a low-density Galactic region typical of thick disk stars, and the stellar Galactic eccentricity is consistent with the thick disk population, but greater than 99.86$\%$ of thin disk stars.

Our CHEOPS photometric images were corrected for instrumental and environmental systematics with the $\sim$300.72\,hours (h) of scientific data extracted using the optimal aperture\cite{note:MM}. Moreover, because there is a background star close to the line of sight towards LHS\,1903, the re-derived CHEOPS light curves account for background-induced noise\cite{note:MM,Wilson2022}. These observations revealed 12, seven, and four transits of LHS\,1903\,b,\,c,and\,d, respectively. Furthermore, the LCOGT, MuSCAT2, MuSCAT3, and SAINT-EX data were processed via standard pipelines\cite{note:MM} with extracted differential photometry covering planetary transits of the LHS\,1903 system. Combining these light curves with the TESS transit data allow for accurate measurement of the physical and orbital properties of the LHS\,1903 planets. The CHEOPS observations showed additional transit-like features (Figs.~S2 and~S3). This prompted us to re-extract the TESS data\cite{note:MM} with a different data reduction pipeline (Fig.~S10, Table~S4). These data show two, three, and two further transits in the TESS, CHEOPS, and LCOGT photometry, all from the same object. After excluding potential false positives\cite{note:MM}, we concluded that these transits are due to a fourth planet (LHS\,1903\,e) on an orbital period of 29.32\,d. 

The HARPS-N spectra were reduced using the standard data reduction software with the optimal stellar cross correlation function mask\cite{note:MM} to produce stellar activity indicators. We utilised a template matching pipeline to optimise RV extraction and planetary mass measurement. Following quality control cuts, we obtained 91 high precision RVs spanning 769\,d. These data provide supporting evidence for the four LHS\,1903 planets (Fig.~S5) and show the stellar rotation period to be $\sim$40.8\,d\cite{note:MM}.

\subsubsection*{Physical properties of the planets}

To determine the properties of the LHS\,1903 planets, we analysed all the transit photometry and RV data simultaneously\cite{note:MM}. We fitted the data with a combined astrophysical model consisting of four transiting planets on Keplerian orbits, using the \textsc{juliet} software\cite{Espinoza2019}. This analysis constrained the planetary orbital period ($P$), transit centre time ($T_0$), planet-to-star radius ratio ($R_{\rm p}/R_\star$), transit impact parameter ($b$), eccentricity ($e$), argument of peristron ($\omega$) in degrees (deg), and RV semi-amplitude ($K$) in metre per second (${\rm m\,s}^{-1}$). We explored the parameter space using the nested sampling technique\cite{note:MM}. From the mass and radius of LHS\,1903 (see Table~1), the stellar density is used to constrain the transit model. We accounted for stellar variability or instrumental signals in the TESS and CHEOPS photometry, and HARPS-N RVs using a {\sc scalpels}\cite{CollierCameron2021}-derived linear noise model\cite{note:MM} and three Gaussian Process (GP) regressions. The assumed Bayesian prior probabilities and best-fitting model parameters of all data are listed in Tables~S5,~S6, and~S7. 

All four planets are detected with statistical significance of $\geq$40$\sigma$ in the transit photometry, and $\geq$4$\sigma$ in the RV data\cite{note:MM}. The model indicates that LHS\,1903\,b has a radius of 1.382$\pm$0.046\,R$_\oplus$, typical of super-Earth planets\cite{Rogers2015}, and an equilibrium temperature ($T_{\rm eq}$) of 796$\pm$20\,K. The corresponding values for planet\,c are 2.046$^{+0.078}_{-0.074}$\,R$_\oplus$, 559$\pm$14\,K, and for planet\,d are 2.500$^{+0.078}_{-0.077}$\,R$_\oplus$, 442$\pm$11\,K; both are sub-Neptune size. LHS\,1903\,e has a radius of 1.732$^{+0.059}_{-0.058}$\,R$_\oplus$, smaller than planets\,c and\,d, but larger than planet\,b. LHS\,1903\,e has an equilibrium temperature of 333$^{+9}_{-8}$\,K, which is colder than most known exoplanets\cite{Southworth2011}. The planet masses ($M_{\rm p}$) are constrained by the RVs from the HARPS-N data; for planets\,b to\,e they are 3.28$\pm$0.42, 4.55$^{+0.73}_{-0.69}$, 5.96$^{+1.15}_{-1.13}$, and 5.79$^{+1.60}_{-1.61}$\,Earth mass (M$_\oplus$), respectively. The fitted and derived stellar and planet parameters are listed in Table~1, including planet transit depth ($\delta_{\rm tr}$) in parts per million (ppm), stellar radius to semi-major axis ratio ($R_\star/a$), planet semi-major axis ($a$) in au, transit duration ($t_{14}$) in h, planet instellation ($S_{\rm p}$) in Earth instellation (S$_\oplus$), and planet surface gravity ($g_{\rm p}$) in metre per square second (${\rm m\,s}^{-2}$), with fitted limb-darkening and noise parameters shown in Tables~S6 and~S7. The fitted transit model is compared to the observed photometric data in Figs.~S1,~S2,~S3, and~S7. The combined fitted Keplerian orbits are compared to the HARPS-N RVs in Fig.~S4. The posterior probability distributions for the main model parameters are shown in Figs.~S11 to~S14. The phase-folded transit photometry and RVs, assuming the best-fitting orbital periods for all four planets, are presented in Figs.~1 and~2. Fig.~3 compares the masses, radii, and densities of the four planets orbiting LHS\,1903 to other precisely measured (planet radius uncertainty ($\sigma\,R_{\rm p}$)$\,<\,$5\% \& planet mass uncertainty ($\sigma\,M_{\rm p}$)$\,<\,$33\%) exoplanets orbiting M-dwarfs.

\subsubsection*{Comparison with radius valley models}

We compare the LHS\,1903 planets to theoretical predictions of the radius valley for exoplanets orbiting M-dwarf stars. The predictions of the T-DML and G-DF models vary as a function of stellar mass. More massive M-dwarfs are expected to have had higher dust masses in their protoplanetary disks\cite{Kokubo2006}, which would increase the maximum mass and radius of a rocky planet in the G-DF framework. In the T-DML model, lower mass M-dwarfs emit a greater fraction of their bolometric flux at X-ray plus extreme ultraviolet (XUV) wavelengths\cite{Shkolnik2014}, causing more rapid atmospheric loss from any orbiting planets. We corrected for these effects by scaling\cite{Wu2019} the radius valley predictions to the mass of LHS\,1903. We find that LHS\,1903\,b is located below the M-dwarf radius valley\cite{Cloutier2020a} predictions ($R_{\rm p, T-DML}$ for the T-DML model and $R_{\rm p, G-DF}$ for the G-DF model) for its orbital period ($R_{\rm p, T-DML}$\,=\,2.21\,R$_\oplus$, $R_{\rm p, G-DF}$\,=\,1.69\,R$_\oplus$). However, planets\,c and\,d have larger radii than the predictions ($R_{\rm p, T-DML}$\,=\,1.99\,R$_\oplus$, $R_{\rm p, G-DF}$\,=\,1.74\,R$_\oplus$, and $R_{\rm p, T-DML}$\,=\,1.84\,R$_\oplus$, $R_{\rm p, G-DF}$\,=\,1.78\,R$_\oplus$, respectively). With a radius of 1.732$^{+0.059}_{-0.058}$\,R$_\oplus$, LHS\,1903\,e lies between the predictions of the two models ($R_{\rm p, T-DML}$\,=\,1.65\,R$_\oplus$, $R_{\rm p, G-DF}$\,=\,1.83\,R$_\oplus$), see Fig.~4.

The four exoplanets orbiting LHS\,1903 span the M-dwarf radius valley within a single system, so can distinguish between these models. The T-DML\cite{Owen2017,Ginzburg2018} model, due to either photo-evaporation or core-powered mass loss, would predict gas-rich exoplanets at greater distances from the host star (longer orbital periods), and rocky planets closer to the star (shorter orbital periods). This would lead to a negative gradient in radius as a function of orbital period. Conversely, the G-DF\cite{Lopez2018} model produces larger rocky bodies at longer orbital periods, and smaller rocky planets closer to their host stars. In this model, the larger rocky bodies form at later times, after the majority of the gas in the protoplanetary disk has been dissipated by irradiation from the star\cite{Lee2014,Lee2016}. A similar process is invoked in formation models of the rocky planets in the Solar System, including Earth\cite{Raymond2009,Morbidelli2012}. This model predicts a higher dust-to-gas ratio at larger distances from the star, leading to a positive radius valley gradient.

Comparing these model predictions to the radius and orbital period of LHS\,1903\,e indicates that this planet is predicted to be gas-rich if formed by T-DML\cite{Owen2017,Ginzburg2018} and rocky if formed by G-DF\cite{Lopez2018} mechanisms. We therefore investigate the composition (gas-rich or rocky) of this object. As a first step, we computed the bulk densities ($\rho_{\rm p}$) of all four planets in the LHS\,1903 system (Fig.~3 and Table~1). We find that planet\,b has a density of 1.24$^{+0.21}_{-0.19}$\,Earth density ($\rho_\oplus$), equivalent to 6.82$^{+1.15}_{-1.04}$\,gram per cubic centimetre (${\rm g\,cm^{-3}}$), which is consistent with being rocky\cite{Rogers2015,Zeng2019}. This contradicts a previous proposal\cite{Kruijssen20} that stars in the Milky Way's thick disk probably do not host terrestrial planets, as does TOI-561\cite{Lacedelli2022}, another thick disk system containing a rocky planet. The densities of LHS\,1903\,c and\,d are 0.53$^{+0.11}_{-0.09}$ and 0.38$^{+0.09}_{-0.08}$\,$\rho_\oplus$ (2.91$^{+0.60}_{-0.52}$ and 2.09$^{+0.47}_{-0.43}$\,${\rm g\,cm^{-3}}$), which is consistent with a primary rocky body with a lower-density envelope, perhaps of water or hydrogen \& helium gas\cite{Zeng2019}. LHS\,1903\,e has a density of 1.11$^{+0.33}_{-0.31}$\,$\rho_\oplus$ (6.10$^{+1.83}_{-1.71}$\,${\rm g\,cm^{-3}}$), which like planet b is consistent with a lack of an extended gaseous atmosphere and being purely rocky. Fig.~S15 shows the densities of the LHS\,1903 planets, normalised by an Earth-like density, compared other planets that orbit M-dwarfs.

\subsubsection*{Modelling the LHS\,1903 planet interior compositions}

To provide a more robust estimate of the LHS\,1903 planet compositions, we conducted an interior structure analysis via comparison of their mass and radius measurements to interior structure models. We used the {\sc planetic}\cite{Leleu2021,Wilson2022,Lacedelli2022,Egger2024} which models planets as consisting of four layers: an iron core, silicate mantle, water, and a hydrogen \& helium atmosphere. We allowed the core and mantle mass fractions to vary between zero and one, and the water mass fraction between zero and 0.5\cite{note:MM}. This model does not include atmospheric water (including steam), due to a lack of observational atmospheric composition constraints of these planets. However, steam could be a notable atmospheric constituent for bodies hotter than 400\,K\cite{Aguichine2021,Burn2024,Venturini2024}, such as LHS\,1903\,b,\,c, and\,d. 

The resulting \textsc{planetic} model outputs have an increasing hydrogen \& helium gas mass fraction with increasing orbital distance for the three innermost planets (Fig.~S15A), from a gas-devoid planet\,b (hydrogen \& helium gas mass fraction $\sim$10$^{-9.5\pm1.2}$) to $\sim$10$^{-3.1\pm1.2}$ (equivalent to $\sim$0.1\%) for LHS\,1903\,c and $\sim$10$^{-1.4\pm0.2}$ (equivalent to $\sim$4\%) for LHS\,1903\,d. However this trend is not followed by the outermost LHS\,1903\,e, which is modelled as gas-depleted ($\sim$10$^{-7.7\pm2.1}$). Using the measured LHS\,1903 planet masses, the gas mass fractions translate into absolute gas masses of $\sim$10$^{-9.0\pm1.3}$, $\sim$10$^{-2.4\pm1.3}$, $\sim$10$^{-0.6\pm0.3}$, and $\sim$10$^{-6.9\pm2.2}$\,M$_\oplus$, respectively. The gas masses of LHS\,1903\,d and\,e therefore differ by $>$3$\sigma$. For all four planets, the water mass fractions are poorly constrained (Fig.~S15B).

\subsubsection*{Testing theory with planetary gas masses}

Our internal structure models of the three inner planets indicate that the gas-poor planet\,b and the gas-rich planets\,c and\,d are consistent with the radius valley predictions from both the T-DML and G-DF scenarios (Fig.~4). However LHS\,1903\,e falls between the two model predictions, so can distinguish between them. Our internal structure model indicates that planet\,e is gas-poor, so should fall below the predicted positions\cite{Rogers2015} of the radius valley in Fig.~4. To assess the compatibility of planet\,e's measured radius (Table~1) with the two theoretical predictions, we integrated the posterior probability distribution of its radius between zero and the radii predicted by the stellar mass-corrected G-DF and T-DML models at the planet's orbital period ($R_{\rm p, G-DF}$\,=\,1.83\,R$_\oplus$ and $R_{\rm p, T-DML}$\,=\,1.65\,R$_\oplus$, respectively). We find that 94\% of the posterior probability distribution is consistent with the G-DF prediction, and 9\% is consistent with the T-DML model. As a check, we also computed the probability\cite{Vysochanskij1980,Pukelsheim1994} that each of the two radius valley predictions is consistent with the true radius posterior probability distribution of LHS\,1903\,e. This metric indicates that, the G-DF prediction has a 100\% probability of being consistent with the radius of LHS\,1903\,e whereas the T-DML prediction consistency probability is $\sim$10\%. We therefore conclude that the bulk densities and compositions of the four planets orbiting LHS\,1903 are most likely explained by the G-DF formation scenario. Both the T-DML and G-DF models assume that only a hydrogen \& helium dominant atmosphere for a rocky planet; other atmospheric compositions are not considered. 

A possible interpretation of these conclusions is that the T-DML and G-DF mechanisms are dominant in different regimes (stellar irradiation, age)\cite{Kubyshkina2022} of a planetary system. Alternatively, the protoplanetary disk of LHS\,1903 might have had an outer edge that receded inwards during planet formation due to the inward radial drift of dust particles ($\sim$megayear (Myr) timescales)\cite{Pinilla2022}. Another possibility is that LHS\,1903\,e could have formed after the gas had dissipated from the protoplanetary disk, but planets\,c and\,d formed before gas dissipation\cite{Chatterjee2014}. 

Previous models of exoplanet atmospheric evolution via core-powered mass loss and photo-evaporation have shown that small planets with low equilibrium temperatures, such as LHS\,1903\,d and\,e, do not undergo substantial atmospheric mass loss\cite{Kubyshkina2022}. Other models that seek to reproduce the radius valley have argued that the gap separates rocky and water-rich planets\cite{Venturini2024}. In combination, these studies indicate that small, cold, high-density planets cannot be produced by the evolution of gaseous bodies, but must be formed rocky or water-rich without a substantial atmosphere. We therefore infer that the radius, density, and internal structure of planet\,e indicates that it formed water-rich in a gas-depleted environment\cite{Luque2022}. 

We tested the potential effect of thermally-driven mass loss on the evolution of the LHS\,1903 planetary system, using the Bayesian atmospheric evolution code {\sc pasta}\cite{note:MM,Bonfanti2021b}. {\sc pasta} models planetary atmospheric loss through internal heating (i.e. core-powered mass loss) and high energy stellar irradiation (i.e. photo-evaporation) over the lifetime of the system to compute the resulting gas mass fractions of planets. By comparing these values to the output of our internal structure model of the LHS\,1903 planets, we find that the current atmospheric gas mass fractions cannot be reproduced by the evolution processes implemented in {\sc pasta}. This supports our conclusion that the planets formed in a gas-depleted environment.

\subsubsection*{Trends within the multi-planet system}

A gas-depleted formation scenario for LHS\,1903\,e can explain the observed trends in radius and gas mass fraction with the system. The stellar gravitation influence on planets decreases with increasing orbital distance that results in larger exoplanet Hill radii. This means that exoplanets orbiting further from their host stars could attract more material. Planet formation theory predicts that the maximum planetesimal mass attained by accretion from the protoplanetary disk is greater at larger orbital distances, and that the critical mass required for planetesimal gas accretion is lower at greater distances\cite{Armitage2020}. This favours larger planets with higher gas mass fractions at longer orbital periods, as found in population synthesis studies\cite{Emsenhuber2021}. From our internal structure modelling, we find that the atmospheric mass of the outer planet\,e is $\sim$3$\sigma$ lower than the inner LHS\,1903\,d, contrary to the predictions of planet formation theory\cite{Armitage2020}. 

To quantify the trends within the LHS\,1903 system, we computed correlation metrics\cite{Mishra2023a} for the radii, masses, bulk densities, and gas mass fractions of the four planets. These correlation parameters evaluate the logarithmic average variation of each measured quantity over multiple planets. We find that the inner three planets of LHS\,1903 have an ordered architecture, as characterised by a coefficient of similarity ($C_{\rm s}$) greater than 0.2, with each parameter are increasing or decreasing uniformly with orbital period ($C_{\rm s}$ = 0.30, 0.30, 0.59, and 0.95, respectively). However LHS\,1903\,e does not follow these trends ($C_{\rm s}$ = 0.08, 0.19, 0.04, and 0.07, respectively), producing a mixed architecture.

This observed trend is not consistent with gas accretion during planet formation. It is possible that LHS\,1903\,e formed later than the other planets, in an evolved radially-truncated disk that had been depleted of its gas. Alternative planet formation models have proposed an inside-out mechanism\cite{Chatterjee2014} whereby the formation of an inner planet produces a local pressure minimum at the planet location. This induces an exterior higher pressure maximum, which subsequently forms another planet. This cycle continues with outer planets forming in sequence. If the first (interior) planet forms $\sim$1\,Myr after formation of the disk, then the repeating nature of this mechanism causes each exterior planet to form $\sim$1\,Myr after the previous planet\cite{Hu2018}. Protoplanetary disks around M-dwarf systems, such as LHS\,1903, have smaller radii than those around higher mass main sequence stars\cite{Andrews2018} so are predicted to have shorter timescales for dust to drift through the protoplanetary disk. For low-mass stars, simulations have predicted that the dust mass within the protoplanetary disk would decrease by 2 orders of magnitude after $\sim$1\,Myr and become fully-depleted after 3 to 4\,Myr, with the disk radial extent reducing by a factor of 3 to 5 over these timescales\cite{Pinilla2022}. Observations have shown that most protoplanetary disks around M-dwarfs dissipate after $\sim$5\,Myr\cite{Pfalzner2024}, which is consistent with the inside-out scenario.

\subsubsection*{Comparison to a planet formation simulation}

We compare the LHS\,1903 system to a M-dwarf planetary population synthesis analysis\cite{Burn2021} constructed from a framework of formation and evolution theory\cite{Emsenhuber2021}. In that simulation, planetesimals form and grow in a gas and dust coupled disk with the internal structure of the planets monitored during the accretion of dust, ice, and gas. The evolution of each planet is calculated including atmospheric escape and migration through gas-driven, tidal forces, and planet-planet scattering modelled by a N-body code that also includes impacts. This includes host star evolution, over the time scale of of the simulation (5\,Gyr). We selected simulated planets with radii, masses, semi-major axes, and gas-mass fractions within 25\% or 3$\sigma$ (whichever was larger)\cite{UlmerMoll2023,Egger2024} of the LHS\,1903 planets. We found 45, 72, 61, and 56 simulated planets that are similar to LHS\,1903\,b,\,c,\,d, and\,e, respectively.

For each of these planet samples, we take the average of the gas and dust disk masses, the gas disk outer radii, and the formation location. This determines the average formation environment of each simulated planet sample that represents the formation environment of the LHS\,1903 planets. We find consistent parameters for the LHS\,1903\,b-like and\,e-like planets: gas disk masses $\sim$0.02\,M$_\odot$, dust disk masses $\sim$100\,M$_\oplus$, gas disk outer radii $\sim$200\,astronomical unit (au), although LHS 1903 e-like planets formed at greater semi-major axes in the simulations. However, the LHS\,1903\,c- and\,d-like simulated planets must have formed in a more extensive and massive dust and gas disk to produce their low densities and inferred higher gas-mass fractions, when compared to the rocky, gas-poor LHS\,1903\,b and\,e. The simulation study\cite{Burn2021} predicted that planets with radii, masses, and semi-major axes similar to LHS\,1903\,e, but with larger gas-mass fractions (i.e. akin to LHS\,1903\,c and\,d) must form in more massive and radially extended gas and dust disks. We therefore suggest that LHS\,1903\,b and\,e might have formed in a lower-mass region of the protoplanetary disk than LHS\,1903\,c and\,d.

For LHS\,1903\,b, this could be because planets at smaller orbital distances have less material to accrete, and therefore lower maximum masses, compared to more distant planets that have a larger reservoir of material\cite{Armitage2020}.
However the opposite effect would occur for LHS\,1903\,e, so it might instead have formed later\cite{Chatterjee2014}, after the protoplanetary disk has experienced substantial material drift\cite{Andrews2018}, resulting in formation in a gas-depleted environment. The rocky Solar System planets formed 10\,Myr after the disk formed, in a gas-depleted environment\cite{Raymond2009,Morbidelli2012}. For LHS\,1903, this implies that LHS\,1903\,c and\,d formed earlier, in gas-rich environments. The LHS\,1903 planets might have migrated from their formation sites to the current orbits, but LHS\,1903\,d and\,e probably experienced similar of thermal evolution due to their similar masses and low equilibrium temperatures, so the observed gas mass fractions of LHS\,1903\,d and\,e are probably unchanged since their formation\cite{Kubyshkina2022}.

\subsubsection*{Dynamical history of the LHS\,1903 system}

Dynamical histories of planetary systems may affect planet properties during formation and early evolution. The orbital period ratios of the LHS\,1903 planets are close to mean motion resonances (MMRs): at 2:1 for planets\,c and\,d, and at 7:3 for planets\,d and\,e. The numerical analysis of fundamental frequencies method\cite{Laskar1990,note:MM} shows that planets\,c and\,d are not in MMR, but planets\,d and\,e might be in the 7:3 MMR. Given this multi-planet architecture, we assess the orbital evolution of the planets\cite{note:MM} and find that the LHS\,1903 system is dynamically stable against destructive excursions over long timescales. 

Impacts within planetary systems might remove gas from the planet atmospheres. However, simulations have shown that impacts with the required energy also cause catastrophic disruption of the planet\cite{Denman2020,Denman2022}. This is because cold, small planets, such as the LHS\,1903 bodies, have high escape velocities\cite{Walker1986} and low initial atmospheric thermal excitation that inhibits atmospheric expansion and therefore loss\cite{Biersteker2019}. Therefore, an impact that would overcome these limitations would likely disrupt the planet. If collisions could remove atmospheres, inner planets would undergo more impacts, due to their shorter orbital periods and therefore more orbit crossing events. We calculated the planet-impactor collision probability using previous methods\cite{JeongAhn2017} with the stellar and planetary properties of the LHS\,1903 system, and find impact probabilities over the system age of 0.37\%, 0.19\%, 0.10\%, and 0.02\% for LHS\,1903\,b,\,c,\,d, and\,e respectively. We conclude that LHS\,1903\,e is unlikely to have experienced a sufficient impact to cause its atmosphere to be stripped. 

\subsubsection*{Summary and conclusions}

The four planets in the LHS\,1903 system do not follow simple trends in their radii and gas masses with orbital distance. LHS\,1903\,e is a gas-depleted long-period planet orbiting a thick disk star, which are rare in previous observations\cite{Kruijssen20}. The properties of these planets within a single dynamically stable system cannot be explained by atmospheric evolution or planetary impacts alone, but instead indicate that at least one of the planets formed in a gas-depleted region of the protoplanetary disk. Therefore, the gas-depleted formation mechanism is the most likely scenario that can result in the innermost rocky planet LHS\,1903\,b, whilst preserving the gas-rich nature of planets LHS\,1903\,c and\,d, and yielding an outer gas-poor planet, LHS\,1903\,e.


\begin{figure}[htbp]
\centerline{\includegraphics[width=\columnwidth]{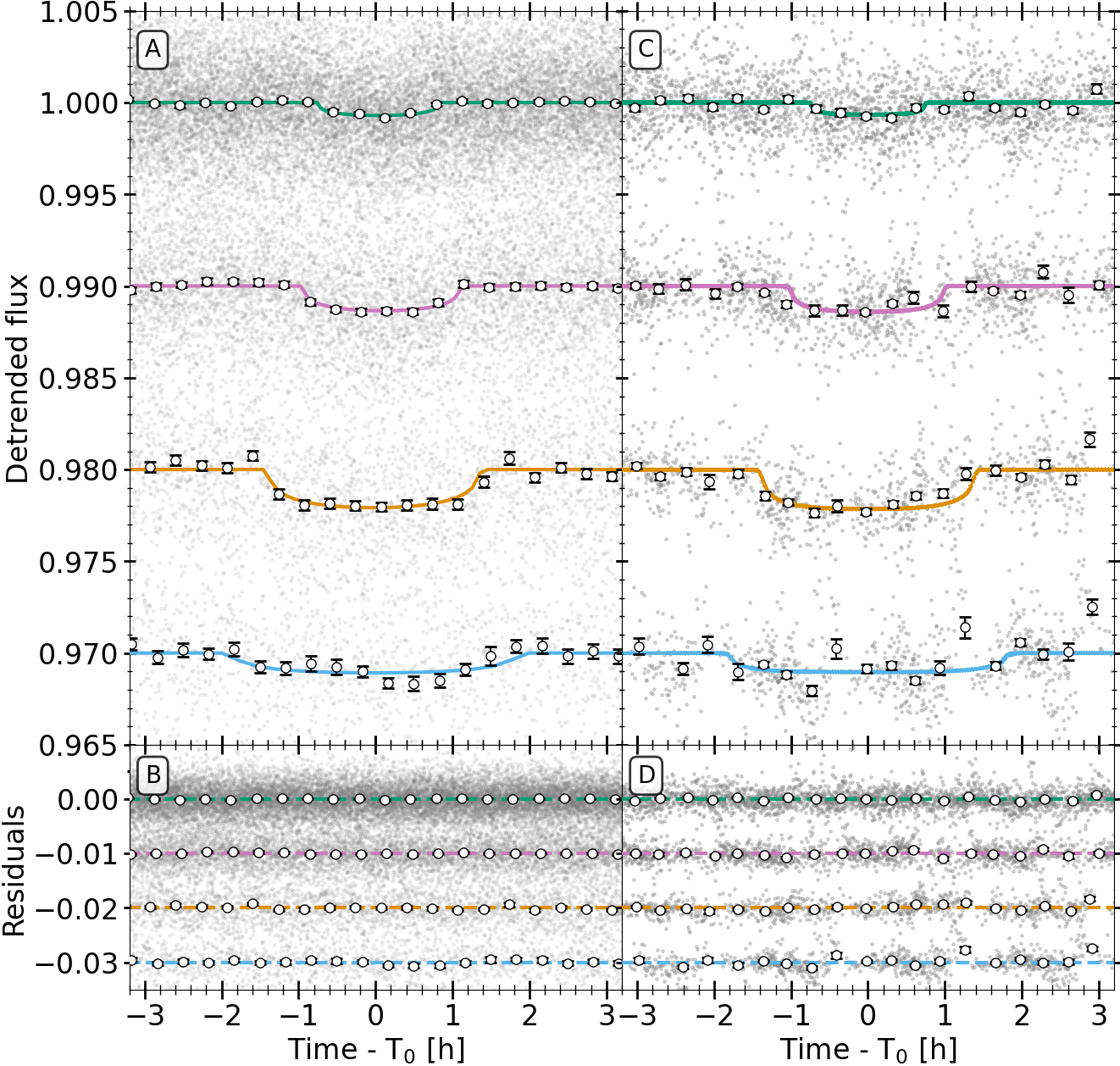}}
\noindent
\begin{small}{\bf Fig. 1: Detrended phase-folded transit photometry of the LHS\,1903\, planets}. TESS and CHEOPS data phase-folded to the orbital periods from our global analysis\cite{note:MM} for the four LHS\,1903 planets; b is green, c is purple, d is orange, and e is cyan, vertically offset by 0.01 for clarity. ({\bf A \& C}) TESS and CHEOPS individual observations are shown as grey points, open circles show the same data binned every 20\,minutes (error bars are 1$\sigma$ uncertainties), and solid lines are the transit models fitted to the data. ({\bf B \& D}) TESS and CHEOPS residuals between the data and the model.\end{small} 
\end{figure}

\begin{figure}[htbp]
    {{\includegraphics[width=0.5\columnwidth]{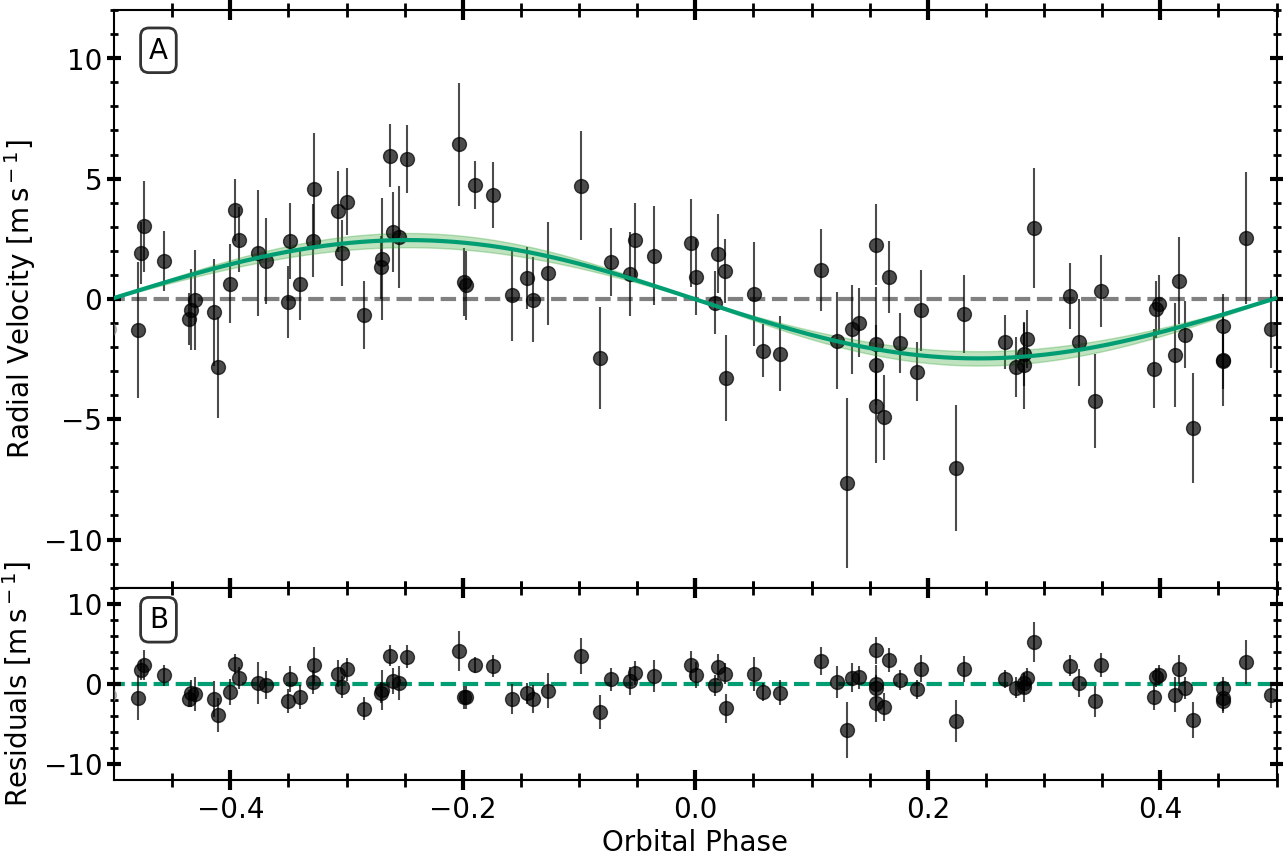} }}\hspace{-0.75cm}\vspace{0.2cm}
    \qquad
    {{\includegraphics[width=0.5\columnwidth]{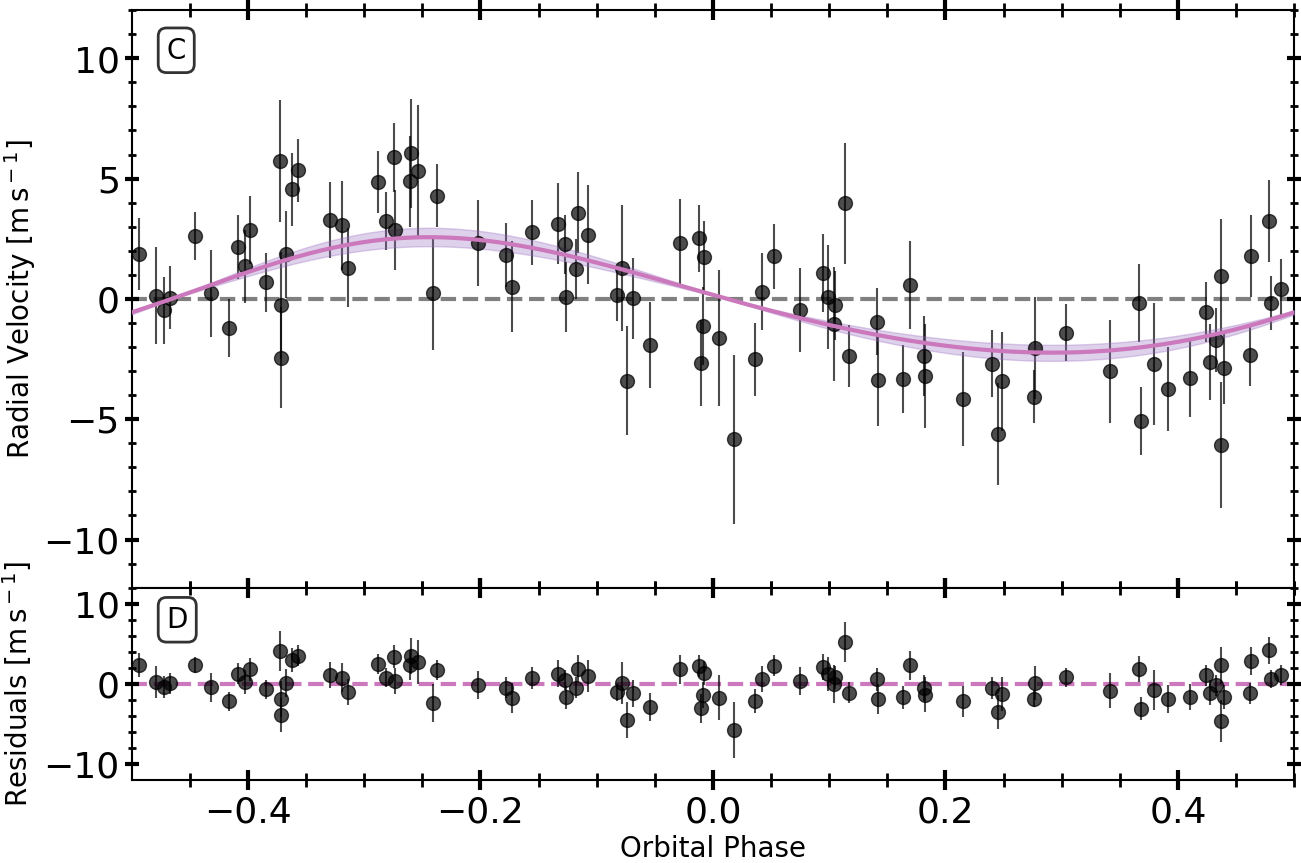} }}\vspace{0cm}\hspace{-0.9cm}
    \qquad
    {{\includegraphics[width=0.5\columnwidth]{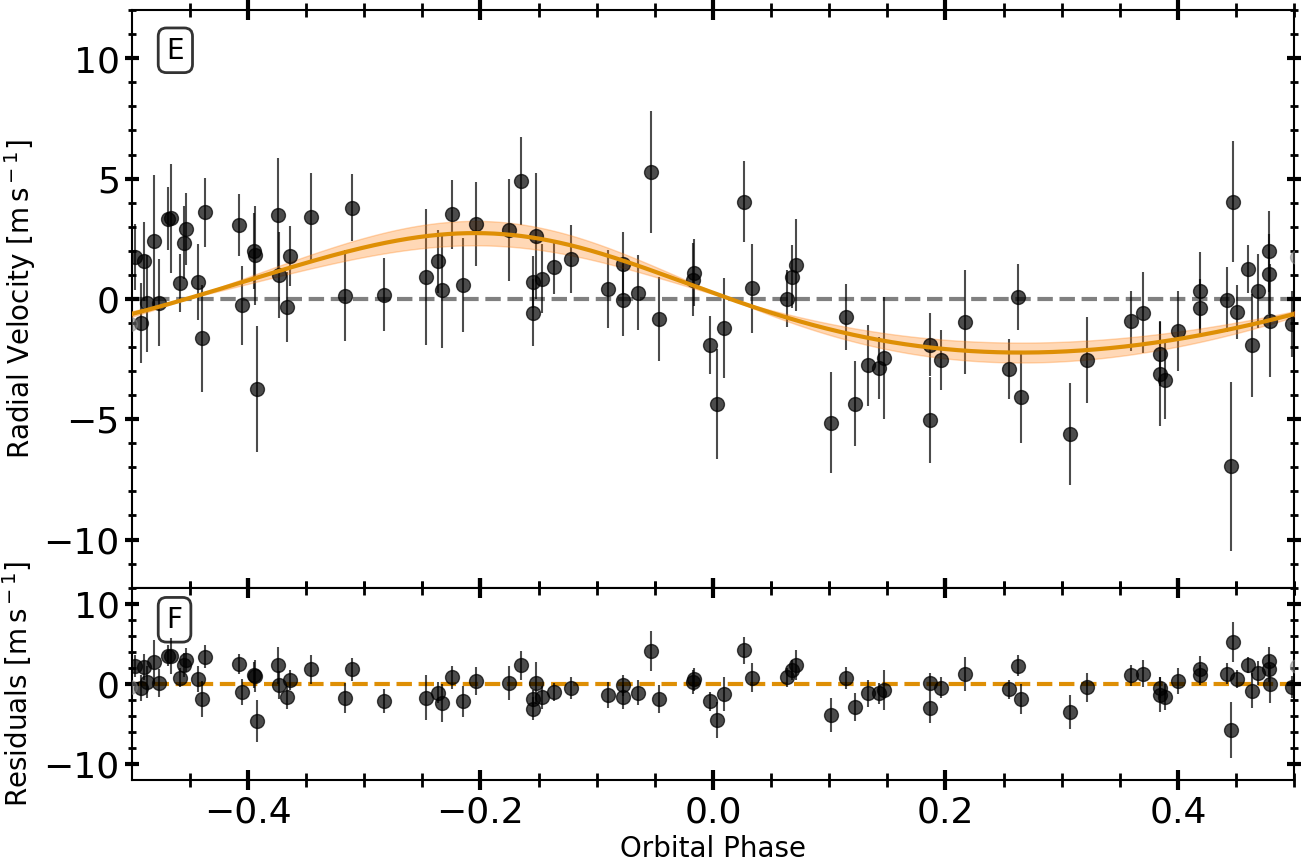} }}\hspace{-0.75cm}
    \qquad
    {{\includegraphics[width=0.5\columnwidth]{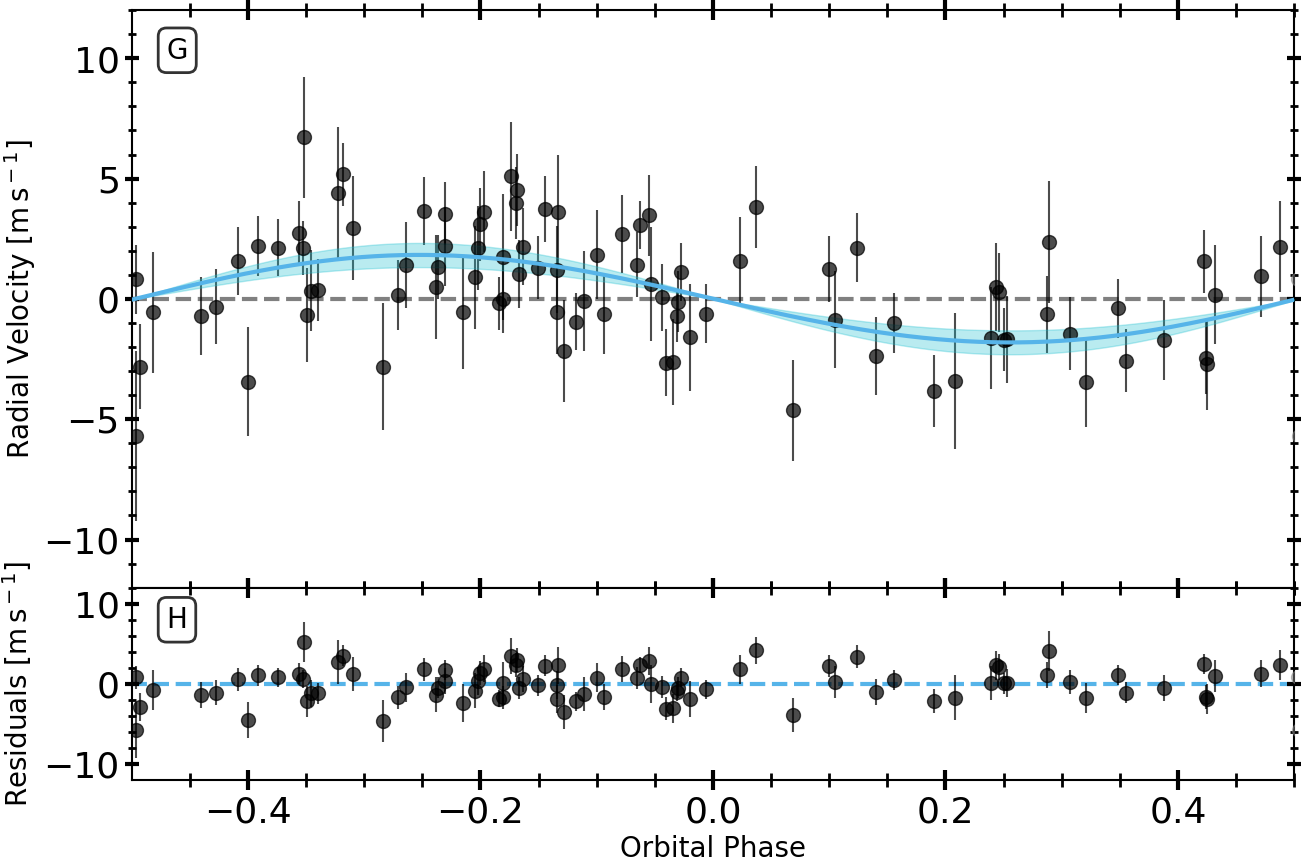} }}\vspace{0cm}
\noindent
\begin{small}{\bf Fig. 2: Detrended phase-folded radial velocity measurements for the LHS\,1903\, planets}. ({\bf A, C, E, G}) Detrended HARPS-N RV data for LHS\,1903\,b,\,c,\,d, and\,e, respectively, as black data points, error bars are 1$\sigma$. The x-axis has been phase-folded to each planets' orbital period (Table~1), determined from the joint fitting\cite{note:MM}. Coloured lines are Keplerian orbit models fitted to the data, with shaded regions indicating the 1$\sigma$ uncertainty. ({\bf B, D, F, H}) Residuals between the data and the model for LHS\,1903\,b,\,c,\,d, and\,e, respectively.\end{small} 
\end{figure} 

\begin{figure}[htbp]
    \centerline{\includegraphics[width=0.78\columnwidth]{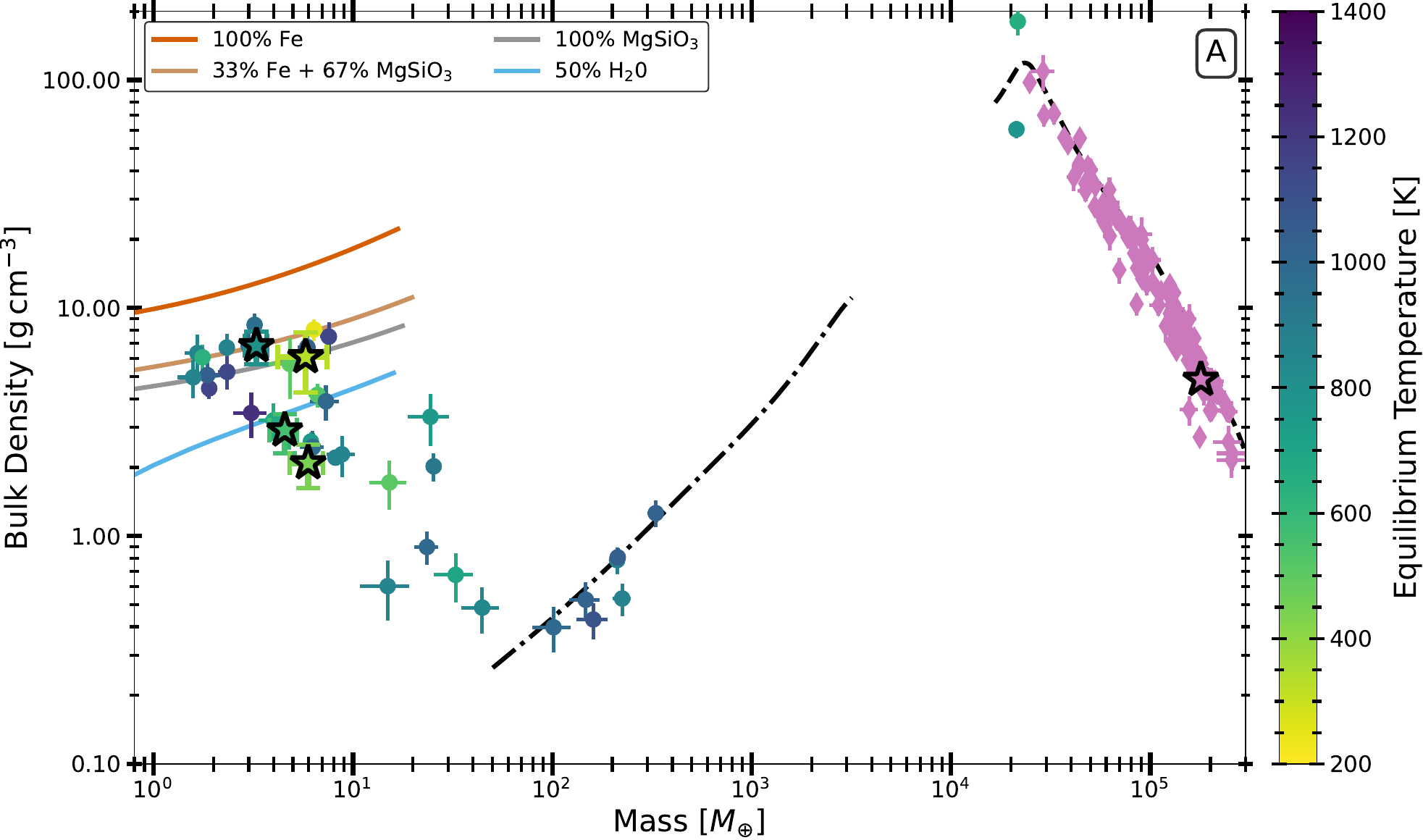}}
    \centerline{\includegraphics[width=0.75\columnwidth]{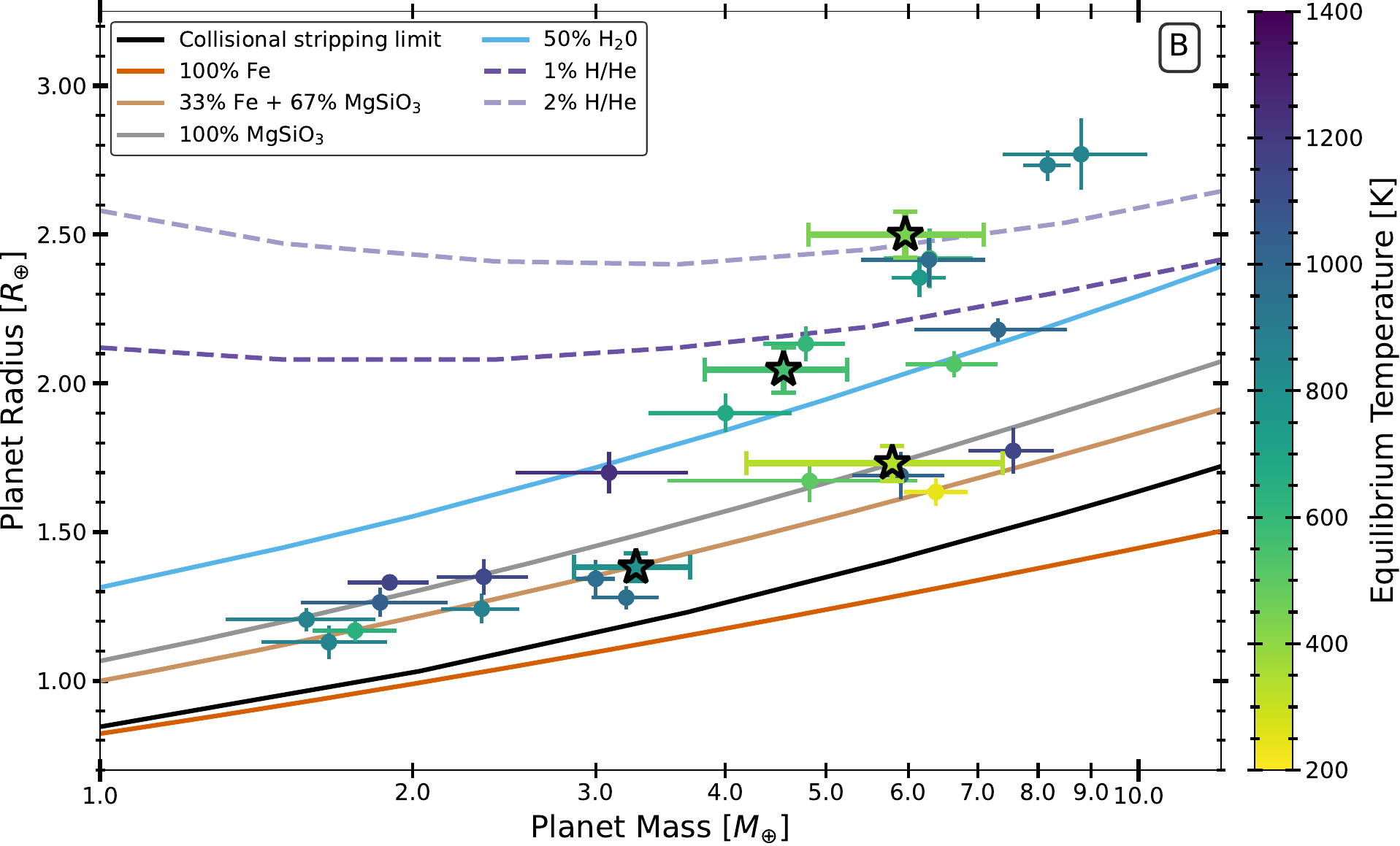}}
\noindent
\begin{small}{\bf Fig. 3: Planet masses as a function of bulk densities and radii for selected planets orbiting M-dwarf stars}. ({\bf A}) Circular data points are well-characterised planets ($\sigma\,R_{\rm p}\,<\,$5\% \& $\sigma\,M_{\rm p}\,<\,$33\%)\cite{Southworth2011} orbiting M-dwarf stars, coloured by their equilibrium temperatures. Purple diamonds are the same quantities for M-dwarf stars\cite{Parsons2018}. Star symbols with black outlines are the four planets and host star in the LHS\,1903 system. Coloured lines are theoretical mass-density curves for small planets (solid lines, see legend)\cite{Zeng2013,Lopez2014}, and giant planets (dot-dashed line)\cite{Baraffe2008}, and M-dwarf stars (black dashed line)\cite{Baraffe2015}. ({\bf B}) The masses and radii of small ($R_{\rm p} <$ 3\,R$_\oplus$) planets orbiting M-dwarf stars, coloured by their equilibrium temperatures. Coloured lines are theoretical mass-radius relations for gas-poor (solid) and gas-rich (dashed)\cite{Zeng2013,Lopez2014} planets. The black curve is the collisional stripping limit\cite{Marcus2010}. The theoretical models provide general guidance but do not determine the specific composition of each individual system due to difference in stellar parameters, irradiation, etc.\end{small}
\end{figure} 

\begin{figure}[htbp]
\centerline{\includegraphics[width=\columnwidth]{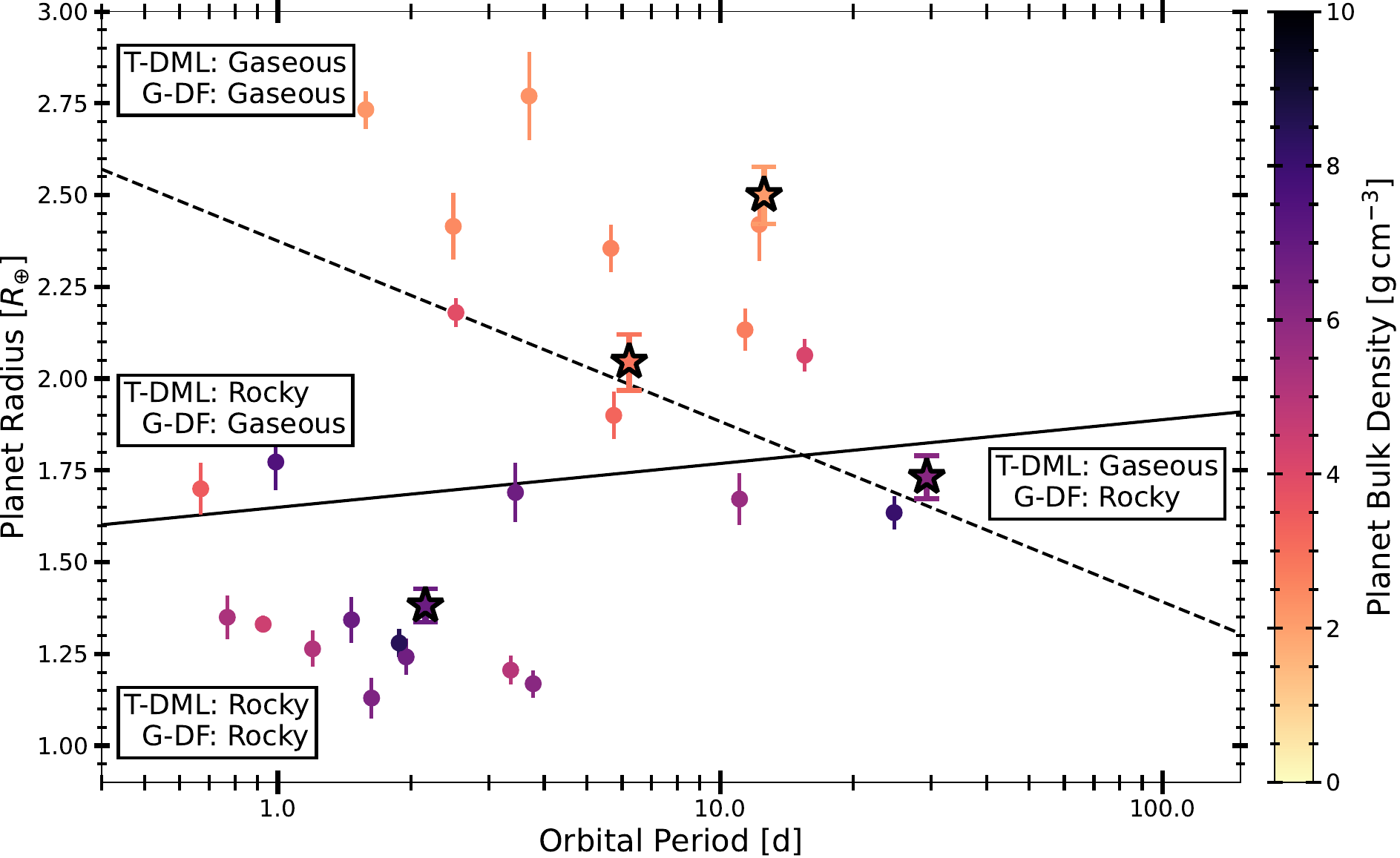}}
\noindent
\begin{small}{\bf Fig. 4: Orbital periods as a function of planet radii for selected planets orbiting M-dwarf stars}. Data points are planets with precise radius and mass measurements ($\sigma\,R_{\rm p}\,<\,$5\% \& $\sigma\,M_{\rm p}\,<\,$33\%), as a function of radius and coloured by their bulk densities.  Error bars are 1$\sigma$ uncertainties. Star symbols indicate the LHS\,1903 planets. The black lines show the LHS\,1903 stellar mass-corrected predictions\cite{Wu2019} for gas-depleted formation (G-DF, solid line)\cite{Lopez2018} and thermally-driven mass loss (T-DML, dashed line)\cite{Owen2017,Ginzburg2018} models of the radius valley\cite{VanEylen2018,Martinez2019,Cloutier2020a,Ho2023}. Planets above or below each line are predicted to be gaseous or rocky by that model, as labelled. The LHS\,1903 planets are located in three different regions. The mass and radius of LHS\,1903\,e indicate that it is rocky (see text), which is contrary to the T-DML prediction but consistent with the G-DF prediction of the radius valley. 
\end{small}
\end{figure}


\begin{table}
\begin{footnotesize}
\begin{small}{\bf Table 1: Stellar and planetary properties of the LHS\,1903 system}. Stellar parameters were determined from the spectroscopy\cite{note:MM}. Fitted and derived stellar and planet parameters were determined by joint fitting of the transit photometry and RV data\cite{note:MM}. Values and uncertainties are the median and 16th/84th percentiles of the posterior probability distribution. Times are given in Barycentric Julian Date (BJD).\end{small}
\vspace{-0.35cm}
\begin{center}
\label{tab:posterior_trrv}
\begin{tabular}{c c c c c}

\hline\hline                 
Parameter (unit) & \multicolumn{3}{c}{Value} \\ 
\hline
\multicolumn{5}{c}{LHS\,1903} \\
\hline
\multicolumn{5}{c}{\textit{Spectroscopic stellar parameters}} \\ 
\hline 
\vspace{-0.06cm}
$T_{\rm eff}$ (K) & \multicolumn{3}{c}{3664$\pm$70} & \\
\vspace{-0.06cm}
$\log{g}$ (cm~s$^{-2}$)  & \multicolumn{3}{c}{4.75$\pm$0.12} & \\
\vspace{-0.06cm}
[Fe/H] (dex) & \multicolumn{3}{c}{-0.11$\pm$0.09} & \\
\vspace{-0.06cm}
$R_\star$ (R$_\odot$) & \multicolumn{3}{c}{0.539$\pm$0.014} & \\
\vspace{-0.06cm}
$M_\star$ (M$_\odot$) & \multicolumn{3}{c}{0.538$^{+0.039}_{-0.030}$} & \\
\vspace{-0.06cm}
$t_\star$ (Gyr)        & \multicolumn{3}{c}{7.08$^{+2.87}_{-1.98}$} & \\
\vspace{-0.06cm}
$\rho_\star$ ($\rho_\odot$) & \multicolumn{3}{c}{3.44$\pm$0.35} & \\
$\rho_\star$ (${\rm kg\,m^{-3}}$) & \multicolumn{3}{c}{4850$\pm$490} & \\
\hline             
\multicolumn{5}{c}{\textit{Fitted stellar parameters}} \\ 
\hline 
\vspace{-0.06cm}
$\rho_\star$ ($\mathrm{\rho_\odot}$) & \multicolumn{3}{c}{3.79$^{+0.17}_{-0.14}$} & \\ 
$\rho_\star$ (${\rm kg\,m^{-3}}$) & \multicolumn{3}{c}{5340$^{+240}_{-190}$} & \\
\hline 
 & LHS\,1903\,b & LHS\,1903\,c & LHS\,1903\,d & LHS\,1903\,e  \\ 
\hline 
\multicolumn{5}{c}{\textit{Fitted planet parameters}} \\ 
\hline 
\vspace{-0.06cm}
$P$ (d) & 2.1555098$^{+0.0000026}_{-0.0000029}$ & 6.226185$^{+0.000028}_{-0.000026}$ & 12.566287$^{+0.000032}_{-0.000028}$ & 29.31773$^{+0.00028}_{-0.00025}$ \\ 
\vspace{-0.06cm}
$T_{0}$ (BJD-2457000) & 1844.5771$^{+0.0012}_{-0.0009}$ & 1844.3692$^{+0.0026}_{-0.0021}$ & 1844.3901$^{+0.0019}_{-0.0021}$ & 1868.8152$^{+0.0071}_{-0.0083}$ \\ 
\vspace{-0.06cm}
$R_{\rm p}/R_\star$ & 0.02351$^{+0.00047}_{-0.00049}$ & 0.03479$^{+0.00097}_{-0.00088}$ & 0.04253$^{+0.00072}_{-0.00071}$ & 0.02947$\pm$0.00062 \\ 
\vspace{-0.06cm}
$b$ & 0.174$^{+0.088}_{-0.091}$ & 0.513$^{+0.088}_{-0.093}$ & 0.329$^{+0.096}_{-0.112}$ & 0.120$^{+0.094}_{-0.079}$ \\ 
\vspace{-0.06cm}
$e$ & 0.015$^{+0.014}_{-0.010}$ & 0.089$^{+0.036}_{-0.030}$ & 0.112$^{+0.055}_{-0.044}$ & 0.014$^{+0.015}_{-0.010}$ \\ 
\vspace{-0.06cm}
$\omega$ (deg) & 216$^{+80}_{-87}$ & 288$^{+38}_{-31}$ & 233$^{+29}_{-19}$ & 263$^{+63}_{-76}$ \\ 
$K$ (${\rm m\,s}^{-1}$) & 2.46$\pm$0.30 & 2.40$^{+0.37}_{-0.35}$ & 2.48$\pm$0.46 & 1.82$\pm$0.50 \\  
\hline 
\multicolumn{5}{c}{\textit{Derived planet parameters}} \\ 
\hline 
\vspace{-0.06cm}
$\delta_{\rm tr}$ (ppm) & 553$^{+22}_{-23}$ & 1210$^{+69}_{-60}$ & 1808$^{+62}_{-60}$ & 868$^{+37}_{-36}$ \\ 
\vspace{-0.06cm}
$R_\star/a$ & 0.09441$^{+0.00320}_{-0.00312}$ & 0.04655$^{+0.00158}_{-0.00154}$ & 0.02914$^{+0.00099}_{-0.00096}$ & 0.01657$^{+0.00056}_{-0.00055}$ \\ 
\vspace{-0.06cm}
\vspace{-0.06cm}
$R_{\rm p}$ (R$_\oplus$) & 1.382$\pm$0.046 & 2.046$^{+0.078}_{-0.074}$ & 2.500$^{+0.078}_{-0.077}$ & 1.732$^{+0.059}_{-0.058}$ \\ 
\vspace{-0.06cm}
$a$ (au) & 0.02656$^{+0.00055}_{-0.00058}$ & 0.05387$^{+0.00112}_{-0.00117}$ & 0.08604$^{+0.00178}_{-0.00186}$ & 0.15135$^{+0.00314}_{-0.00338}$ \\ 
\vspace{-0.06cm}
$t_{14}$ (h) & 1.562$^{+0.058}_{-0.057}$ & 1.986$^{+0.125}_{-0.134}$ & 2.762$^{+0.127}_{-0.134}$ & 3.780$^{+0.134}_{-0.133}$ \\ 
\vspace{-0.06cm}
$S_{\rm p}$ (S$_\oplus$) & 66.66$^{+7.08}_{-6.46}$ & 16.20$^{+1.72}_{-1.57}$ & 6.35$^{+0.67}_{-0.62}$ & 2.05$^{+0.22}_{-0.20}$ \\ 
\vspace{-0.06cm}
$T_{\rm eq}$ (K)$^{\rm *}$ & 796$\pm$20 & 559$\pm$14 & 442$\pm$11 & 333$^{+9}_{-8}$  \\ 
\vspace{-0.06cm}
$M_\mathrm{p}$ (M$_\oplus$) & 3.28$\pm$0.42 & 4.55$^{+0.73}_{-0.69}$ & 5.96$^{+1.15}_{-1.13}$ & 5.79$^{+1.60}_{-1.61}$ \\ 
\vspace{-0.06cm}
$\rho_{\rm p}$ ($\rho_\oplus$) & 1.24$^{+0.21}_{-0.19}$ & 0.53$^{+0.11}_{-0.09}$ & 0.38$^{+0.09}_{-0.08}$ & 1.11$^{+0.33}_{-0.31}$ \\ 
\vspace{-0.06cm}
$\rho_{\rm p}$ (${\rm g\,cm^{-3}}$) & 6.82$^{+1.15}_{-1.04}$ & 2.91$^{+0.60}_{-0.52}$ & 2.09$^{+0.47}_{-0.43}$ & 6.10$^{+1.83}_{-1.71}$ \\ 
$g_\mathrm{p}$ (${\rm m\,s}^{-2}$) & 16.8$\pm$2.4 & 10.7$\pm$1.8 & 9.3$\pm$1.9 & 18.9$\pm$5.4 \\ 

\hline\hline    
\end{tabular}
\end{center}
\vspace{-0.35cm}
$^{\rm *}$ Computed assuming zero Bond albedos and perfect energy redistribution\cite{note:MM}.
\end{footnotesize}
\end{table}


\clearpage 

%
\bibliography{science_template} 
\bibliographystyle{sciencemag}

%
%
%
%
%
%


\section*{Acknowledgments}
T.G.W thanks Farzana Meru for useful discussions on planet formation within protoplanetary disks.
CHEOPS is an ESA mission in partnership with Switzerland with important contributions to the payload and the ground segment from Austria, Belgium, France, Germany, Hungary, Italy, Portugal, Spain, Sweden, and the United Kingdom. The CHEOPS Consortium would like to gratefully acknowledge the support received by all the agencies, offices, universities, and industries involved. Their flexibility and willingness to explore new approaches were essential to the success of this mission. This work is based on observations made with the Italian Telescopio Nazionale Galileo (TNG) operated on the island of La Palma by the Fundación Galileo Galilei of the INAF at the Spanish Observatorio del Roque de los Muchachos of the Instituto de Astrofisica de Canarias (GTO programme). The HARPS-N project was funded by the Prodex Program of the Swiss Space Office (SSO), the Harvard-University Origin of Life Initiative (HUOLI), the Scottish Universities Physics Alliance (SUPA), the University of Geneva, the Smithsonian Astrophysical Observatory (SAO), the Italian National Astrophysical Institute (INAF), the University of St. Andrews, Queen’s University Belfast and the University of Edinburgh. Funding for the TESS mission is provided by NASA's Science Mission Directorate. We acknowledge the use of public TESS data from pipelines at the TESS Science Office and at the TESS Science Processing Operations Center. This research has made use of the Exoplanet Follow-up Observation Program website, which is operated by the California Institute of Technology, under contract with the National Aeronautics and Space Administration under the Exoplanet Exploration Program. Resources supporting this work were provided by the NASA High-End Computing (HEC) Program through the NASA Advanced Supercomputing (NAS) Division at Ames Research Center for the production of the SPOC data products. This work makes use of observations from the LCOGT network. Part of the LCOGT telescope time was granted by NOIRLab through the Mid-Scale Innovations Program (MSIP). MSIP is funded by NSF. This article is based on observations made with the MuSCAT2 instrument, developed by ABC, at Telescopio Carlos Sánchez operated on the island of Tenerife by the IAC in the Spanish Observatorio del Teide. This paper is based on observations made with the MuSCAT3 instrument, developed by the Astrobiology Center and under financial supports by JSPS KAKENHI (JP18H05439) and JST PRESTO (JPMJPR1775), at Faulkes Telescope North on Maui, HI, operated by the Las Cumbres Observatory. This work has made use of data from the European Space Agency (ESA) mission Gaia (https://www.cosmos.esa.int/gaia), processed by the Gaia Data Processing and Analysis Consortium (DPAC, https://www.cosmos.esa.int/web/gaia/dpac/consortium). Funding for the DPAC has been provided by national institutions, in particular the institutions participating in the Gaia Multilateral Agreement. This work is based upon observations carried out at the Observatorio Astron\'{o}mico Nacional on the Sierra de San Pedro M\'{a}rtir (OAN-SPM), Baja California, M\'{e}xico. SAINT-EX observations and team were supported by the Swiss National Science Foundation (PP00P2-163967 and PP00P2-190080), the Centre for Space and Habitability (CSH) of the University of Bern, the National Centre for Competence in Research PlanetS, supported by theSwiss National Science Foundation (SNSF), and UNAM PAPIIT-IG101321 and PAPIIT-IG101224. 

\paragraph*{Funding:}
T.G.W. acknowledges support from STFC consolidated grant numbers ST/R000824/1 and ST/V000861/1, and UKSA grant ST/R003203/1, and the University of Warwick and UKSA. 
A.C.C. and K.H. acknowledge support from STFC consolidated grant numbers ST/R000824/1 and ST/V000861/1, and UKSA grant ST/R003203/1. 
A.M.S. acknowledges support from the SAO REU program which is funded in part by the National Science Foundation REU and Department of Defense ASSURE programs under NSF Grant no.\ AST 1852268 and 2050813, and by the Smithsonian Institution. 
V.A. is supported by Funda\c{c}\~ao para a Ci\^encia e Tecnologia (FCT), Portugal, through national funds by grants (UIDB/04434/2020 and UIDP/04434/2020) and work contract through the Scientific Employment Incentive program with reference 2023.06055.CEECIND/CP2839/CT0005.
Y.A. acknowledges the support of the Swiss National Fund under grant 200020\_172746. 
M.S. acknowledges financial support from the Belgian Federal Science Policy Office (BELSPO), in the framework of the PRODEX Programme of the European Space Agency (ESA) under contract number C4000140754.
J.A.E. acknowledges the support of the Swiss National Fund under grant 200020\_172746. 
P.M. acknowledges support from STFC research grant number ST/M001040/1. 
S.G.S. acknowledge support from FCT through FCT contract nr. CEECIND/00826/2018 and POPH/FSE (EC). 
M.F. and C.M.P. gratefully acknowledge the support of the Swedish National Space Agency (DNR 65/19, 174/18). 
S.S. have received funding from the European Research Council (ERC) under the European Union’s Horizon 2020 research and innovation program (grant agreement No 833925, project STAREX). 
V.V.G. is an F.R.S-FNRS Research Associate. 
A.J.M. acknowledges the support of the Swedish National Space Agency (Career grants 120/19C and 2023-00146). 
This work has been carried out within the framework of the NCCR PlanetS supported by the Swiss National Science Foundation under grants 51NF40\_182901 and 51NF40\_205606. 
This project has received funding from the European Research Council (ERC) under the European Union’s Horizon 2020 research and innovation programme (project GPRV. grant agreement No 865624).
L.B., G.B., V.N., I.P., G.P., R.R., G.S., V.S., and T.Z. acknowledge support from CHEOPS ASI-INAF agreement n. 2019-29-HH.0. 
We acknowledge support from the Spanish Ministry of Science and Innovation/State Agency of Research MCIN/AEI/ 10.13039/501100011033  and the European Regional Development Fund through grants ESP2016-80435-C2-1-R, ESP2016-80435-C2-2-R, PGC2018-098153-B-C33, PGC2018-098153-B-C31, ESP2017-87676-C5-1-R, PID2019-107061GB-C61 and PID2023-150468NB-I00  and MDM-2017-0737Unidad de Excelencia Maria de Maeztu-Centro de Astrobiología (INTA-CSIC), as well as the support of the Generalitat de Catalunya/CERCA programme. The MOC activities have been supported by the ESA contract No. 4000124370. 
S.C.C.B. acknowledges support from FCT through FCT contracts nr. IF/01312/2014/CP1215/CT0004. 
X.B., S.C., D.G., M.F. and J.L. acknowledge their role as ESA-appointed CHEOPS science team members. 
A.Br. was supported by the SNSA. 
S.C.G. acknowledges support from UNAM PAPIIT-IG101321. 
K.A.C. acknowledges support from the TESS mission via subaward s3449 from MIT. 
P.E.C. is funded by the Austrian Science Fund (FWF) Erwin Schroedinger Fellowship, program J4595-N. 
D.B. has been funded by grants No. PID2019-107061GB-C61 and PID2023-150468NB-I00 by the Spain Ministry of Science, Innovation/State Agency of Research MCIN/AEI/ 10.13039/501100011033.
J.R.A.D. acknowledges support from the DiRAC Institute in the Department of Astronomy at the University of Washington. The DiRAC Institute is supported through generous gifts from the Charles and Lisa Simonyi Fund for Arts and Sciences, and the Washington Research Foundation. 
This project was supported by the CNES. 
The Belgian participation to CHEOPS has been supported by the Belgian Federal Science Policy Office (BELSPO) in the framework of the PRODEX Program, and by the University of Liège through an ARC grant for Concerted Research Actions financed by the Wallonia-Brussels Federation. 
L.D. is an F.R.S.-FNRS Postdoctoral Researcher. 
This work was supported by FCT - Fundação para a Ciência e a Tecnologia through national funds and by FEDER through COMPETE2020 - Programa Operacional Competitividade e Internacionalizacão by these grants: UID/FIS/04434/2019, UIDB/04434/2020, UIDP/04434/2020, PTDC/FIS-AST/32113/2017 \& POCI-01-0145-FEDER- 032113, PTDC/FIS-AST/28953/2017 \& POCI-01-0145-FEDER-028953, PTDC/FIS-AST/28987/2017 \& POCI-01-0145-FEDER-028987, O.D.S.D. is supported in the form of work contract (DL 57/2016/CP1364/CT0004) funded by national funds through FCT. 
B.-O. D. acknowledges support from the Swiss State Secretariat for Education, Research and Innovation (SERI) under contract number MB22.00046. 
D.D. acknowledges support from the TESS Guest Investigator Program grant 80NSSC19K1727 and NASA Exoplanet Research Program grant 18-2XRP18\_2-0136. 
This work has been carried out within the framework of the National Centre of Competence in Research PlanetS supported by the Swiss National Science Foundation under grants 51NF40\_182901 and 51NF40\_205606. The authors acknowledge the financial support of the SNSF.  This project has received funding from the European Research Council (ERC) under the European Union’s Horizon 2020 research and innovation programme (grant agreement SCORE No 851555). 
This project has received funding from the European Research Council (ERC) under the European Union’s Horizon 2020 research and innovation programme (project {\sc Four Aces}. 
grant agreement No 724427). It has also been carried out in the frame of the National Centre for Competence in Research PlanetS supported by the Swiss National Science Foundation (SNSF). DE acknowledges financial support from the Swiss National Science Foundation for project 200021\_200726. 
E. E-B. acknowledges financial support from the European Union and the State Agency of Investigation of the Spanish Ministry of Science and Innovation (MICINN) under the grant PRE2020-093107 of the Pre-Doc Program for the Training of Doctors (FPI-SO) through FSE funds. 
D.G. gratefully acknowledges financial support from the CRT foundation under Grant No. 2018.2323 ``Gaseous or rocky? Unveiling the nature of small worlds''. 
M.G. is an F.R.S.-FNRS Research Director.
M.N.G. is the ESA CHEOPS Project Scientist and is responsible for the ESA CHEOPS Guest Observers Programme. He does not participate in, or contribute to, the definition of the Guaranteed Time Programme of the CHEOPS mission through which observations described in this paper have been taken, nor to any aspect of target selection for the programme. 
R.D.H. is funded by the UK Science and Technology Facilities Council (STFC)'s Ernest Rutherford Fellowship (grant number ST/V004735/1). 
S.H. gratefully acknowledges CNES funding through the grant 837319. 
K.G. is the ESA CHEOPS Project Scientist and is responsible for the ESA CHEOPS Guest Observers Programme. She does not participate in, or contribute to, the definition of the Guaranteed Time Programme of the CHEOPS mission through which observations described in this paper have been taken, nor to any aspect of target selection for the programme. 
This work was granted access to the HPC resources of MesoPSL financed by the Region Ile de France and the project Equip@Meso (reference ANR-10-EQPX-29-01) of the programme Investissements d'Avenir supervised by the Agence Nationale pour la Recherche. 
M.L. acknowledges support of the Swiss National Science Foundation under grant number PCEFP2\_194576. 
R.L. acknowledges funding from University of La Laguna through the Margarita Salas Fellowship from the Spanish Ministry of Universities ref. UNI/551/2021-May 26, and under the EU Next Generation funds. 
This work is partly supported by MEXT/JSPS KAKENHI Grant Numbers 15H02063, JP17H04574, JP18H05439, JP18H05442, JP24H00017, 22000005, and JST CREST Grant Number JPMJCR1761. 
B.A.N. acknowledges support from STFC Consolidated Grant ST/S000488/1 (PI Balbus).
Funding from the University of La Laguna and the Spanish Ministry of Universities is acknowledged. H.P. acknowledges support by the Spanish Ministry of Science and Innovation with the Ramon y Cajal fellowship number RYC2021-031798-I. 
F.P. and C.L. would like to acknowledge the Swiss National Science Foundation (SNSF) for supporting research with HARPS-N through the SNSF grants nr. 140649, 152721, 166227 and 184618. The HARPS-N Instrument Project was partially funded through the Swiss ESA-PRODEX Programme. 
M.P. acknowledges the financial support from the ASI-INAF Addendum n.2018-24-HH.1-2022 ``Partecipazione italiana al Gaia DPAC  - Operazioni e attivit\`a di analisi dati''. 
This work was also partially supported by a grant from the Simons Foundation (PI Queloz, grant number 327127). 
D.R. was supported by NASA under award number NNA16BD14C for NASA Academic Mission Services. 
F.R. is funded by the University of Exeter's College of Engineering, Maths and Physical Sciences, UK. 
I.R. acknowledges support from the Spanish Ministry of Science and Innovation and the European Regional Development Fund through grant PGC2018-098153-B- C33, as well as the support of the Generalitat de Catalunya/CERCA programme. 
The material is based upon work supported by NASA under award number 80GSFC21M0002. 
We acknowledge support from the Swiss National Science Foundation (PP00P2-163967 and PP00P2-190080). We acknowledge support from the Centre for Space and Habitability (CSH) of the University of Bern. Part of this work received support from the National Centre for Competence in Research PlanetS, supported by the Swiss National Science Foundation (SNSF). 
This work was supported by FCT - Fundação para a Ciência e a Tecnologia through national funds and by FEDER through COMPETE2020 - Programa Operacional Competitividade e Internacionalização by these grants: UIDB/04434/2020. 
UIDP/04434/2020." Funded/Co-funded by the European Union (ERC, FIERCE, 101052347). Views and opinions expressed are however those of the author(s) only and do not necessarily reflect those of the European Union or the European Research Council. Neither the European Union nor the granting authority can be held responsible for them. A.M.S acknowledges support from the Fundação para a Ciência e a Tecnologia (FCT) through the Fellowship 2020.05387.BD. 
G.M.S. acknowledges the support of the Hungarian National Research, Development and Innovation Office (NKFIH) grant K-125015, a PRODEX Experiment Agreement No. 4000137122, the Lend\"ulet LP2018-7/2021 grant of the Hungarian Academy of Science and the support of the city of Szombathely. 
M.Sta. acknowledges financial support from the Belgian Federal Science Policy Office (BELSPO), in the framework of the PRODEX Programme of the European Space Agency (ESA) under contract number C4000140754.
A.T. acknowledges funding support from the STFC via a PhD studentship. 
M.T. is supported by JSPS KAKENHI grant No.24H00242.
N.A.W. acknowledges UKSA grant ST/R004838/1, and is a member of the UK Space Agency’s Discovery Advisory Committee for Science.

\paragraph*{Author contributions:}
T.G.W. analyzed the transit photometry and radial velocity data, performed the stellar characterisation, additional planet search analyses, and lead the manuscript writing.
A.M.S., A.C.C., R.Cl., \& A.A.J. performed radial velocity analysis and stellar activity correction.
M.Sta. performed the orbital dynamical and stability analysis.
Y.A., J.A.E., A.Bo., \& L.F. performed the internal structure and atmospheric escape analyses.
V.A., A.Bo., S.G.S., M.F., S.Sa., A.J.M., \& A.A.-K., performed the stellar analysis. 
H.P.O, T.Z., \& M.J.H. analysed the transit photometry data.
A.Be., M.B., N.B., P.Gut., L.L.K., T.Ba., T.Be., D.B.N., A.L., R.L., F.B. E.Pa., I.R., N.C.S., I.P., R.R., H.R., A.E., M.B.D., A.L.dE., M.Ste., W.Ba., N.A.W., A.T., D.P., M.De., H.P., M.N.G., G.O., W.Be., S.U., N.T., M.Gi., V.V.G., J.V., F.V., P.E.C., R.O., P.F.L.M., N.H., T.dR., B.-O.D., S.C.C.B., A.E.S., L.B., M.L., A.Br., A.M.S.S., A.Fo., C.M., C.B., D.S., D.B., D.M., D.E., D.G., D.Q., K.G.I., G.S., G.Pi., G.Pe., G.A., G.M.S., J.L., J.C., L.D., M.Gu., O.D.S.D., R.D.W., R.A., S.H., S.Ch., F.Ra., S.Cs., V.N., N.R., \& X.B., contributed to CHEOPS operations, observations, and data extraction.
K.R., A.M., R.D.H., F.A.P., C.L., M.M., E.Po., A.M.S., X.D., A.G., A.F.M.F., R.Co., A.H., M.Pe., L.A.B., A.So., B.A.N., B.S.L., F.Re., C.A.W., S.D., F.L., V.D., L.P., L.A., M.Pi., S.A., L.M., A.V., M.L.-M., D.D.S., D.C., \& M.Da. contributed to HARPS-N operations, observations, and data extraction.
G.R.R., R.K.V., D.W.L., S.Se., J.N.W., J.M.J., K.A.C., S.N.Q., D.R.C., D.D., A.Sh., J.R.A.D., J.A.G.J., P.R., R.H., D.J.-H., A.W.M., C.X.H., \& D.R. contributed to TESS operations, observations, and data extraction.
K.I.C., P.Gue., K.H., E.L.N.J., R.P.S, R.S., R.Z., \& C.Z. contributed to LCOGT observations, and data extraction.
E.E.-B., A.Fu., I.F., Y.H., M.I., Y.K., J.dL., F.M., N.N., M.T., \& Y.Z. contributed to MuSCAT2 and MuSCAT3 observations, and data extraction. 
S.C.G., Y.G.M.C., L.S, N.S., \& U.S contributed to SAINT-EX observations, and data extraction. 
M.B.L contributed PHARO observations, and data extraction. 
C.D.D., H.G., S.G., \& A.B.S. contributed to ShARCS observations, and data extraction. 

\paragraph*{Competing interests:}
The authors declare that they have no competing financial interests. 
A.M.S. is also affiliated with the Department of Physics and Kavli Institute for Astrophysics and Space Research, Massachusetts Institute of Technology, Cambridge, USA.
A.A.J. is also affiliated with the School of Physics \& Astronomy, University of Birmingham, Birmingham, UK.
M.B. is also affiliated with the Eidgen\"ossische Technische Hochschule Zurich Centre for Origin and Prevalence of Life, Zurich, Switzerland.
T.Be. is also affiliated with OHB System AG, Weßling, Germany.
N.H. is also affiliated with the Laboratoire d’Astrophysique de Marseille, Marseille, France.
A.T. is also affiliated with the Department of Astronomy, University of Maryland, College Park, USA and NASA Goddard Space Flight Center, Greenbelt, USA.
F.L. is also affiliated with the Department of Physics, Eidgen\"ossische Technische Hochschule Zurich, Zurich, Switzerland.
A.W.M. is also affiliated with the Department of Physics and Astronomy, San Francisco State University, San Francisco, USA.
D.R. is also affiliated with the Center for Astrophysics and Space Astronomy, Department of Astrophysical and Planetary Sciences, University of Colorado Boulder, Boulder, USA.
M.I. is also affiliated with the Astrobiology Center, Tokyo, Japan.
S.G. is also affiliated with the Department of Astronomy, California Institute of Technology, Pasadena, USA.
N.A.W. is the Deputy Chair of the UK Space Agency's Discovery Advisory Committee for Science.

\paragraph*{Data and materials availability:}
The TESS Pre-search Data Conditioning Simple Aperture Photometry (PDCSAP) observations are available in the Mikulski Archive for Space Telescopes (MAST), \url{https://exo.mast.stsci.edu} by searching for the target name LHS\,1903. The CHEOPS data are available from \url{https://cheops-archive.astro.unige.ch/archive_browser/} using the file names listed in Table~S1. Our reduced and detrended TESS, CHEOPS, LCOGT, MuSCAT2, MuSCAT3, and SAINT-EX transit photometry, {\sc s-bart}  extracted HARPS-N RV, and PHARO and ShARCS imaging data of LHS\,1903 are available at CDS via anonymous ftp to cdsarc.u-strasbg.fr (130.79.128.5) or via \url{https://cdsarc.cds.unistra.fr/viz-bin/cat/J/other/Sci/}. Gaia data for the analysis of Galactic kinematics are available here: \url{https://gea.esac.esa.int/archive}. The synthetic population of M-dwarf planets used for comparison to the LHS\,1903 bodies is available from the Data \& Analysis Center for Exoplanets \url{https://dace.unige.ch/populationAnalysis/?populationId=1}. Our TPFED/FFIED pipeline used to produce the TESS data is available at \url{https://github.com/ThomasGWilson/TPFED-FFIED} and archived at Zenodo\cite{note:TPFEDFFIED}. Our HARPS-N RV simulation code is available at \url{https://github.com/ThomasGWilson/PredictRVs} and archived at Zenodo\cite{note:PredictRVs}. Planet and stellar parameters are listed in Tables~1 and~S3. The results of our internal structure models are listed in Table~S8.


\subsection*{Supplementary materials}
Materials and Methods\\
Figs. S1 to S25\\
Tables S1 to S8\\
References (75–212)


\newpage


\renewcommand{\thefigure}{S\arabic{figure}}
\renewcommand{\thetable}{S\arabic{table}}
\renewcommand{\theequation}{S\arabic{equation}}
\renewcommand{\thepage}{S\arabic{page}}
\setcounter{figure}{0}
\setcounter{table}{0}
\setcounter{equation}{0}
\setcounter{page}{1} 


\begin{center}
\section*{Supplementary Materials for\\ \scititle}

{\small Thomas G. Wilson$^{1,2,\ast}$, 
Anna M. Simpson$^{3}$, 
Andrew Collier Cameron$^{1}$, 
Ryan Cloutier$^{4,5}$,}\and 
{\small Vardan Adibekyan$^{6,7}$, 
Ancy Anna John$^{1}$, 
Yann Alibert$^{8,9}$, 
Manu Stalport$^{10,11}$, 
Jo Ann Egger$^{8}$,}\and 
{\small Andrea Bonfanti$^{12}$
% ...
}\and 
\\
	\small$^\ast$Corresponding author. Email: thomas.g.wilson@warwick.ac.uk\and

\end{center}

\subsubsection*{This PDF file includes:}
Materials and Methods\\
Figs. S1 to S25\\
Tables S1 to S8

\newpage


\subsection*{Materials and Methods}

\subsection*{TESS observations and data reduction}

By conducting photometric observations of sectors for $\sim$27\,d each, the TESS spacecraft\cite{Ricker2015} searches for transiting exoplanets around stars in the TESS Input Catalogue (TIC\cite{Stassun2018,Stassun2019}). LHS\,1903, indexed as TIC 318022259 in the TIC, was observed in Sector 20, camera 1, charge-coupled device (CCD) 2, between 2019-December-24 and 2020-January-21 that yielded 23.18\,d of science observations. Presearch data conditioning simple aperture photometric (PDCSAP) light curves\cite{Smith2012,Stumpe2012,Stumpe2014} were extracted from the 2\,min cadence calibrated pixel files that were produced by the Science Processing Operations Center (SPOC\cite{Jenkins2016}) at the National Aeronautics and Space Administration (NASA) Ames Research Center. The SPOC conducted a transit search with an adaptive, noise-compensating matched filter\cite{Jenkins2002,Jenkins2010a,Jenkins2020}, producing two threshold crossing events (TCEs) that passed planet diagnostic tests\cite{Twicken2018,Li2019}. The transits were also detected by the Quick Look Pipeline (QLP) at the Massachusetts Institute of Technology (MIT\cite{Huang2020a,Huang2020b}). This system was announced as TESS Object of Interest (TOI\cite{Guerrero2021}) TOI-1730, hereafter LHS\,1903.

TESS re-observed LHS\,1903 in Sector 47, camera 1, CCD 2, from 2021-December-30 to 2022-January-28 that were processed by the SPOC into 20\,s cadence photometry covering 23.04\,d (in General Investigator (GI) Cycle 4 programs G04039, G04139, G04148, G04171, and G04242). A further 23.62\,d of TESS data were taken in Sector 60, camera 1, CCD 2, from 2022-December-25 to 2023-January-18 that were processed into 2\,min cadence photometry (in the GI Cycle 5 programs G05002). The transit signature of TOI-1730.03 was identified by a QLP pipeline search\cite{Huang2020a,Huang2020b} which flagged a TCE and alerted a TOI\cite{Guerrero2021}. In total, there are 28, 10, and three transits of LHS\,1903\,b,\,c, and\,d, detected with signal-to-noise ratios (SNRs) of 13.1, 18.7, and 17.0 indicating planetary signals with orbital periods of 2.156, 6.222, and 12.566\,d.

We retrieved the 2\,min and 20\,s systematics-corrected PDCSAP light curves\cite{Smith2012,Stumpe2012,Stumpe2014} from the Mikulski Archive for Space Telescopes (MAST) using the default quality bitmask. We rejected photometry with not-a-number fluxes or flux errors and poor-quality flagged data ({\tt QUALITY} $>$ 0) that resulted in 128,842 data points, see Fig.~S1. 

\subsection*{CHEOPS observations and data reduction}

To confirm and photometrically characterise the LHS\,1903 system and search for additional planets, we observed the system in the CHEOPS Guaranteed Time Observations (GTO) X-Gal programme. This programme aims to characterise small, transiting planets around kinematic thick disk and halo stars, and chemically metal-poor objects. Our stellar kinematic pipeline (see Section ``Host star properties'') computes that LHS\,1903 has a thick disk membership probability of 63\%. We conducted 22 CHEOPS visits with an exposure time of 60\,s yielding 300.72\,hr of photometry between 2021-November-19 and 2023-February-7. These data cover 12, seven, and four transits of LHS\,1903\,b,\,c, and\,d, (see Table~S1). 

All visits were processed using version 14 of the CHEOPS Data Reduction Pipeline (DRP \cite{Hoyer2020}). Two contaminating sources, Gaia DR3 978086481343776128 (difference in Gaia magnitude ($\Delta\,G$)\,=\,+3.9\,mag) and Gaia DR3 978086477047534720 ($\Delta\,G$\,=\,+2.2\,mag), are located 16.2 and 17.1\,arcsec away so we subtracted contamination estimates created by simulating CHEOPS point spread functions (PSF) of Gaia identified stars in the FoV. Using {\sc pycheops}\cite{Maxted2022}, we retrieved the light curve produced by the aperture that minimised the root mean square (RMS) for each visit. The apertures used are listed in Table~S1 alongside the observational efficiency that reports the percentage of an observation that contains usable data.

Substantial flux variation systematics remained in the CHEOPS data, e.g. Fig.~S2A. Nearby background sources, or illumination or thermal changes over the timescale of the roll of the spacecraft can induce flux variations in CHEOPS data due to PSF shape changes affecting the flux and flux surface density within an aperture. Therefore, we applied an existing PSF-based principal component analysis (PCA) detrending method \cite{Wilson2022} to remove these effects. The number of required principal components was selected using a leave-one-out cross validation method and ranged from 3 to 49. An example of the linear model produced by the PSF-based PCA method is shown in Fig.~S2A. We utilised {\sc pycheops} to determine if further decorrelation was needed. The identified instrumental basis vectors are listed in Table~S1 yielding 7-59 vectors per visit resulting in 590 parameters in total when combined with PSF components. This represents $\sim$5.9\% of available vectors. To be conservative and propagate detrending model uncertainties, we first conducted a linear noise model detrending on each visit then inflated the flux errors following previous studies \cite{Vanderburg2019,Delrez2021,Wilson2022}, see Figs.~S2 and~S3.

\subsection*{HARPS-N observations and data reduction}

To determine the masses of the LHS\,1903 planets and characterise the host star, we obtained 108 high-resolution spectra with the HARPS-N optical \'{e}chelle spectrograph\cite{Cosentino2012,Cosentino2014} (resolving power ($R$)\,=\,115,000) on the Telescopio Nazionale Galileo (TNG) in La Palma. These data span 870\,d from 2020-October-2 to 2023-February-19 with a gap due to the Tajogaite volcano eruption in 2021. We obtained observations within the HARPS-N collaboration GTO and in an Optical Infrared Coordination Network for Astronomy (OPTICON) programme (OPT22A\_38; PI: Wilson) following RV signal existence simulations\cite{note:PredictRVs}.

To simulate additional HARPS-N RVs and determine how much additional data we required, we first performed an initial fit of the HARPS-N collaboration GTO data using the {\sc juliet}\cite{Espinoza2019} software suite to produce a model of the RV modulation due to stellar activity and orbiting planets. We found that four Keplerian orbits with ephemerides taken from our transit photometry data combined with a quasi-periodic kernel GP applied to the RVs with a $P_{\rm rot}$ hyper-parameter set to the stellar rotation period (see spectral activity indicator analysis below) are able to well-model observed RV variations and result in white-noise residuals. To produce simulated RVs we propagated this model and its uncertainty to the 2022A observing season window. The RV value was then obtained by randomly drawing from a distribution centred on the system RV model at a given epoch with a width of the model uncertainty summed in quadrature with the white-noise jitter term from our initial analysis. The errors on the simulated RVs were drawn randomly from a cumulative distribution function (CDF) of the uncertainties on the HARPS-N collaboration GTO data. To preserve observational scheduling by taking into account telescope accessibility constraints and poor-weather conditions, we determined the epochs at which to produce the simulated RV by randomly drawing from a CDF of the time between epochs of the HARPS-N collaboration GTO data. We performed a Stacked Bayesian Generalised Lomb-Scargle (BGLS) periodogram\cite{Mortier2017} on the combined real data and simulated RVs to determine the SNR of Keplerian orbits with our RV time series as a function of orbital period and number of RV data points. Therefore, by constraining the periodogram to the periods regions of the transiting planets, we determined the number of additional RVs needed to significantly (semi-amplitude $>$3$\sigma$) detect the LHS\,1903 planets. These RVs were taken in the OPT22A\_38 OPTICON programme at a similar cadence as the HARPS-N GTO observations.

All observations were taken with an exposure time of 1800\,s that resulted in a median SNR per \'{e}chelle order at 550\,nm of 32. These data were reduced with the HARPS-N Data Reduction Software (DRS v2.3.5\cite{Dumusque2021}) with a M2 cross correlation function (CCF) mask that determined the CCF full width at half maximum (FWHM), bisector span (BIS), and contrast, and Ca\,{\sc ii} H\,\&\,K (S-index), and H$\alpha$ activity indicators. To extract the RVs we used the template matching\cite{AngladaEscude2012,AstudilloDefru2015,Cloutier2021} software, {\sc s-bart}\cite{Silva2022}. We excluded data taken in poor conditions (high airmass or seeing) that result in large RV uncertainties. This yielded 91 RVs that are shown in Fig.~S4, with RV and activity indicator periodograms presented in Fig.~S5. The strongest CCF contrast and S-index periodogram peaks are at $\sim$40\,d, with peaks also found in the CCF FWHM and H$\alpha$ activity indicators. In our CCF-{\sc scalpels}\cite{CollierCameron2021} analysis below, we identify this as the LHS\,1903 rotation period. In Fig.~S6, we show how removal of this stellar signal with a Quasi-periodic GP kernel strengthens the planetary RV signals.

\subsection*{Ground-based photometric and imaging observations}

TESS photometric apertures typically extend to $\sim$1\,arcminute from the target covering multiple pixels of scale $\sim$21\,arcsec\,pixel$^{-1}$, causing multiple stars to blend with the target. To support the CHEOPS observations, verify the source of transit detections, and to monitor for potential transit timing variations, we conducted ground-based photometric observations within the TESS Follow-up Observing Program\cite{Collins2019}.

\paragraph{LCOGT}

We observed three, three, and one transits of LHS\,1903\,b,\,c, and\,e, using the LCOGT\cite{Brown2013} 1.0\,m network, see Table~S2, using the software {\sc tapir tess transit finder}\cite{Jensen2013} to schedule observations. The 1\,m telescopes have $4096\times4096$ Sinistro cameras with an image scale of 0.389\,arcsec\,pixel$^{-1}$, resulting in a 26$\times$26\,arcminute$^2$ field of view (FoV). The images were calibrated using the LCOGT {\sc banzai} pipeline\cite{McCully2018} with differential photometric data extracted using {\sc astroimagej}\cite{Collins2017}. We detected transit-like signals in all light curves using photometric apertures that exclude all Gaia-identified neighbours of LHS\,1903, indicating that the transits are from that star. These data are shown in Fig.~S7.

\paragraph{MuSCAT2}

A transit of LHS\,1903\,b was observed on 2020-March-17 using MuSCAT2\cite{Narita2019} mounted on the 1.5\,m Telescopio Carlos S\'{a}nchez (TCS) at the Teide Observatory. MuSCAT2 is equipped with four CCDs that obtains simultaneous images in $g$, $r$, $i$, and $z_s$ bands. Each CCD has 1024$\times$1024\,pixels with a FoV of 7.4$\times$7.4\,arcminute$^2$. During observations the telescope was defocused with exposure times of 40, 20, 10 and 6\,s in the four bands, see Table~S2. Data reduction (i.e., dark and flat field correction) and aperture photometry using a radius of 17.8\,pixels (7.83\,arcsec) was performed using the MuSCAT2 pipeline\cite{Parviainen2019}. These data are shown in Fig.~S7.

\paragraph{MuSCAT3}

We observed a transit of LHS\,1903\,b on 2022-March-14 using MuSCAT3\cite{Narita2020} on the 2\,m Faulkes Telescope North (FTN) at the Haleakala Observatory. MuSCAT3 has four 2k$\times$2k\,pixel CCDs, each with a pixel scale of 0.27\,arcsecond\,pixel$^{-1}$, allowing simultaneous imaging in $g$, $r$, $i$, and $z_s$ bands. The telescope was defocused with exposure times set to 23, 8, 12, and 11\,s in the four bands. After applying dark and flat-field corrections, we performed aperture photometry using a pipeline\cite{Fukui2011} with aperture radii of 24, 24, 24, and 26\,pixels, see Table~S2. These data are shown in Fig.~S7.

\paragraph{SAINT-EX}

We obtained a partial transit of LHS\,1903\,d with SAINT-EX on 2022-March-09 in the I+z bands. SAINT-EX is a 1\,m telescope located at the Observatorio Astronómico Nacional, San Pedro Mártir\cite{Demory2020}. The telescope was defocused during observations that consisted of 571 images with an exposure time of 10\,s, covering 188\,minutes in total, see Table~S2. The data were reduced using the {\sc prince} pipeline\cite{Demory2020}, with the final light curve obtained from an aperture radius of 11\,pixels (3.8\,arcsec). These data are shown in Fig.~S7.

\paragraph{PHARO}

The Palomar/PHARO\cite{Hayward2001} observations of LHS\,1903 were made using the natural guide star adaptive optics (AO) system P3K\cite{Dekany2013} on 2020-November-05 in a standard 5-point quincunx dither pattern with steps of 5\,arcsec. We used the narrow-band Br$\gamma$ filter (central wavelength ($\lambda_0$)\,=\,2.1686\,micron (µm); wavelength range ($\Delta\lambda$)\,=\,0.0326\,µm). Each dither position was observed three times, offset by 0.5\,arcsec for a total of 15 frames; with an integration time of 9.912\,seconds per frame totalling 148\,s. PHARO has a pixel scale of 0.025\,arcsecond\,pixel$^{-1}$ and a FoV of $\sim$25\,arcsec. The sky-subtracted, flat-fielded science frames were combined into a single image with a final PSF FWHM resolutions of 0.14\,arcsec. The $5\sigma$ limit at each separation was determined from the average values determine from injection and recovery anlyses at that separation and the uncertainty was set by the RMS dispersion at a given radial distance\cite{furlan2017}. The final sensitivity curve is shown in Fig.~S8; no additional stellar companions were detected.

\paragraph{ShARCS}

We observed LHS\,1903 on 2020-December-01 and 02 using ShARCS on the Shane 3\,m telescope at Lick Observatory\cite{Kupke2012,Gavel2014,McGurk2014} with the adaptive optics system in natural guide star mode. We collected observations with a $K_s$ filter ($\lambda_0$\,=\,2.150\,µm, $\Delta\lambda$\,=\,0.320\,µm) and reduced the data using the {\sc simmer} pipeline\cite{Savel2020,Savel2022}. The reduced image and corresponding contrast curve is shown in Fig.~S8. We find no nearby stellar companions within our detection limits.

\subsection*{Host star properties}

To aid in the analysis and interpretation of the LHS\,1903 planetary system, we determined physical properties of the host star. The main derived parameters are listed in Tables~1 and~S3.

\paragraph{Galactic Kinematics}

To characterise the Galactic population membership of LHS\,1903, we computed the right-handed, heliocentric Galactic space velocities\cite{Johnson1987}, $U$, $V$, and $W$, using the Gaia DR3 coordinates, proper motions, offset-corrected parallax\cite{Lindegren2021b}, and RV\cite{GaiaCollaboration2023}. These values are reported in Table~S3. Using the Galactic velocities, and the positions of the north Galactic pole and Galactic longitude of the first intersection of the Galactic plane\cite{GaiaCollaboration2023}, we determined kinematic Galactic family probabilities\cite{Reddy2006}. We computed the thin disk, thick disk, and halo membership probabilities for four sets of velocity dispersion standards\cite{Bensby2003, Bensby2014, Reddy2006, Chen2021} corrected for the Local Standard of Rest\cite{Koval2009}. The resulting 100,000 samples were combined via a weighted average to determine the kinematic Galactic family probabilities. We find that LHS\,1903 has thin disk, thick disk, and halo membership probabilities of 36.9\%, 63.0\%, and 0.1\%. Therefore, LHS\,1903 likely belongs to the kinematic thick disk.

We calculated the Galactic eccentricity of LHS\,1903 by integrating the Galactic orbit over our stellar age estimates (4.90, 7.08, and 13.02\,Gyr, see below) using {\sc galpy}\cite{Bovy2015} and Gaia DR3 coordinates, proper motions, offset-corrected parallax, and RV\cite{GaiaCollaboration2023}. We find an eccentricity of 0.36, higher than 99.86\% of thin disk stars, but consistent with the thick disk eccentricity distribution\cite{Yan2019}. We also calculated the local phase space density of LHS\,1903 using the Mahalanobis distance metric for the 6D phase space of position and velocity for LHS\,1903 and neighbouring stars\cite{Winter20}. Stars were assigned to a low- or high-density population using a Gaussian mixture model. This indicates that LHS\,1903 belongs to the low-density population (Fig.~S9) which is a mixture of kinematically hot thin disk, thick disk, and halo stars\cite{Mustill22}. These analyses are consistent with LHS\,1903 being a thick disk star. 

\paragraph{Stellar Atmospheric Properties and Abundances}

We performed a spectral analysis of LHS\,1903 using {\sc{specmatch-emp}}\cite{Yee2017} on our co-added HARPS-N high resolution spectroscopic data. Via comparison of the observed spectrum with library spectra of stars with known physical properties through a minimization algorithm\cite{Hirano2018, Fridlund2020}, we find the stellar effective temperature, surface gravity, and iron abundance of LHS\,1903 are $T_{\rm eff}\,=\,3664\pm70$\,K, $\log{g}\,=\,4.75\pm$0.12\,cm~s$^{-2}$, and [Fe/H]$\,=\,-0.11\pm0.09$\,dex. We also used the {\sc odusseas} code\cite{Antoniadis2020,Antoniadis2024} to infer the effective temperature and metallicity, finding $T_{\rm eff}\,=\,3712\pm$91\,K and [Fe/H]$\,=\,-0.08\pm0.11$\,dex. Due to the smaller effective temperature uncertainty, we adopt the {\sc{specmatch-emp}} values reported in Table~1.

The determination of M-dwarf individual elemental abundances from visible spectra is challenging due to extensive line blending\cite{Maldonado2020}. Therefore, we estimated the Mg and Si abundances by relating the Fe abundance to Mg and Si using 15000 red giant stars spanning a wide range of metallicities (-1.0\,dex to ~0.5\,dex)\cite{Demangeon2021} from the Apache Point Observatory Galactic Evolution Experiment (APOGEE) Data Release 17 (DR17)\cite{Abdurrouf2022}. For a sample of stars with similar metallicities to LHS\,1903, we calculated the Mg and Si mean abundances and standard deviation to be -0.05$\pm$0.13 and -0.08$\pm$0.12 that we adopt as the empirical Mg and Si abundances and uncertainty ranges LHS\,1903.

\paragraph{Stellar Radius, Mass, and Age}

We computed the stellar radius of LHS\,1903 using a modified Markov-Chain Monte Carlo (MCMC) infrared flux method (IRFM\cite{Blackwell1977,Schanche2020}). We built spectral energy distributions (SED) using \textsc{atlas}\cite{Kurucz1993,Castelli2003} and \textsc{phoenix}\cite{Allard2014} atmospheric models using our measured spectral parameters and conducted synthetic photometry in the following bandpasses; Gaia {\em G}, {\em G$_{\rm BP}$} (blue photometer), and {\em G$_{\rm RP}$} (red photometer), Two Micron All-Sky Survey (2MASS) {\em J}, {\em H}, and {\em K$_s$}, and Wide-field Infrared Survey Explorer (WISE) {\em W1} and {\em W2}\cite{Skrutskie2006,Wright2010,GaiaCollaboration2023}. This was compared to observed broadband fluxes\cite{Skrutskie2006,Wright2010,GaiaCollaboration2023} to yield the effective temperature and angular diameter, that was converted to the radius using the Gaia DR3 offset-corrected parallax\cite{Lindegren2021b}. We accounted for atmospheric modeling uncertainties using a Bayesian modeling averaging of stellar radius posterior probability distributions. We find $R_\star=0.539\pm0.014\,R_\odot$.

We used $T_{\rm eff}$, [Fe/H], and $R_\star$ to derive the isochronal mass, $M_\star$, and age, $t_\star$, from two sets of stellar evolutionary models; Padova and Trieste stellar evolution code (PARSEC) v1.2s\cite{Marigo2017} and Code Liegeois d'Evolution Stellaire (CLES\cite{Scuflaire2008}). We utilised placement\cite{Bonfanti2015,Bonfanti2016} and minimisation\cite{Salmon2021} algorithms, along with a $\chi^2$-based criterion mutual consistency check\cite{Bonfanti2021a} to obtain $M_\star=0.538_{-0.030}^{+0.039}$\,M$_\odot$ and $t_\star=4.9\pm4.0$ Gyr. Given that physical parameters of M-dwarfs evolve very slowly and spectra contain blended lines, it is challenging to constrain the isochronal (and gyrochronological, chemical, or astroseismological) stellar age.
As stars age, there is an increasing probability of kinematic disturbances that results in an increase in stellar Galactic velocity\cite{Wielen1977,Nordstrom2004,Casagrande2011}, so we can estimate stellar age based on kinematics alone\cite{Maciel2011,AlmeidaFernandes2018}. By comparing the Galactic $U$, $V$, and $W$ velocities, and eccentricity of LHS\,1903 to kinematic-age probability distributions of 9000 stars\cite{AlmeidaFernandes2018}, we find ages of 13.02$^{+0.03}_{-5.13}$ and 7.08$^{+2.87}_{-1.98}$\,Gyr via two methods using the $U$, $V$, and $W$ velocities and eccentricity, and $V$ and $W$ velocities, respectively. 

\subsection*{Orbital architecture of the system}

Due to the short TESS observing windows that also contain data gaps, long orbital period planets may have missed transits potentially precluding their detection. Therefore, we inspected the TESS PDCSAP data to investigate the LHS\,1903 system architecture. In addition to the 41 transits of the three planets announced as TOIs (see above), we identified a full transit in TESS Sector 47 at $\sim$2459601.8\,BJD that cannot be attributed to LHS\,1903\,b,\,c, or\,d, see Fig.~S1. To verify the transit and ascertain the nature of the occulting body we conducted custom extractions and detrendings of the TESS observations, and analysed the resulting light curves and our RV, imaging, and Gaia astrometric data as detailed below.

\paragraph{The TPFED/FFIED pipeline}

We developed the Target Pixel File Extraction and Detrending/Full Frame Image Extraction and Detrending (TPFED/FFIED) pipeline\cite{note:TPFEDFFIED}. For both Target Pixel Files (TPFs) and Full Frame Images (FFIs), the tool retrieves the calibrated data using {\sc tesscut}\cite{Brasseur2019} and the default quality bitmask then conducts noise-optimised aperture photometry. This is done by extracting target fluxes for a range of custom aperture masks with radii of two to four pixels, in steps of 0.1\,pixels, centred on the target. Because the target does not fall in the exact centre of a pixel, increasing the aperture mask radius by 0.1\,pixels can result in non-circular masks. The optimal aperture is chosen by minimising the Combined Differential Photometric Precision (CDPP)\cite{Jenkins2010b,Christiansen2012} noise after removal of data points with not-a-number flux or flux uncertainty values. All light curves were background-corrected after determining the sky level using corresponding custom background masks, built by selecting pixels whose flux values do not deviate by more than 1$\sigma$ from the flux median across the field of view and time series.

The TPFED/FFIED pipeline performs initial detrending by conducting principal component analyses on the custom background masks to determine the scattered-light flux contribution to the light curves and removes these systematics using the {\sc lightkurve} package\cite{Lightkurve2018}. The tool corrects flux modulation due to spacecraft jitter by retrieving the co-trending basis vectors (CBVs) and two-second cadence engineering quaternion measurements for the cameras and CCDs that the targets are observed in. The averages of the quaternions over the scientific observational cadences of the TPFs and FFIs are computed following previous methods\cite{Vanderburg2019,Delrez2021}. TPFED/FFIED produces additional detrending vectors using the PSF-{\sc scalpels} approach\cite{Wilson2022} to remove flux systematic variations due to PSF shape changes. In brief, PSF-{\sc scalpels} conducts a PCA on the outputs of auto-correlation function of photometric images to produce components that measure modulations in the shape of PSFs over a time series of observations. The PCA output is dimensionally reduced using a leave-one-out cross validation method to select components that most contribute to PSF shape variation\cite{Parviainen2022,Wilson2022,Hawthorn2023}.

We produced two TPFED/FFIED light curves for each sector of TESS observations; one with detrending conducted using only the CBVs and quaternions, and a second that also includes the PSF-{\sc scalpels} components. Thus, this procedure yields three sets of TESS photometry when also considering the PDCSAP data. Fig.~S10 shows the transit feature appears in all three photometric extraction and detrending methods, so we conclude it is astrophysical in nature.

\paragraph{Additional transits}

We also identified a partial transit at the end of TESS Sector 20, $\sim$2458868.8\,BJD see Fig.~S1, that is apparent regardless of extraction and detrending method. We inspected our CHEOPS and ground-based lightcurves, and found a full transit at $\sim$2459572.4\,BJD in visit 5 of the CHEOPS data (file key CH\_PR120054\_TG001501\_V0300), see Fig.~S2E, and partial transits at $\sim$2459249.9 and $\sim$2459894.9\,BJD in LCOGT photometry, see Fig.~S7. We obtained further CHEOPS photometry that revealed three full transits at $\sim$2459924.2, $\sim$2459953.6, and $\sim$2459982.9\,BJD in visits 12, 18, and 22, see Figs.~S2L,\,R, and\,V. We fitted these transits with the {\sc juliet}\cite{Espinoza2019} software and found similar transit depths (881$\pm$74, 908$\pm$62, 949$\pm$34, 839$\pm$55, 871$\pm$72, 883$\pm$74, 895$\pm$65, and 810$\pm$45\,ppm, respectively) corresponding to SNRs of 12 to 28. This is evidence that these flux deficits are caused by the same orbiting body. The time between the CHEOPS and TESS Sector 47 transits indicated an orbital period of $\sim$29.32\,d with shorter periods ruled out by the photometric coverage of TESS Sector 47. This is supported by the transit features seen in the TESS Sector 20 and LCOGT data. We conclude that there is an additional body in the LHS\,1903 system.

\paragraph{Binarity assessments}

The transit signals of planets b, c, and d passed the TESS vetting criteria, which exclude the possibility that they are due to an eclipsing binary. Statistical studies\cite{Latham2011,Lissauer2012} have shown that the multiplicity of a planetary system provides evidence that additional transits are due to another planet in the system. This is supported by the lack of a nearby source in our high-resolution imaging, in which we find that LHS\,1903 is a single star with no contaminating companion of brightness $\Delta$ Br$\gamma$ and $\Delta$ K$_{\rm s} > $6\,mag outside a separation of 0.5\,arcsec. Fig.~S8 excludes a main sequence star at spatial limits of 8.9 to 71\,au. This is supported by the lack of a RV trend in our HARPS-N data, see Fig.~S4, over our 769\,d, which was determined by comparing the Bayesian evidence for a model consisting of four Keplerian orbits plus a linear trend, compared to a model of only four Keplerian orbits. 

The Gaia astrometric data also constrain binarity, because any unknown companion would produce excess astrometric noise. The Gaia DR3 astrometric excess noise and renormalised unit weight error (RUWE) values of LHS\,1903 are 0.133\,mas and 1.137\cite{GaiaCollaboration2023}, respectively, which are slightly higher than the median values for single stars in the G$=$9 to 12 magnitude range, that includes LHS\,1903, with an astrometric solution (0.113 and 1.039, respectively\cite{Lindegren2021a}), but is consistent with being a member of that population\cite{GaiaCollaboration2023}. Working under the assumption that the RUWE indicates deviations from a single source, we computed the amplitude of photocentre angular perturbation of the Gaia observations for LHS\,1903 using the Gaia DR3 RUWE and parallax error as a proxy for observational along-scan error\cite{Belokurov2020}. We then calculated the physical perturbation of LHS\,1903 due to a potential companion by multiplying the photocentre perturbation by the distance, taken as the inverse of the parallax. For LHS\,1903 this value is 0.004\,au. Assuming that a potential companion contributes only negligible flux to the photocentre variation, we rule out Jupiter-mass planets and brown dwarfs out to semi-major axes of 1 and 10\,au, respectively. Main sequence star companions at these orbital distances were ruled out by our high-spatial imaging analysis above. Studies that have combined Gaia data with astrometry from Hipparcos\cite{Brandt2021,Kervella2022} or adaptive optics imaging\cite{Lamman2020} find no evidence for LHS\,1903 being a binary system. 

We used the {\sc triceratops}\cite{Giacalone2021} validation tool to fit photometric models of 18 planetary and false positive (FP) scenarios to our combined transit photometry dataset, taking into account the stellar properties and our high spatial resolution imaging of LHS\,1903. We find that LHS\,1903\,c,\,d, and\,e have FP probabilities (FPPs) of 1.00\%, 0.06\%, and 0.48\%, respectively, whereas planet b has a transit photometry FPP of 4.4\% likely due to the shallow nature of the transit. This FPP is higher than the typical 1\% value used to validate planets\cite{Fressin2013}, however for $R_{\rm p} < 6\,{\rm R_\oplus}$ planets observed with TESS, the detection of multiple planets in the same system reduces the FPP by a factor 54\cite{Guerrero2021}; after applying this correction, the FPPs for the LHS\,1903 planets are 0.08\%, 0.02\%, 0.001\%, and 0.009\%. The presence of each of these planets is confirmed by the detected RV signal at the orbital periods, so we consider them to be validated. Therefore, we conclude that the additional observed transit signals arise from a fourth orbiting planet, LHS\,1903\,e.

\paragraph{BLS and TIP transit reality checks}

To assess the reality of the transit signals, constrain the orbital period of LHS\,1903\,e, and search for additional signals we conducted a box least squares (BLS) analysis using the {\sc lightkurve} package\cite{Lightkurve2018}. For all three TESS extracted lightcurves (PDCSAP, TPFED/FFIED, and TPFED/FFIED \& PSF-{\sc scalpels}), there were remnant long term systematic trends. We removed these trends via two separate methods resulting in six light curves. These methods are fitting a GP with a {\sc celerite}\cite{ForemanMackey2017} Mat\'{e}rn-3/2 kernel to the out-of-transit data or modelling the baseline flux using the biweight method from the {\sc wotan} package\cite{Hippke2019}. For each of the six TESS light curves variants (in combination with the CHEOPS and ground-based photometry), we ran iterative BLS analyses. This was done by removing the most likely (delta log likelihood ($\Delta$log(L)) in favour of a transit model compared to a null detection) transit model identified by the BLS from the data and repeating the process until no strong detections were detected. For all six light curves, we first detect LHS\,1903\,d with an orbital period of $\sim$12.56\,d and $\Delta$log(L) of $>$940. After removing planet d from the data, LHS\,1903\,c (period of $\sim$6.23\,d) was always the strongest detection, with $\Delta$log(L)$>$350. After subtracting planets d and c, we detect a transit an orbital period of $\sim$29.317\,d with $\Delta$log(L)$>$240, which is LHS\,1903\,e. After also removing this planet, we recover the signal of LHS\,1903\,b on an orbital period of $\sim$2.16\,d with $\Delta$log(L)$>$210. No further likely signals were detected after the removal of planet b. Table~S4 lists the $\Delta$log(L) values derived from this analysis. 

As a final assessment of our photometric data, we utilise a transit existence determination method to search for transit-like features within our space- and ground-based transit photometry light curves. The true inclusion probability (TIP\cite{Hara2022}) of a signal in a data set is defined as the convolution of the posterior probability of a fitted planet model parameter and a null planet model. For example, a photometric light curve with a baseline noise model can be compared to a baseline noise plus transit model to compute the TIP that a transit with a given centre time is present in the data set\cite{Ehrenreich2023}. The TIP therefore statistically determines whether transits exist within a light curve in a similar manner to likelihood searches\cite{Gill2020}.

Due to the multiple planets orbiting LHS\,1903 and the large photometric baseline, we divide our combined TESS, CHEOPS, LCOGT, MuSCAT2, MuSCAT3, and SAINT-EX data into sections with a maximum length of one day, then fit them with zero, one, two, three, and four transit models. We initially use unconstrained priors to agnostically search for transit signals. We recover evidence of all (eight and eight) transits of LHS\,1903\,d and\,e, 13 (out of 17) transits of planet\,c, and 15 (out of 42) of LHS\,1903\,b. The majority of the transits that do not pass this test are from the lower cadence data, which has reduced photometric quality\cite{Huber2022}.

We conduct a second analysis using uniform transit centre time and period priors for all four planets, centred on the values obtained by the BLS method with 20$\sigma$ widths. With this focused search we recover all the LHS\,1903\,c,\,d, and\,e transits, and 41 out of 42 transits of planet\,b. In this dataset, we do not find any further transit signals beyond those attributable to the four detected planets.

\paragraph{Blind RV planet searches}

We performed a blind search of the radial velocities extracted with {\sc s-bart}, using a model comprising up to five Keplerian signals and a quasi-periodic stellar activity model, using the {\sc kima} nested-sampling package\cite{Faria2018}. Time-domain activity-decorrelation vectors were calculated with CCF-{\sc scalpels}\cite{CollierCameron2021}, which entails principal-component analysis of the autocorrelation function of the CCF. Some shift-like signals can elude {\sc scalpels} analysis\cite{John2022}, so any remaining rotationally-modulated signals were modelled with GP regression applied to the RVs. For this analysis using {\sc kima}, the quasi-periodic kernel used log-uniform priors on the amplitude ($\eta_1$) (0$<$ln($\eta_1$)$<$2.7) and active-region evolution timescale ($\eta_2$) (60 $<\eta_2<$ 120\,d). The priors on the rotation period ($\eta_3$) and log harmonic complexity ($\eta_4$) were set to a uniform distribution with 35 $<\eta_3<$ 50\,d and -1.6$<$ln($\eta_4$)$<$0, respectively.

This analysis found detections of Keplerian signals at 2.15, 6.22, 12.6, and 29.3\,d, with $\eta_3$, a proxy for the stellar rotation period, constrained to 41.8\,d. A possible additional Keplerian signal is present near P=53.9\,d. A second {\sc kima} analysis, treating the four transiting planets as known signals, but with the same CCF-{\sc scalpels} decorrelation and GP regression model of the rotational modulation, yielded orbital parameters consistent with those found in the joint {\sc juliet} analysis of the photometry and RV data.

We used the Gaussian mixture modelling method {\sc tweaks}\cite{John2023} to analyse the results of the first {\sc scalpels} and GP model fitting detailed above. The {\sc tweaks} results show $>$3$\sigma$ detections of the three inner planets and a marginal signal of the 29.3\,d planetary object. The relative weakness of this signal might be due to the proximity of the orbital period to the mean lunar synodic period of 29.52\,d, exacerbated by the observational epochs of LHS\,1903. Due to the scheduling availability of HARPS-N, our observations were taken in two or three blocks per month with one run typically centred around the full Moon every cycle. Moon contamination is likely to be minimal due to the high absolute stellar RV and distance to the ecliptic plane ($\sim$25$^\circ$) that means that Solar spectral lines from reflected Lunar light are sufficiently separated from the absorption features of LHS\,1903. However, stronger contamination could affect the CCF shape that would then be corrected by {\sc scalpels}. To test this, we performed an additional analysis with a single decorrelation step using a GP alone. From this analysis, we find detections ($\geq$3$\sigma$) for all four transiting planets, with radial velocity semi-amplitudes 3.16$\pm$0.66, 3.03$\pm$0.58, 3.07$\pm$1.08 and 1.79$\pm$0.39\,${\rm m\,s}^{-1}$, respectively.

\subsection*{Joint transit photometry and RV analysis}

To characterise the planetary system around LHS\,1903, we simultaneously analysed all transit photometry and RV data. We selected the TESS PDCSAP data due to lower average flux uncertainties and the detrended CHEOPS photometry, described above. We used the {\sc juliet}\cite{Espinoza2019} software that performs transit and RV fitting using {\sc batman}\cite{Kreidberg2015} and {\sc radvel}\cite{Fulton2018}, and simultaneous correlated noise modelling with GPs (implemented in {\sc celerite}\cite{ForemanMackey2017} and {\sc george}\cite{Ambikasaran2015}). We used a nested sampling algorithm with 1000 live points and the stopping criterion as the change in log Bayes Evidence ($\Delta$logZ)\,=\,0.3 using the nested sampling algorithm {\sc dynesty}\cite{Speagle2020}. The model parameters and assumed priors (Tables~S5,~S6, and~S7) were:

\begin{itemize}
    \item The orbital periods, $P$, and transit centre times, $T_{\rm 0}$, for each of the four planets were taken as wide uniform priors centred on the TESS PDCSAP plus GP BLS results, see above.
    \item The planet-to-star radius ratios, $R_{\rm p}/R_\star$, transit impact parameters, $b$, and argument of periastrons, $\omega$, assumed wide uniform priors for all planets.
    \item The RV semi-amplitudes, $K$, assumed log-uniform priors for all planets that were modified to be uniform from 0 to 1\,${\rm m\,s}^{-1}$ and log-uniform from 1 to 100\,${\rm m\,s}^{-1}$.
    \item The eccentricities, $e$, for all planets were constrained with half-Gaussian zero-mean priors\cite{Lacedelli2021,Lacedelli2022} with the width determined by the eccentricity distribution of previously observed compact multi-planet systems\cite{VanEylen2019}.
    \item The stellar density, $\rho_*$, was taken with a Gaussian prior determined from our stellar analysis, see above, which allows a single value to constrain the scaled semi-major axes for all planets included in the model.
    \item The $q$ parameterisation\cite{Kipping2013} of the limb-darkening coefficients was used with wide priors for all transit photometry. We described limb-darkening using the quadratic law, with two limb-darkening coefficients ($q_1$,$q_2$), for TESS and CHEOPS following previous work\cite{Espinoza2016} as this parameterisation provides non-uninformative priors\cite{Kipping2013}. We used the linear law, with a single limb-darkening coefficient ($q_1$), for LCOGT, MuSCAT2, MuSCAT3, and SAINT-EX to aid convergence for these lower precision data.
\end{itemize}

For the TESS and CHEOPS datasets we modelled any remaining long time-scale systematic noise using GPs with an approximate Mat\'{e}rn kernel with wide priors on the hyper-parameters (amplitude ($\sigma_{\rm GP}$) and length-scale ($\rho_{\rm GP}$) of the GP). To model stellar activity in the HARPS-N RVs, we included a linear noise model with eight basis vectors ($\theta_{\rm N}$) from our CCF-{\sc scalpels}\cite{CollierCameron2021} analysis and a GP with a quasi-periodic exponential-sinusoidal-squared kernel ($\sigma_{\rm GP}$, inverse (squared) length-scale ($\alpha_{\rm GP}$), amplitude of the sinusoidal component ($\Gamma_{\rm GP}$), period of the quasi-periodic component (${\rm Prot}_{\rm GP}$) of the GP)\cite{Aigrain2012,Haywood2014,Barragan2022,Delisle2022,Nicholson2022,Aigrain2023,Hara2023} with wide uniform priors centred on the results from our blind RV analysis, see above. The formulations of the approximate Mat\'{e}rn and quasi-periodic exponential-sinusoidal-squared GP kernels ($k_i$($x_l$,$x_m$)) used within {\sc juliet} are:

\begin{align}
  k_i(x_l,x_m) = \sigma_{{\rm GP},i}^2 \left[(1+1/\epsilon)e^{-(1-\epsilon)\sqrt{3}\lvert x_l-x_m \rvert/\rho_{{\rm GP},i}}+(1-1/\epsilon)e^{-(1+\epsilon)\sqrt{3}\lvert x_l-x_m \rvert/\rho_{{\rm GP},i}}\right]\label{eq:gp_mat}\tag{S1}
\end{align}
\begin{align}
  k_i(x_l,x_m) = \sigma_{{\rm GP},i}^2\,{\rm exp}\left(-\alpha_{{\rm GP},i}\,\lvert x_l-x_m \rvert^2-\Gamma_{{\rm GP},i}\,{\rm sin}^2\left[\frac{\pi\lvert x_l-x_m \rvert}{{\rm Prot}_{{\rm GP},i}}\right]\right)\label{eq:gp_qpo}\tag{S2}
\end{align}

with $x_l$ and $x_m$ representing pairs of data to be fitted with the GPs. For the approximate Mat\'{e}rn kernel, the celerite parameter ($\epsilon$) was set to 0.01\cite{Espinoza2019}. We included jitter terms ($\sigma_{\rm jitter}$) to account for any residual noise in the light curves and RVs, and zero-point flux and RV offsets ($\mu$). 

The combined photometric analysis detects planets\,b,\,c,\,d, and\,e with statistical significances 50.0$\sigma$, 39.5$\sigma$, 59.9$\sigma$, and 47.5$\sigma$, respectively. The RV analysis detects four Keplerian orbits at 8.2$\sigma$, 6.9$\sigma$, 5.4$\sigma$, and 3.6$\sigma$. The results are given in Tables~1 and~S6. The fitted transit models with phase-folded, detrended TESS and CHEOPS data, and ground-based photometry are presented in Figs.~1 and~S7. The phase-folded RVs and planet models are reported in Fig.~2 and the HARPS-N RVs with the fitted summed Keplerian orbits presented in Fig.~S4. Posterior probability distributions for selected model parameters are shown in Figs.~S11 to~S14. 

To place these results in context, we selected well-characterised ($\sigma\,R_{\rm p}\,<\,$5\% \& $\sigma\,M_{\rm p}\,<\,$33\%) bodies orbiting M-dwarfs\cite{Southworth2011}, which are plotted in Fig.~3A. From this selection, we constructed a sample of small ($R_{\rm p} <$ 3\,R$_\oplus$) planets, shown in Fig.~3B, which compares them to mass-radius models\cite{Zeng2013,Lopez2014}. We plot the radii and Earth-like normalised bulk densities of these planets against their orbital periods in Figs.~4 and~S15 for comparison with the LHS\,1903 planets and observed and theoretical models for the radius and density valleys.

\subsection*{Planetary internal structure analysis}

The possible interior structure of the four planets in the system are inferred using a Bayesian analysis\cite{Dorn2015,Dorn2017}. Each planet is assumed to be composed of four layers: a central iron and sulfur core, a silicate mantle (containing Si, Mg, and Fe), a water layer, and a H and He gas layer. We adopted equations of state for water\cite{Haldemann2020}, the iron core\cite{Hakim2018} (which can also contain sulfur), and the silicate mantle\cite{Sotin2007} which depends on the mole fractions of Si, Mg, and Fe. We assumed that the planetary Si/Mg/Fe molar ratio is equal to the central star. There is observational evidence that these compositions are linked\cite{Adibekyan2021b}, however the correlation might not be one-to-one. The thickness of the gas envelope is adopted from previous work\cite{Lopez2014}; it depends on the age, planetary mass, etc. We did not consider the effect of the gas layer on the inner layers (core, mantle, water), because the mass of this atmospheric layer is negligible (see below). The assumed prior distributions of the mass fractions of core, mantle, and water layer are uniform on the simplex (the surface defined by the sum of the three mass fractions equal to one). This allows the core and mantle layers to have mass fractions between zero and one. We also constrained the mass fraction of water in the planet to be smaller than 0.5\cite{Thiabaud2014,Marboeuf2014}. The prior of the gas mass is assumed to be uniform in logarithmic space. The posterior probability distributions of the main internal structure parameters are shown in Figs.~S17 to~S20 for the four planets. 

We find the gas mass varies between the four planets (see Fig.~S16). The gas mass fraction increases from $\sim$10$^{-7.7}$ to 4\% for the three innermost planets, it is substantially reduced for the outermost planet\,e which is similar, in terms of gas fraction, to planet\,b. We find a similar result for the water fraction (see Fig.~S16), which is smaller in LHS\,1903\,e than in planet\,d. The derived parameters are shown in Figs.~S17 to~S20 and Table~S8.

\subsection*{Atmospheric escape analysis}

We used the system parameters and atmospheric mass fraction posterior probability distributions obtained above to constrain the planetary atmospheric mass at the dispersal of the protoplanetary disk, $f_{\rm atm}^{\rm start}$, using the {\sc pasta} planetary atmospheric evolution code\cite{Kubyshkina2019a,Kubyshkina2019b,Bonfanti2021b}. {\sc pasta} models the atmospheric evolution of close-in planets by assuming that the planets were not subject to substantial migration following the dispersal of the protoplanetary disk and that all planets hosted, or still host, a hydrogen-dominated atmosphere. We assumed that the planetary atmosphere is affected by escape due to both internal thermal heating and high energy (XUV) stellar irradiation\cite{Kubyshkina2018,Krenn2021}, where the latter evolves following the evolution of the stellar rotation period. {\sc pasta} models the evolution of the atmosphere of all planets in a system simultaneously, combining a model predicting planetary atmospheric mass-loss rates based on hydrodynamic simulations\cite{Kubyshkina2018}, a model of the evolution of the stellar XUV flux\cite{Bonfanti2021b}, a model relating planetary parameters and atmospheric mass\cite{Johnstone2015a}, and stellar evolutionary tracks\cite{Choi2016}. The free parameters of the model are $f_{\rm atm}^{\rm start}$ and the parameters of the power law controlling the evolution of the stellar rotation period\cite{Bonfanti2021b}. These are constrained within a Bayesian framework\cite{Cubillos2017}, employing the observed system parameters, with their uncertainties, as priors. Our analysis follows previous work\cite{Bonfanti2021b} except our fitted model parameter is the planetary atmospheric mass fraction, instead of the planetary radii. This avoids the continuous conversion of the atmospheric mass fraction into planetary radius, given the other system parameters\cite{Delrez2021}.

The results are shown in Fig.~S21. {\sc pasta} returns broad distributions that are unable to constrain $f_{\rm atm}^{\rm start}$. If atmospheric escape due to thermal processes was dominant in this system, we would expect narrow and defined initial atmospheric mass fraction posterior probability distributions as this would indicate such processes have occurred. However, the broad distributions that we find indicate that these processes did not dominate the atmospheric evolution of the LHS\,1903 planets.

\subsection*{Orbital stability}

Because LHS\,1903 contains four known planets with orbital periods below 30\,d, we test the long-term stability of the system and search for MMR pairs in the system.

We computed the orbital evolution of the planets over 100\,Myr ($\sim$1.23$\times10^9$ orbits of planet\,e) with the {\sc whfast} symplectic integrator\cite{Rein2015_whfast} implemented in the {\sc rebound} package\cite{Rein2012}, together with a symplectic corrector of order 17. The time step used for the integration is $10^{-4}$\,yr, or about 1/59\, of the orbital period of LHS\,1903\,b. General relativity effects due to the strong gravitational potential close to the star are accounted for, via a correction implemented in {\sc reboundx}\cite{Anderson1975, Tamayo2020}. We assumed the best-fitting orbital parameters from Table 1 and neglect their uncertainties.

Throughout the integration timespan, the system displays apparently stable and regular behaviour. This is illustrated in Fig.~S22, which presents the temporal evolution of the semi-major axes and orbital eccentricities of the four planets in the system. Fig.~S21A displays apparent straight lines, indicating the absence of substantial semi-major axis variations, and therefore, very limited exchange of energy between the planetary orbits. Fig.~S21B shows no any large excursions of eccentricity which could lead to destructive configurations.

From our global analysis, we obtained the period ratios of planet\,c to b ($P_c/P_b$)\,=\,2.889, planet\,d to c ($P_d/P_c$)\,=\,2.018, and planet\,e to d ($P_e/P_d$)\,=\,2.333 between successive planet pairs. The two last pairs are close to the 2/1 and 7/3 commensurabilities, respectively, so we explored the potential for those planet pairs to be in MMR. We investigated the chaotic dynamics around the best-fitting solution by computing a chaos map, a two-dimensional section of the parameter space in which chaos is locally estimated through numerical simulations. We built a 151$\times$151 grid in the parameter space ($P_d/P_c$, eccentricity of planet\,d ($e_d$)), and initially fixed all the other dynamical parameters to their best-fitting values. This produces 22801 unique system configurations that are used as initial conditions to short-term numerical integrations, which employed the high-order adaptive timestep {\sc ias15} integrator\cite{Rein2015_ias15} implemented in {\sc rebound}. Each system configuration is integrated on a 10\,kyr timespan, long enough for rapid chaos to develop. Throughout the integrations, we recorded the mean longitudes of the planets at regular time intervals for a total of 20\,000 output times. We applied a frequency analysis technique\cite{Laskar1988} on the simulations that terminated without a close encounter or escape of one body. With this technique, we derived the principal orbital frequency for each planet, i.e. the mean-motion. Frequency analysis was applied twice, on each half of the integration. We then computed the Numerical Analysis of Fundamental Frequencies (NAFF) fast chaos indicator\cite{Laskar1990, Laskar1993, Stalport2022}: 

\begin{align}
  \mathrm{NAFF} ~ = ~ \max_{j} ~ \left[ {\rm log}_{10} \dfrac{\mid n_{j,2} - n_{j,1} \mid}{n_{j,0}} \right] \tag{S3}
\end{align}

where $n_{j,1}$ and $n_{j,2}$ are the mean-motions of planet $j$ over the first and second half of the integration, respectively, and $n_{j,0}$ is the mean-motion at time $0$ which serves as a normalisation. The larger the drift in mean-motion over the two halves of the orbital evolution, the stronger the chaotic behaviour of the system. 

Fig.~S23 shows the resulting chaos map. The period ratio $P_d/P_c$ is explored by successively changing $P_d$ while fixing $P_c$ to its best-fitting value. The large feature around $P_d/P_c$=2 is associated with the 2:1 MMR. The planet\,c and\,d orbital pair is overall chaotic, likely because of the initial difference between the longitudes of periastron of the planets, yet there is a zone of high regularity at small eccentricity. We find planets\,c and\,d are outside the 2:1 MMR. Planets\,d and\,e are inside another, smaller resonant structure corresponding to the 7:3 MMR. 

The 7:3 MMR is a fourth-order resonance, so its dynamical influence occurs at moderate to high eccentricities. This is illustrated in Fig.~S23, where the width of the resonance appears very small at low eccentricity $e_d$, and the zone of high regularity exists at substantial eccentricity $e_d\sim$0.15. Close to the 7/3 commensurability, the specific integer combination of mean-longitudes ($\lambda$) of planets d and e (7$\lambda_e$-3$\lambda_d$) varies slowly. To test for resonant behaviour, we check if any of the resonant angles ($\Theta_{\rm N}$) associated with the 7:3 MMR oscillates around some value, i.e. libration. The resonant angles are:

\begin{align}
\begin{split}
\Theta_1 ~ & = ~ 7\lambda_e - 3\lambda_d - 4\varpi_d \\ 
\Theta_2 ~ & = ~ 7\lambda_e - 3\lambda_d - 3\varpi_d - \varpi_e \\ 
\Theta_3 ~ & = ~ 7\lambda_e - 3\lambda_d - 2\varpi_d - 2\varpi_e \\
\Theta_4 ~ & = ~ 7\lambda_e - 3\lambda_d - \varpi_d - 3\varpi_e \\
\Theta_5 ~ & = ~ 7\lambda_e - 3\lambda_d - 4\varpi_e. 
\end{split}\tag{S4}
\end{align}

They differ from each other by the combinations of longitudes of periastron ($\varpi$) of planets d and e. We searched for the libration of any resonant angle in the posterior probability distribution by defining a sub-sample of 25\,000 solutions drawn from the posterior, then computede their orbital evolution for 10kyr. Again, we employed the {\sc ias15} integrator of {\sc rebound} to perform the integrations. We find that 2.7$\%$ of the solutions have at least one of the 7:3 MMR resonant angles that librates over the course of the dynamical evolution. We also computed the NAFF chaos indicator for these simulations, with the same procedure as described above. The results are presented in Fig.~S24A, which shows our posterior probability distribution projected onto NAFF and the eccentricity of planet d. The resonant solutions are confined in the region where $\Theta_1$ librates, and are found at moderate eccentricity $e_d$. Fig.~S24B shows the temporal evolution of one resonant solution in the space ($e_d\cos\Theta_1$, $e_d\sin\Theta_1$) which shows libration: the system evolves on arcs and does not circulate. While most of the systems in our posterior do not harbour any libration of the resonant angles, this analysis shows that the data are compatible with the 7:3\,MMR but does not require it. 

We also used the NAFF results to set constraints on the orbital parameters from orbital stability. We first calibrated the NAFF chaos metric to orbital stability of a planetary system, following previous work\cite{Stalport2022}, therefore validating the use of the NAFF to define the planet system stability. This is based on the heuristic observation that better constrained systems tend to be less chaotic. We set a stability threshold at NAFF=-4, as shown in Fig.~S24A. We find that 60$\%$ of our posterior sample passes the stability criterion (NAFF $<$ -4), which means that the remaining 40$\%$ of our posterior either falls above the threshold line, or does not have any NAFF value computed because those posterior solutions did not reach the end of the simulation (due to a strong instability such as close encounter between two bodies). The stable sub-sample is used to update the estimations of the orbital parameters. The only substantial update is $e_d$, for which we find the value of 0.090$\substack{+0.045 \\ -0.035}$, which is still within 1$\sigma$ of the measurement reported in Table~1. All the resonant systems passed our stability criterion. 

We conclude that the best-fitting orbital parameters describe a stable system that is not in an MMR state. However, we do find alternative solutions within the posterior probability distribution for which the d-e planet pair is inside the 7:3 MMR. Therefore, the dynamical state of this planet pair is uncertain. 

\subsection*{Search for transit timing variations}

To search for the presence of TTVs in the LHS\,1903 system we conducted individual transit analyses for each planet using our combined photometry and {\sc juliet}\cite{Espinoza2019}. We fitted observed transit centre times for 38, 14, 6, and 5 transits of planet\,b,\,c,\,d, and\,e, respectively, then compared them to calculated values computed from the ephemerides obtained from our best-fitting orbital parameters. The results are shown in Fig.~S25. We find no TTVs for LHS\,1903\,b and\,c, though there are small variations in the planet\,d and\,e timings. From these data alone, we cannot determine whether a 7:3 MMR between these planets is the cause of these variations. Our individual transit analyses also yielded planet-to-star radius ratios and impact parameters, which were consistent with the values listed in Table~1. \\



\begin{figure}[htbp]
\centerline{\includegraphics[width=\columnwidth]{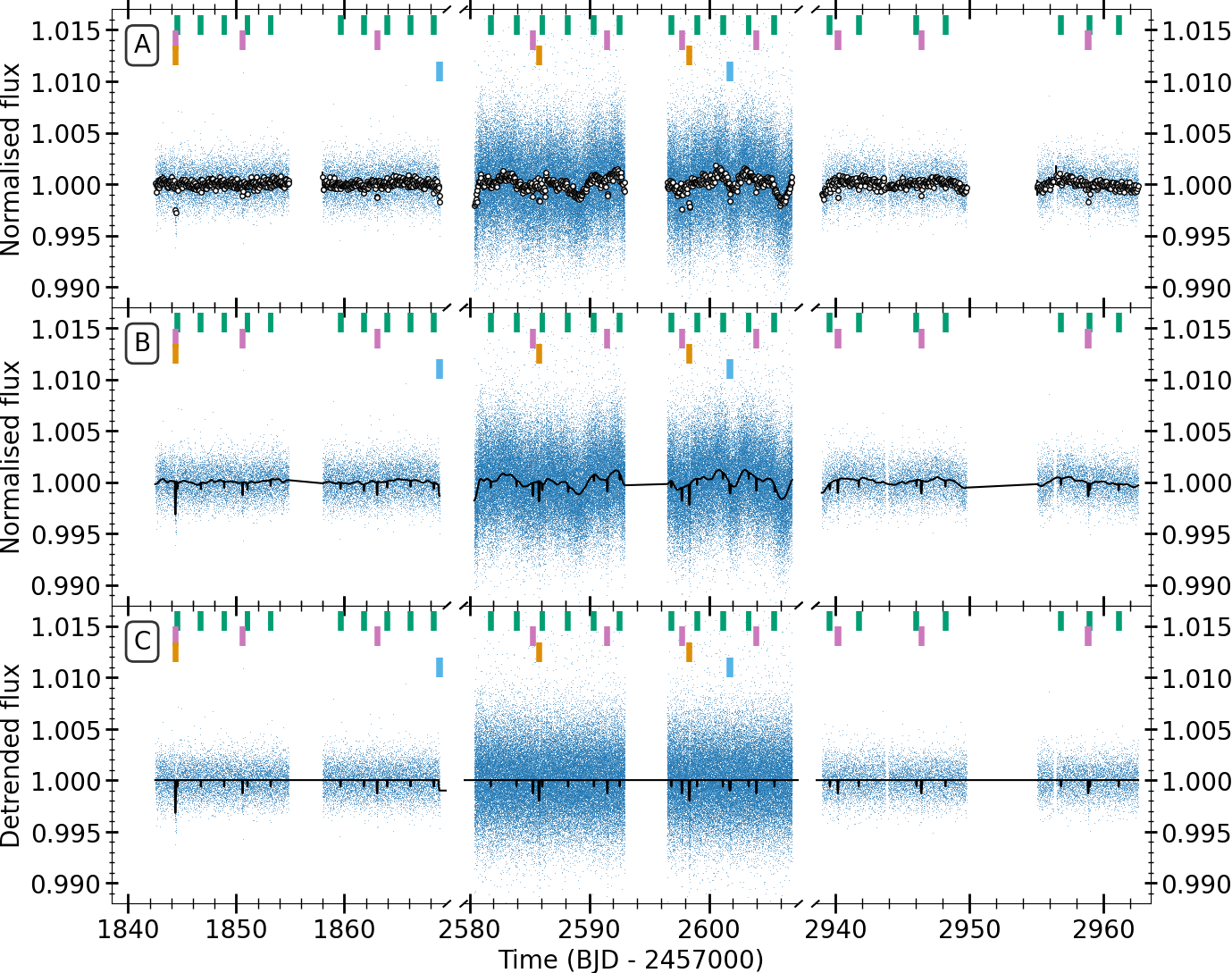}}
\noindent
\begin{small}{\bf Fig. S1: TESS transit photometry of the LHS\,1903 system}. TESS Sector 20 (left), 47 (centre), and 60 (right) light curves with transits of LHS\,1903\,b,\,c,\,d, and\,e indicated by green, purple, orange, and cyan bars. ({\bf A}) Blue points are PDCSAP 2\,min and 20\,s photometry and white circles with black outlines are the same data binned to 1\,hr. The larger photometric scatter in Sector 47 is due to the higher cadence observations and therefore poorer cosmic ray mitigation, combined with spacecraft jitter. ({\bf B}) The same data overlain with a black line showing the fitted transit and GP model from the simultaneous fitting. ({\bf C}) Detrended TESS photometry (blue points) overlain with a black line showing the transit model fitted to the data. \end{small} 
\end{figure}

\begin{figure}[htbp]

    {{\includegraphics[width=0.329\columnwidth]{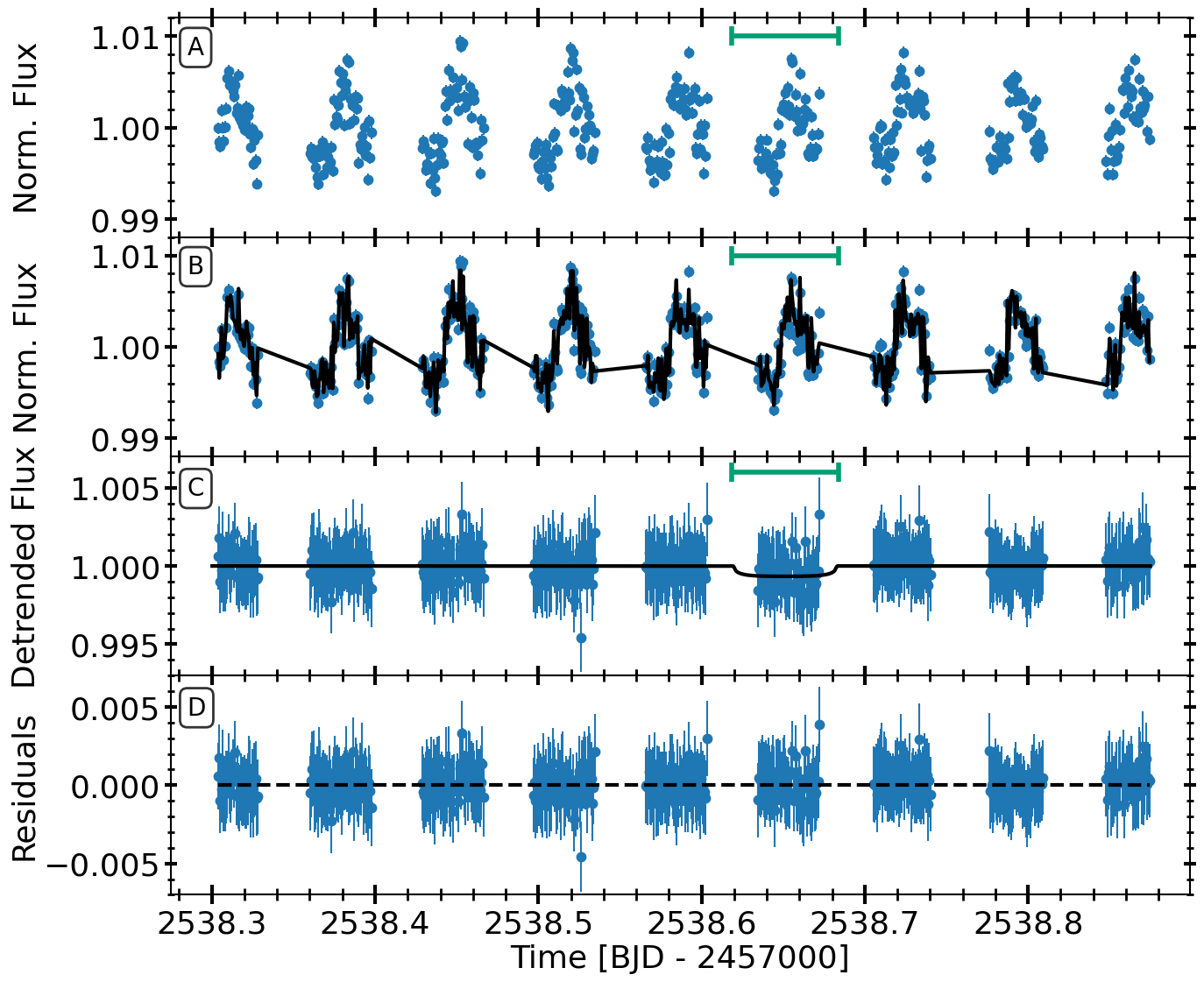} }}
    \hspace{-0.25cm}
    {{\includegraphics[width=0.329\columnwidth]{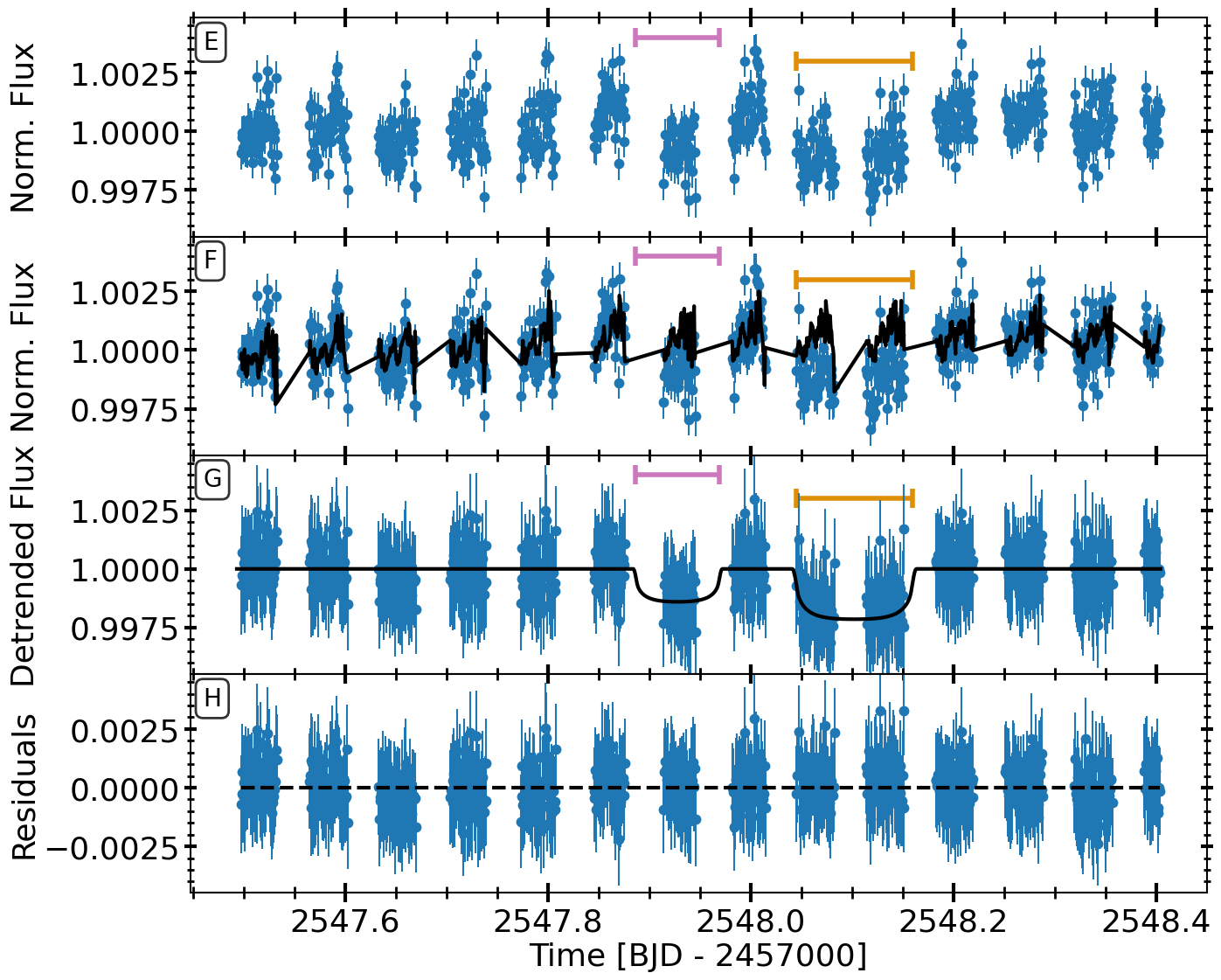} }}
    \hspace{-0.25cm}
    {{\includegraphics[width=0.329\columnwidth]{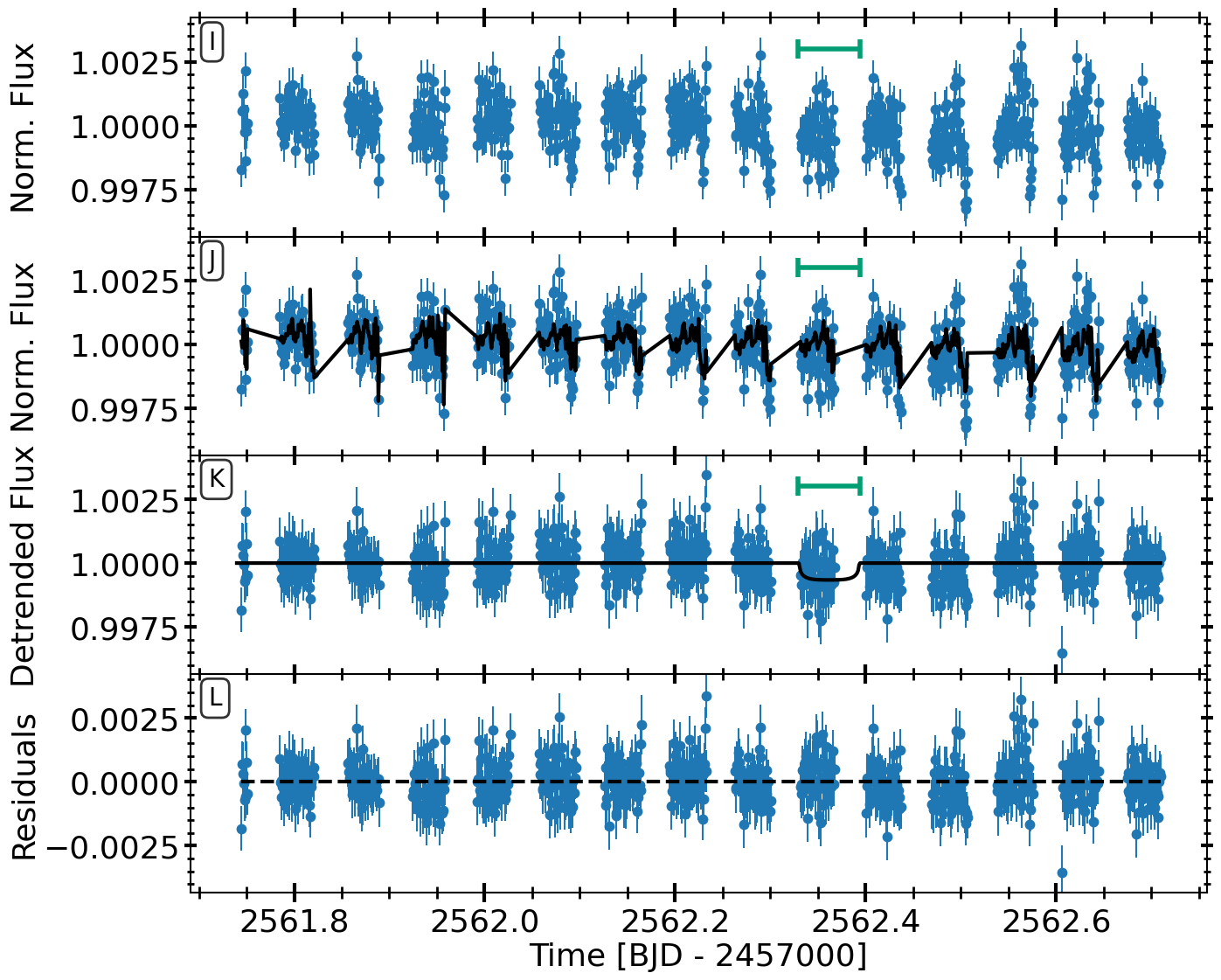} }}
    
    {{\includegraphics[width=0.329\columnwidth]{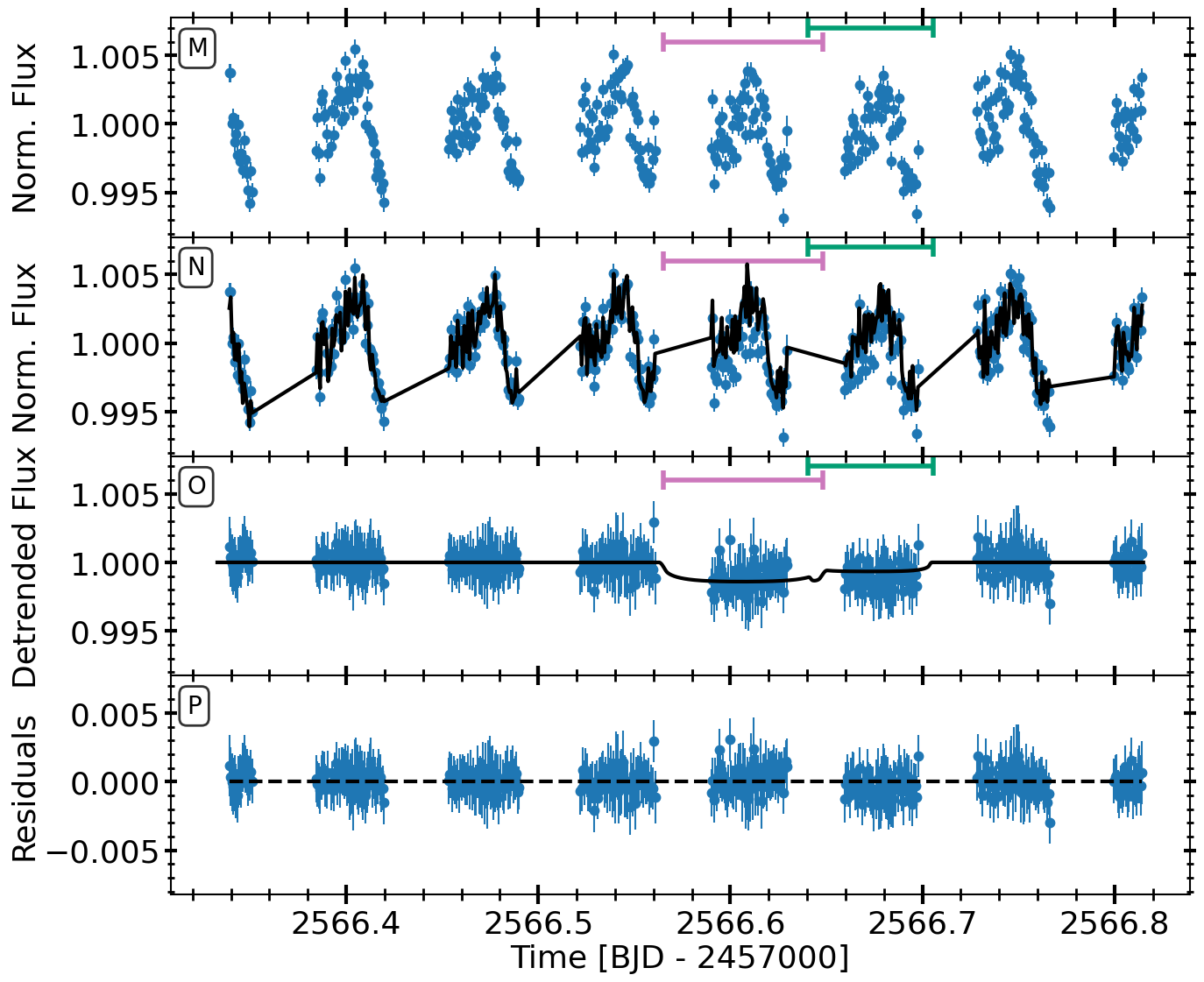} }}
    \hspace{-0.25cm}
    {{\includegraphics[width=0.329\columnwidth]{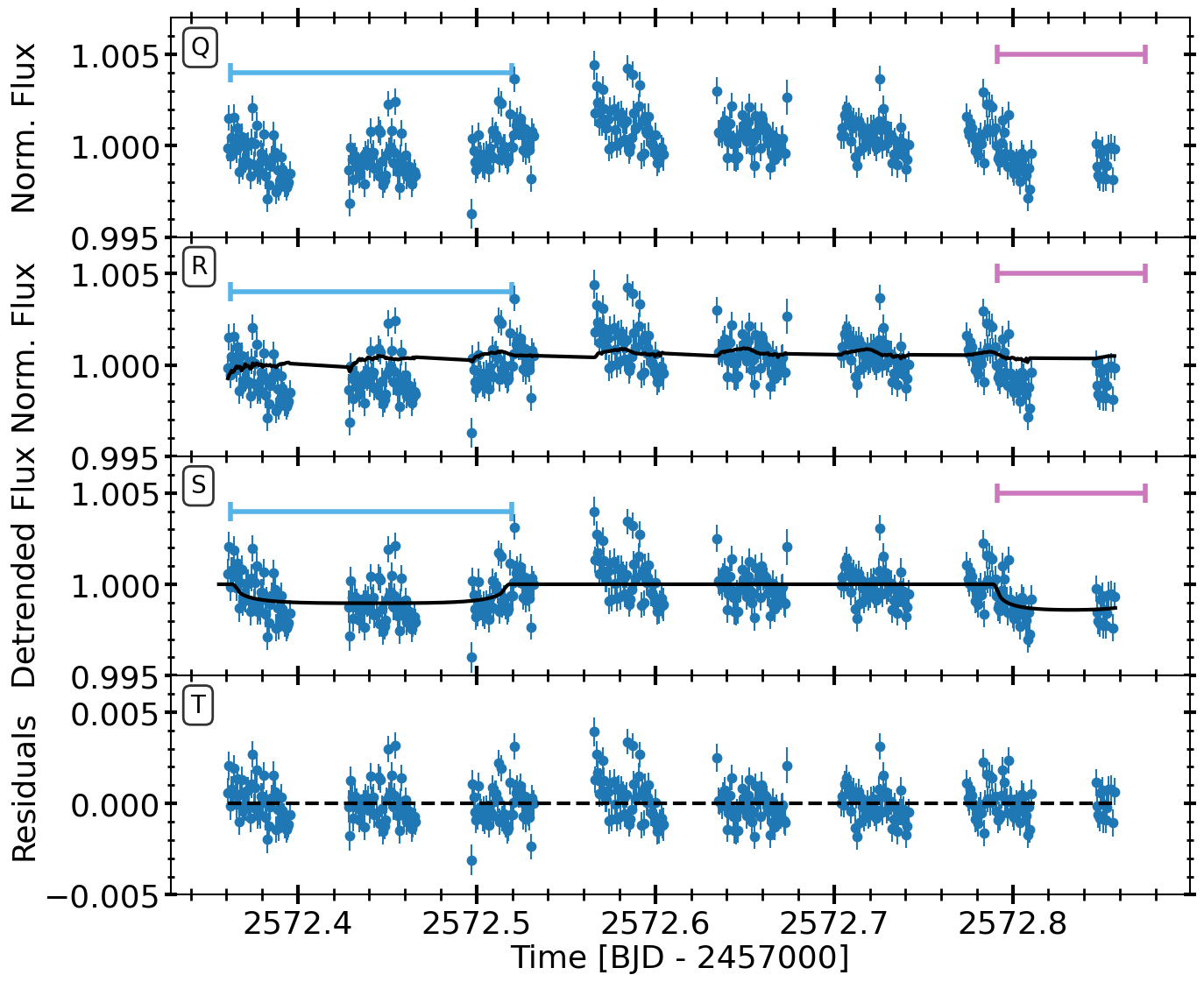} }} 
     \hspace{-0.25cm}
    {{\includegraphics[width=0.329\columnwidth]{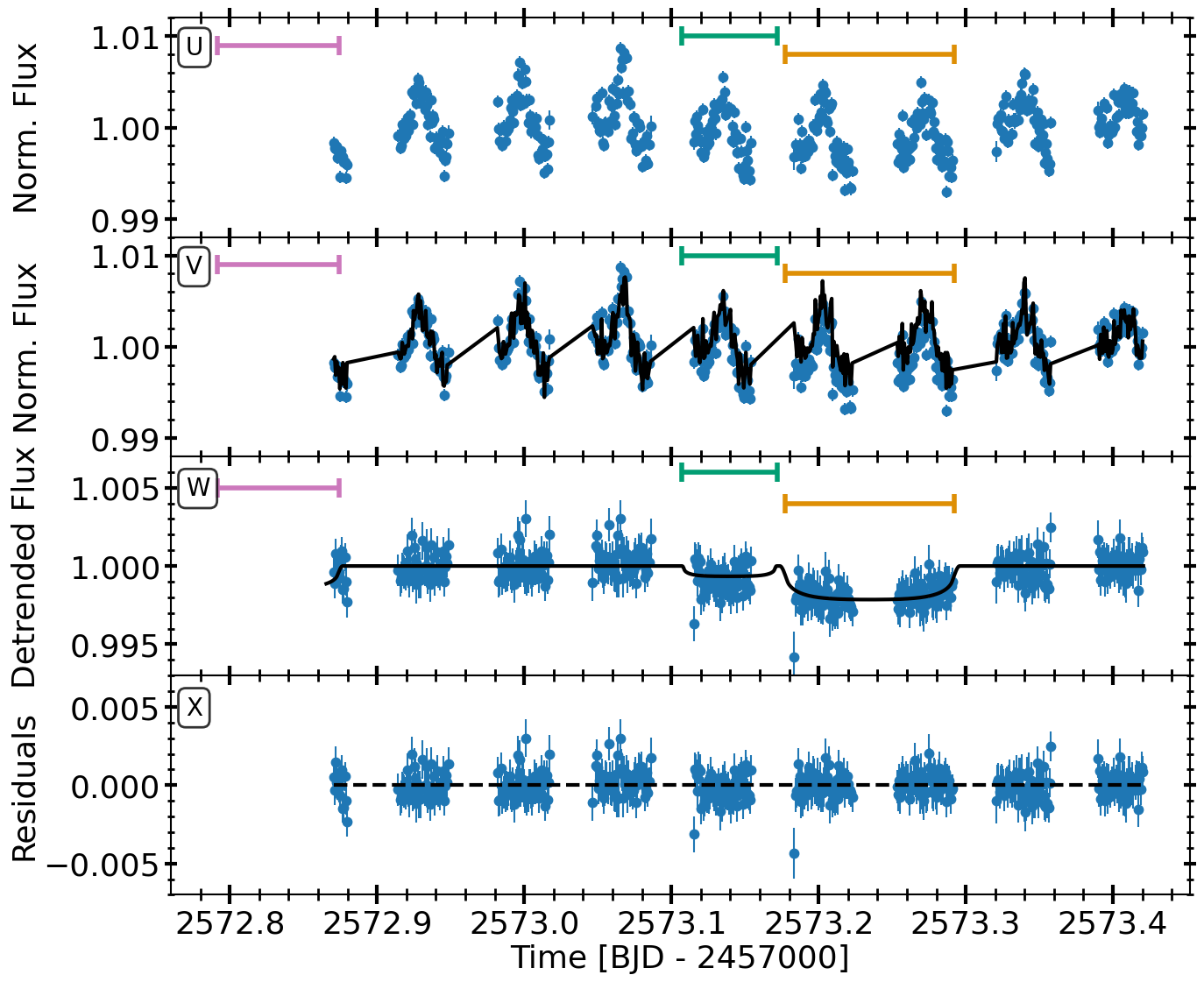} }}

    {{\includegraphics[width=0.329\columnwidth]{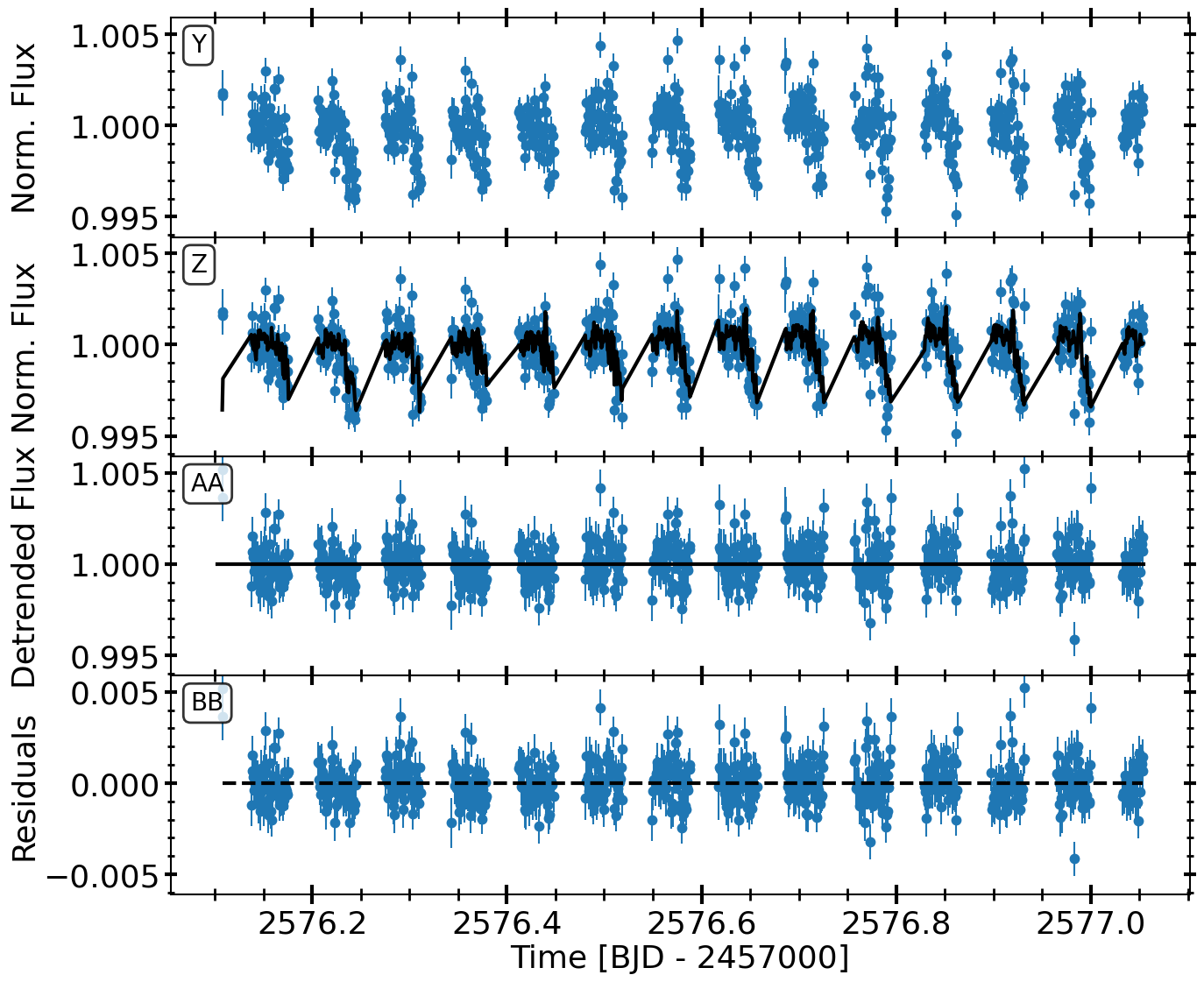} }}
    \hspace{-0.25cm}
    {{\includegraphics[width=0.329\columnwidth]{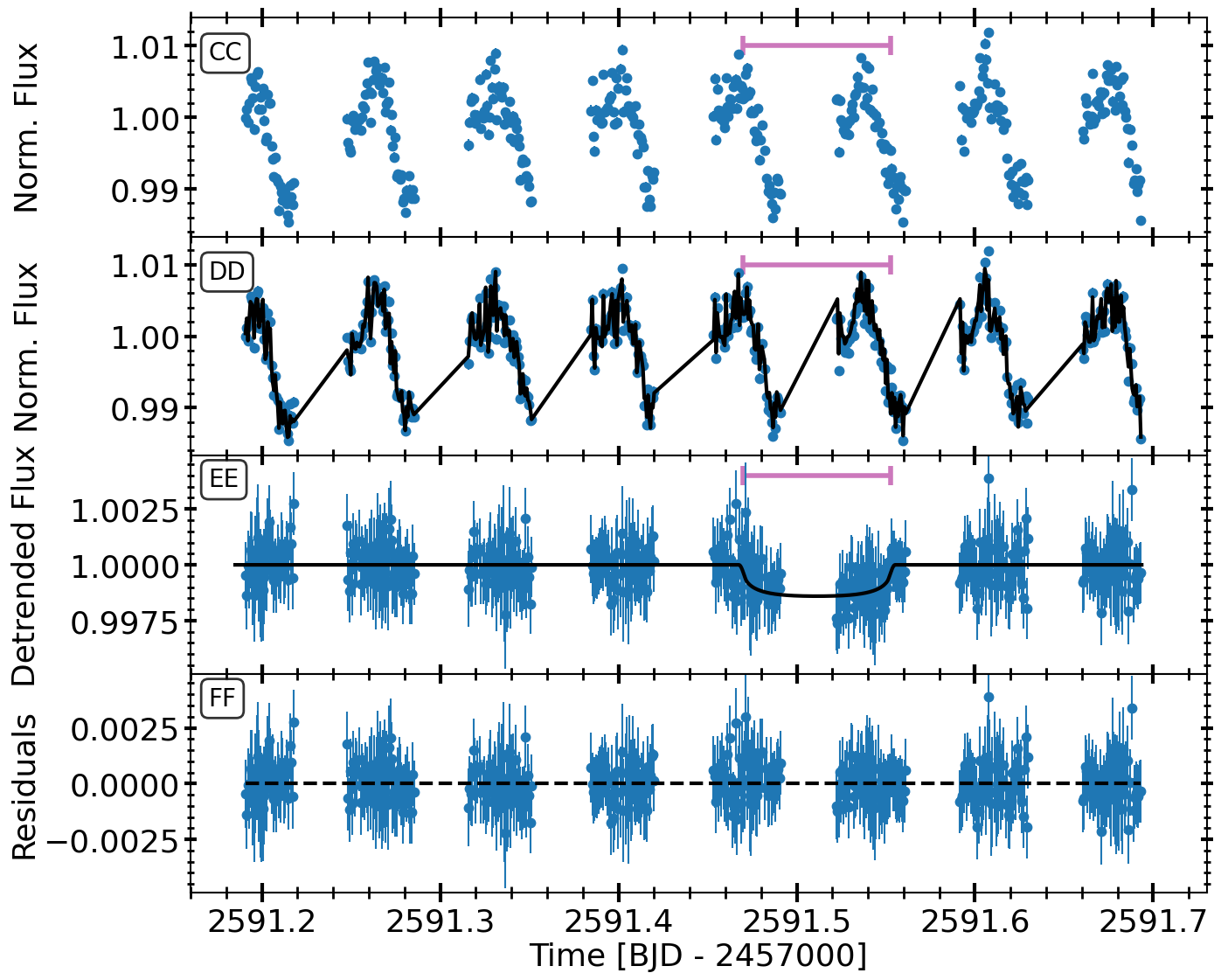} }}
    \hspace{-0.25cm}
    {{\includegraphics[width=0.329\textwidth]{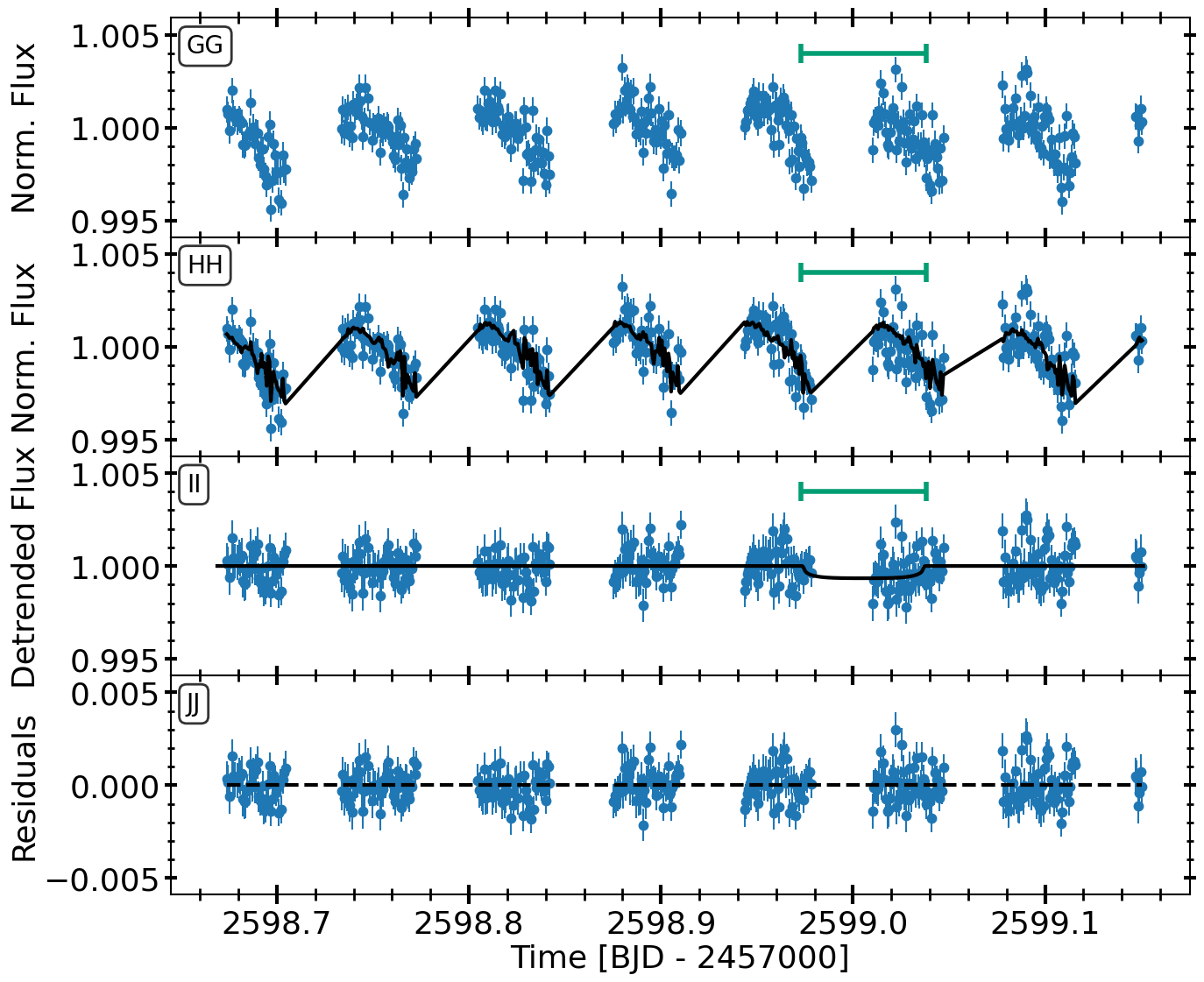} }}
    
    {{\includegraphics[width=0.329\textwidth]{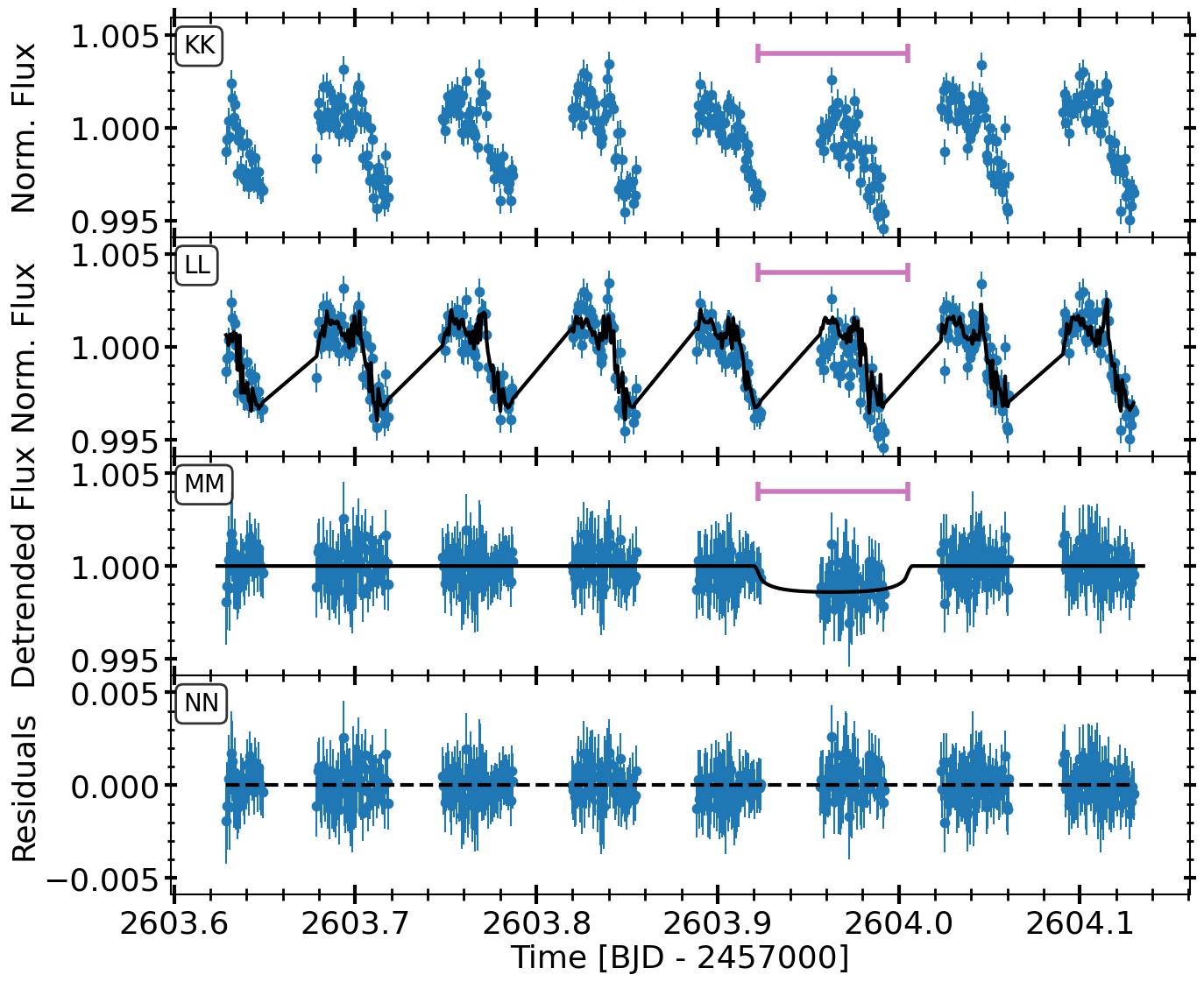} }}
    \hspace{-0.25cm}
    {{\includegraphics[width=0.329\textwidth]{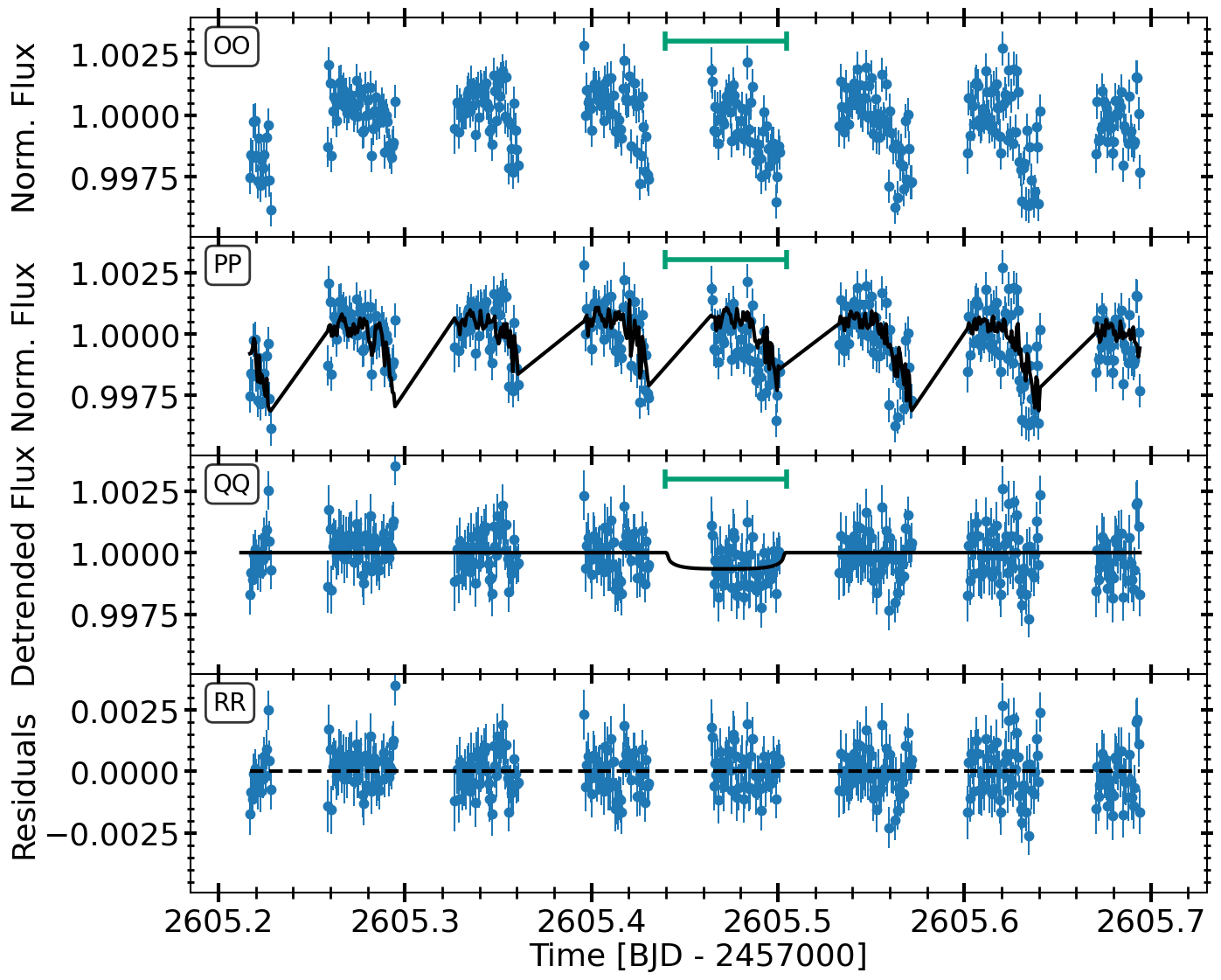} }}
    \hspace{-0.25cm}
    {{\includegraphics[width=0.329\columnwidth]{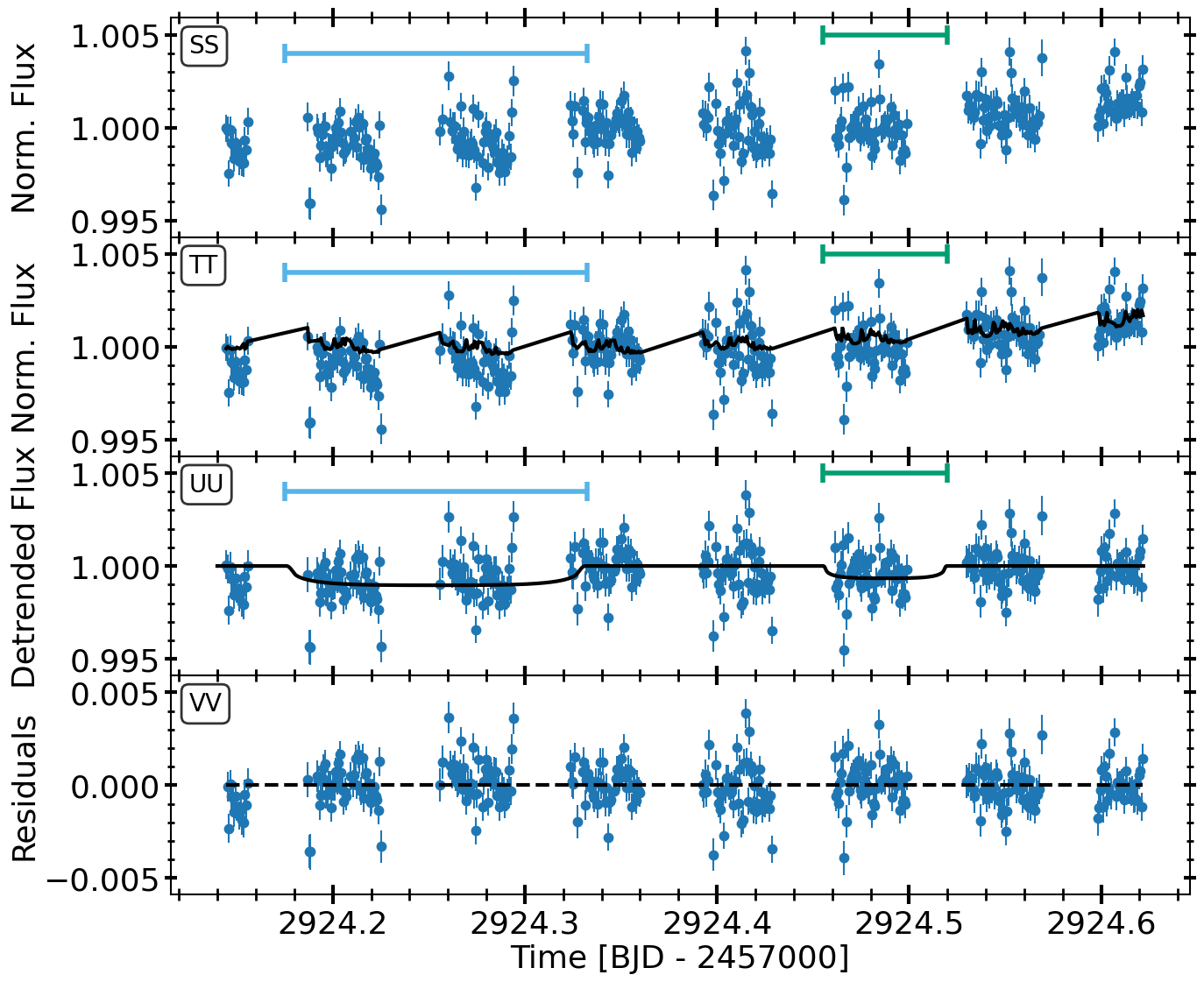} }}
    \vspace{-1.15cm}

\noindent
\begin{small}{\bf Fig. S2: CHEOPS transit photometric data and 1$\sigma$ uncertainties of the LHS\,1903 system}. Time-ordered normalised light curves with truncated visit file keys are listed in Table~S1. Transits of LHS\,1903\,b,\,c,\,d, and\,e are indicated by green, purple, orange, and cyan bars. ({\bf A, E, I, M, Q, U, Y, CC, GG, KK, OO, SS}) The flux produced by the CHEOPS DRP. ({\bf B, F, J, N, R, V, Z, DD, HH, LL, PP, TT}) The linear noise model from the instrument and PSF-{\sc scalpels} components fitted to the DRP fluxes. ({\bf C, G, K, O, S, W, AA, EE, II, MM, QQ, UU}) The detrended data and transit model with the flux errors inflated by the linear model uncertainties. ({\bf D, H, L, P, T, X, BB, FF, JJ, NN, RR, VV}) The residuals to the fit. \end{small} 
\end{figure}

\begin{figure}[htbp]

    {{\includegraphics[width=0.329\textwidth]{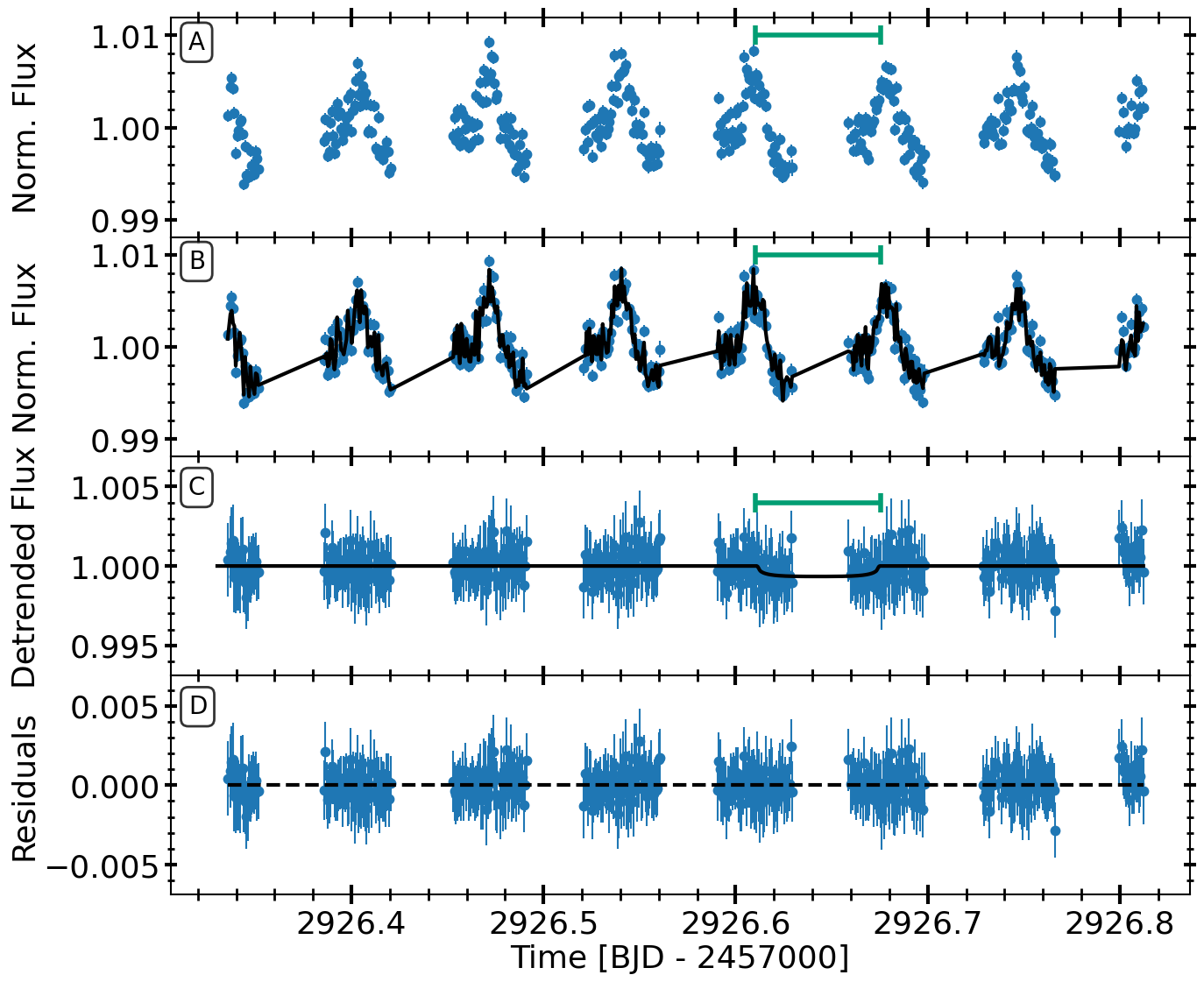} }}
    \hspace{-0.25cm}
    {{\includegraphics[width=0.329\textwidth]{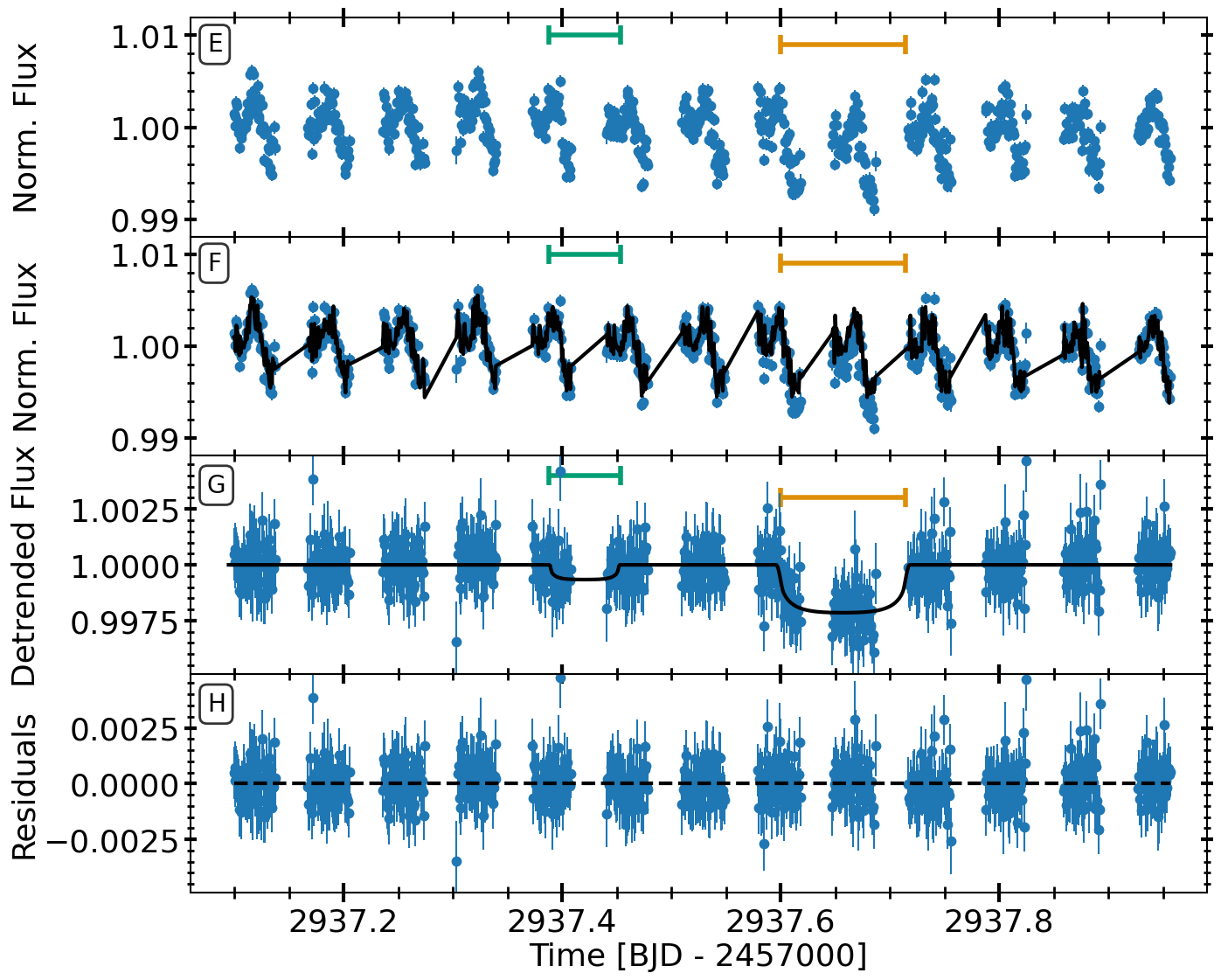} }}
    \hspace{-0.25cm}
    {{\includegraphics[width=0.329\textwidth]{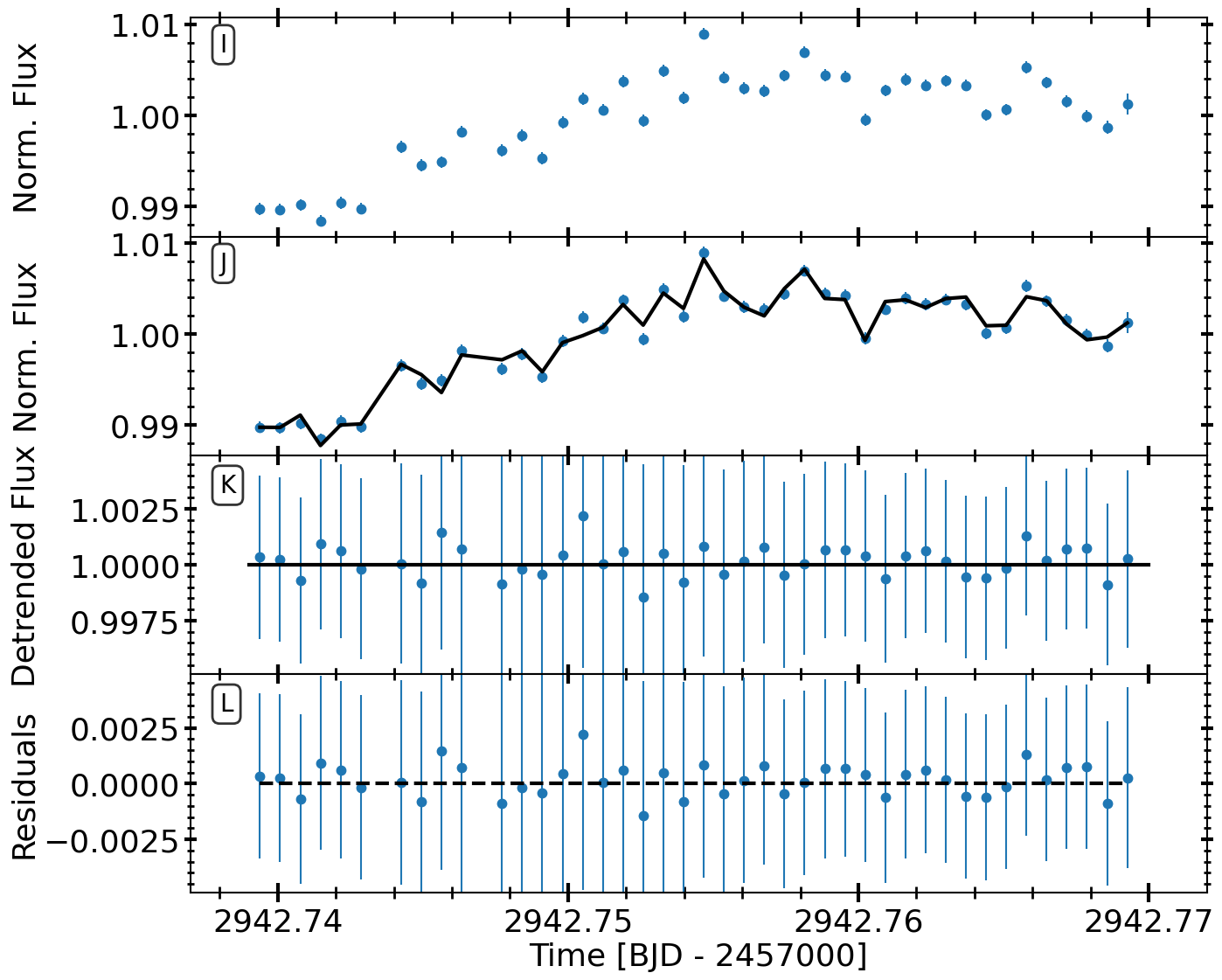} }}

    {{\includegraphics[width=0.329\columnwidth]{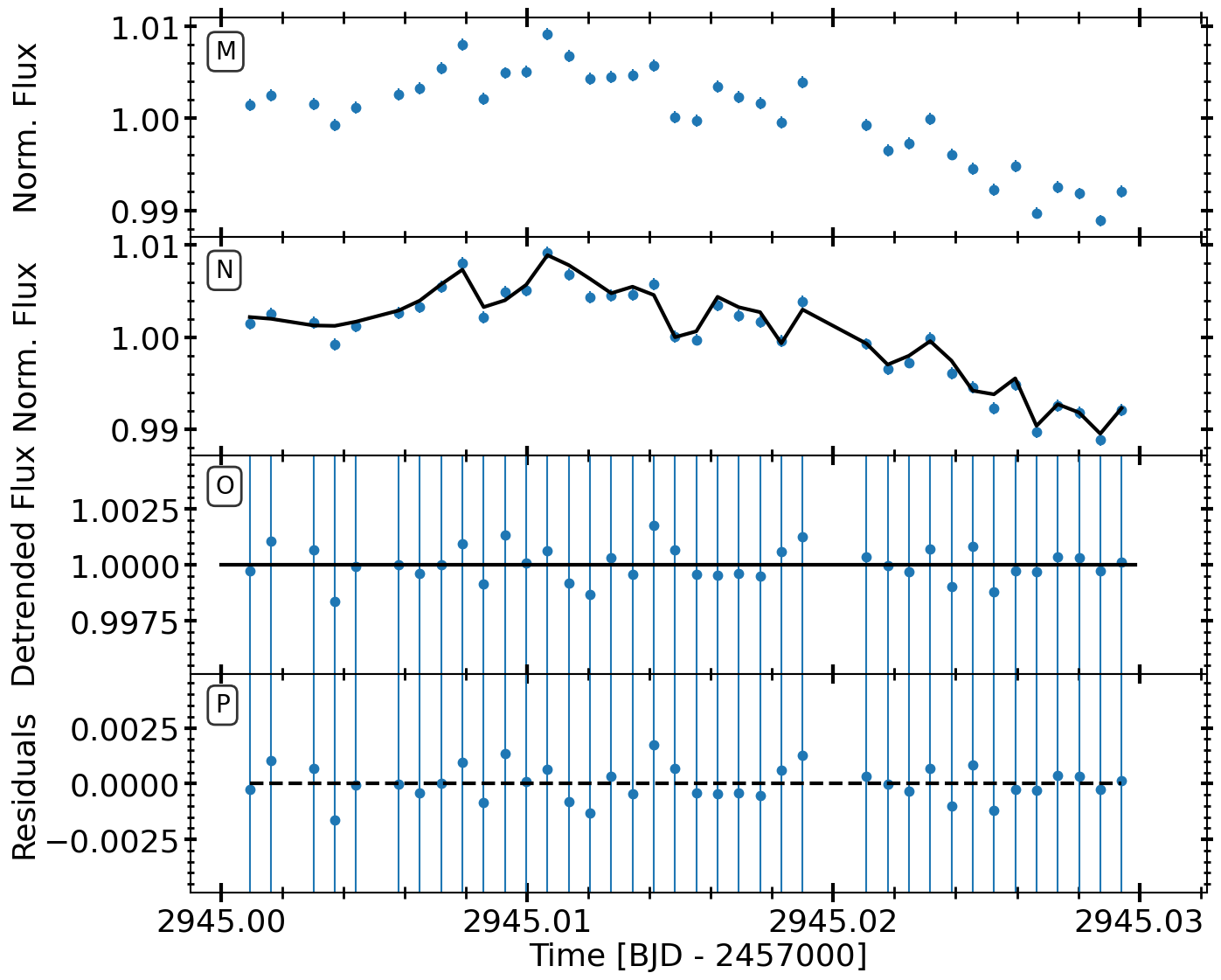} }}
    \hspace{-0.25cm}
    {{\includegraphics[width=0.329\textwidth]{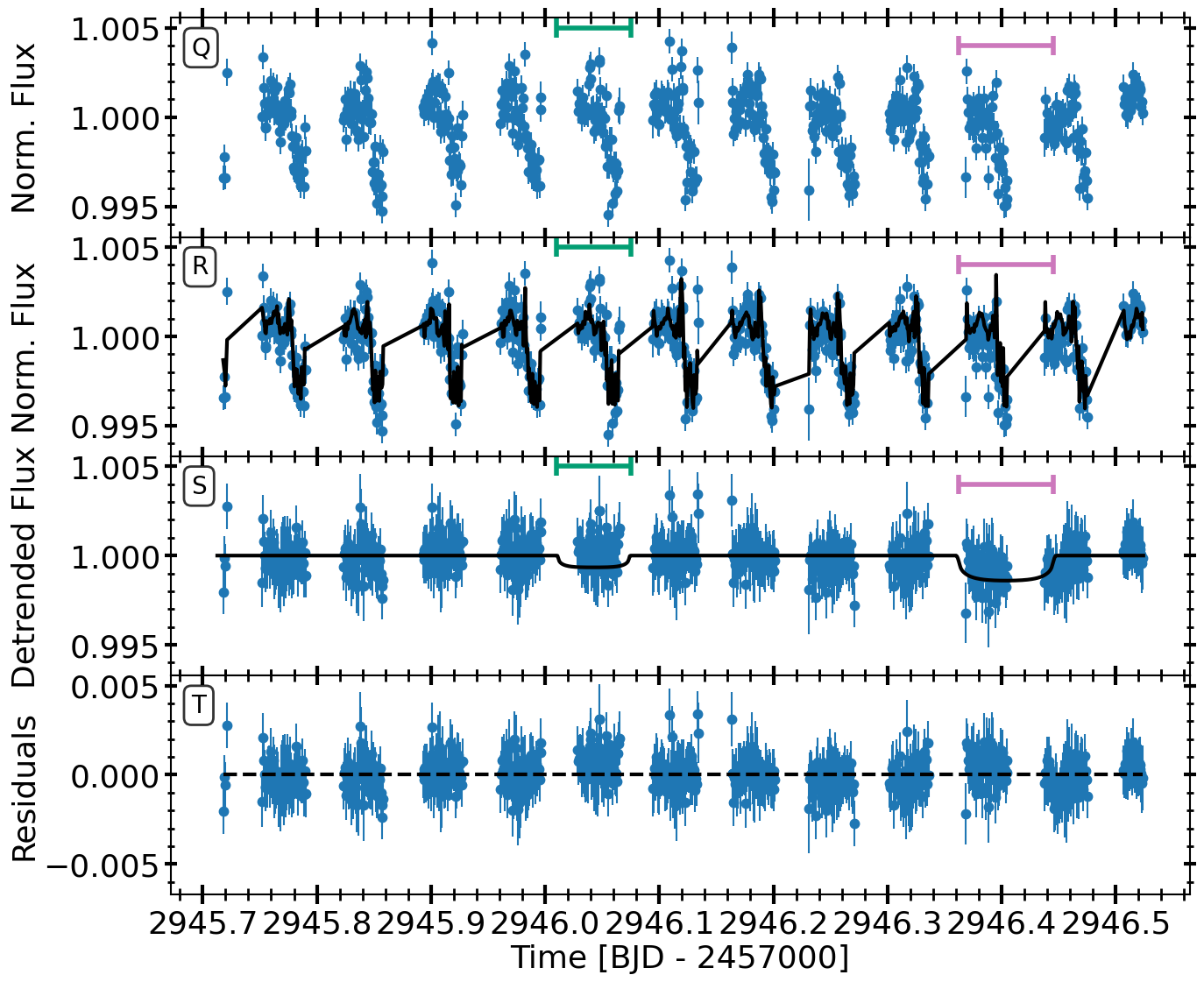} }}
    \hspace{-0.25cm}
    {{\includegraphics[width=0.329\textwidth]{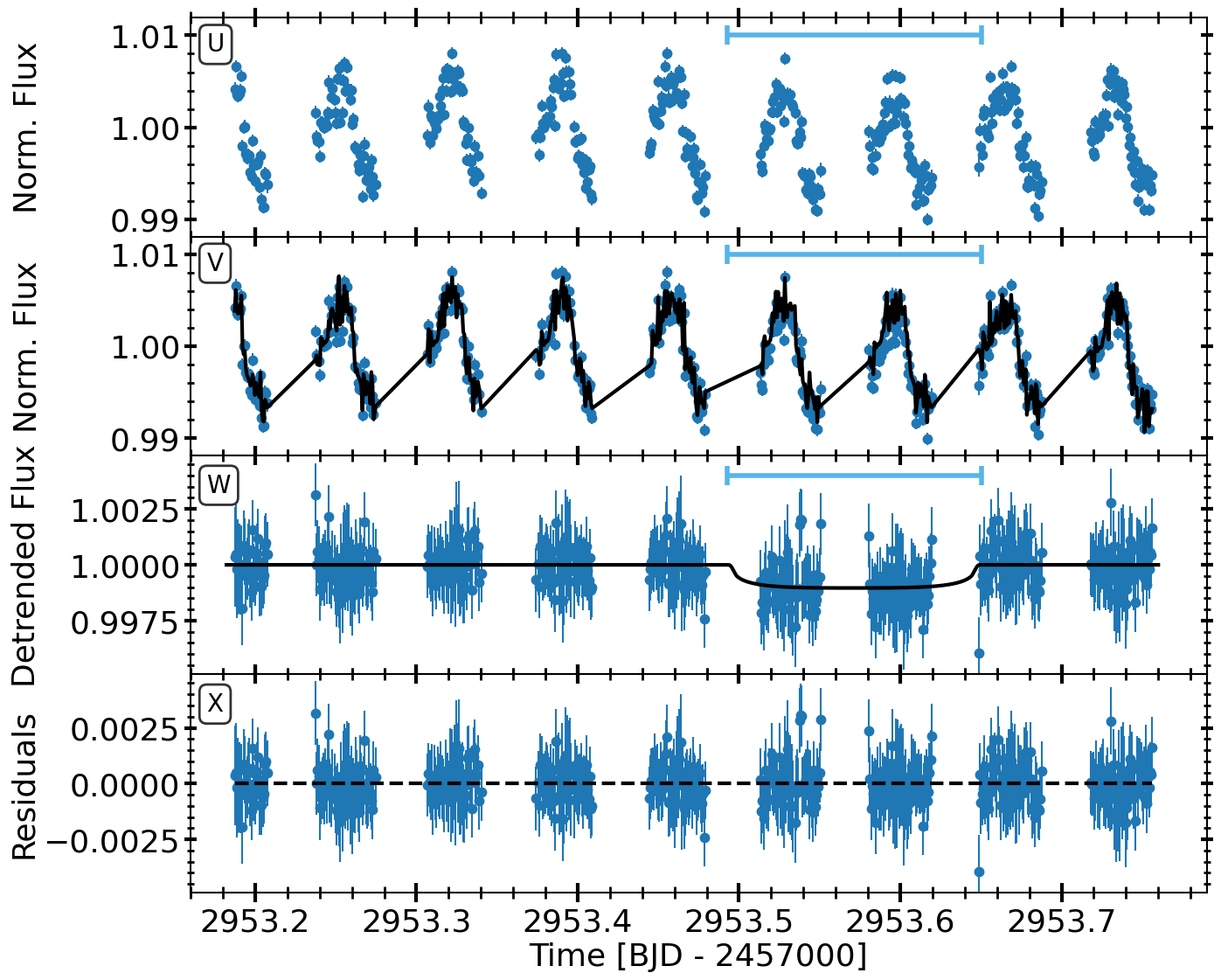} }}

    {{\includegraphics[width=0.329\textwidth]{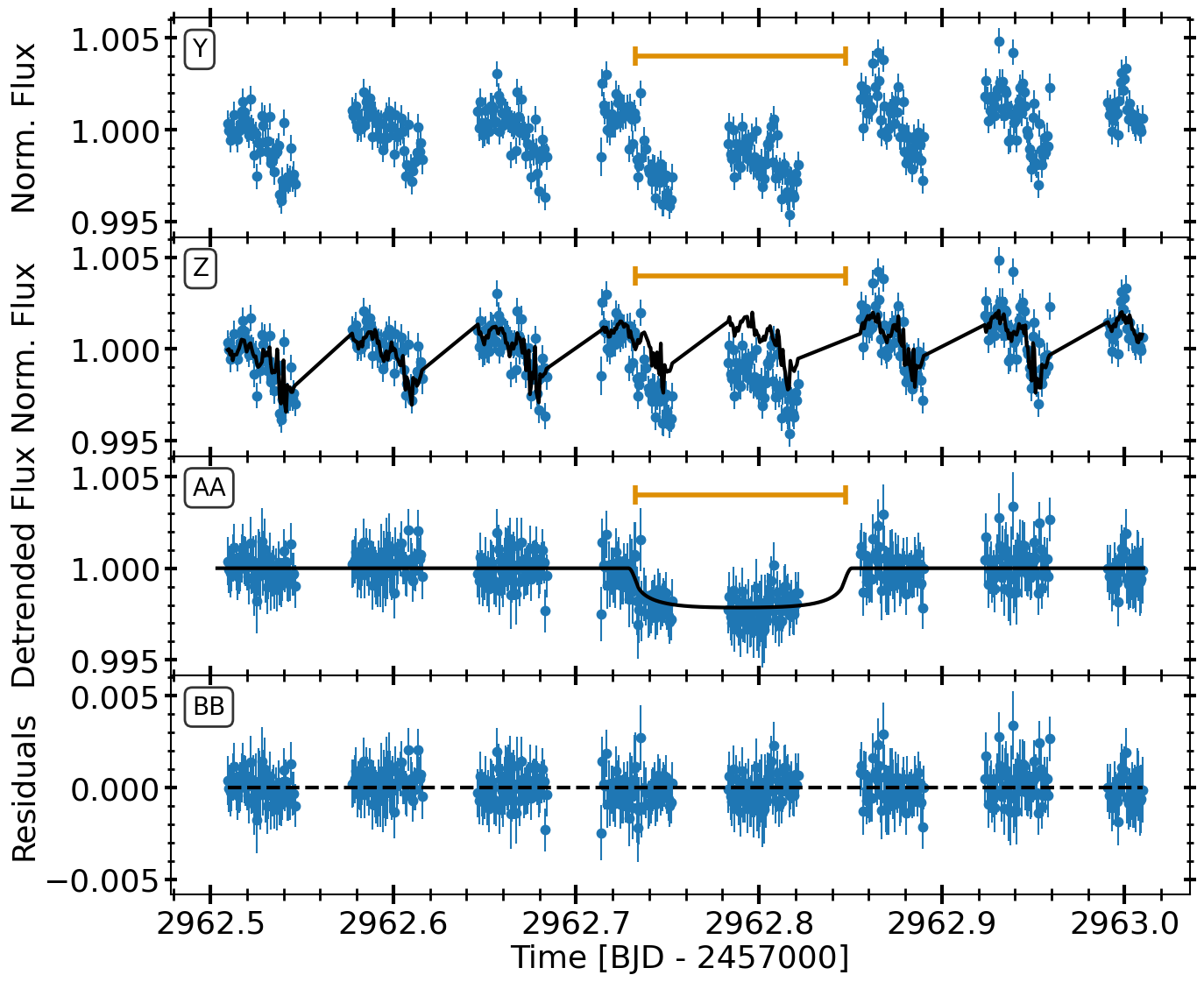} }}
    \hspace{-0.25cm}
    {{\includegraphics[width=0.329\columnwidth]{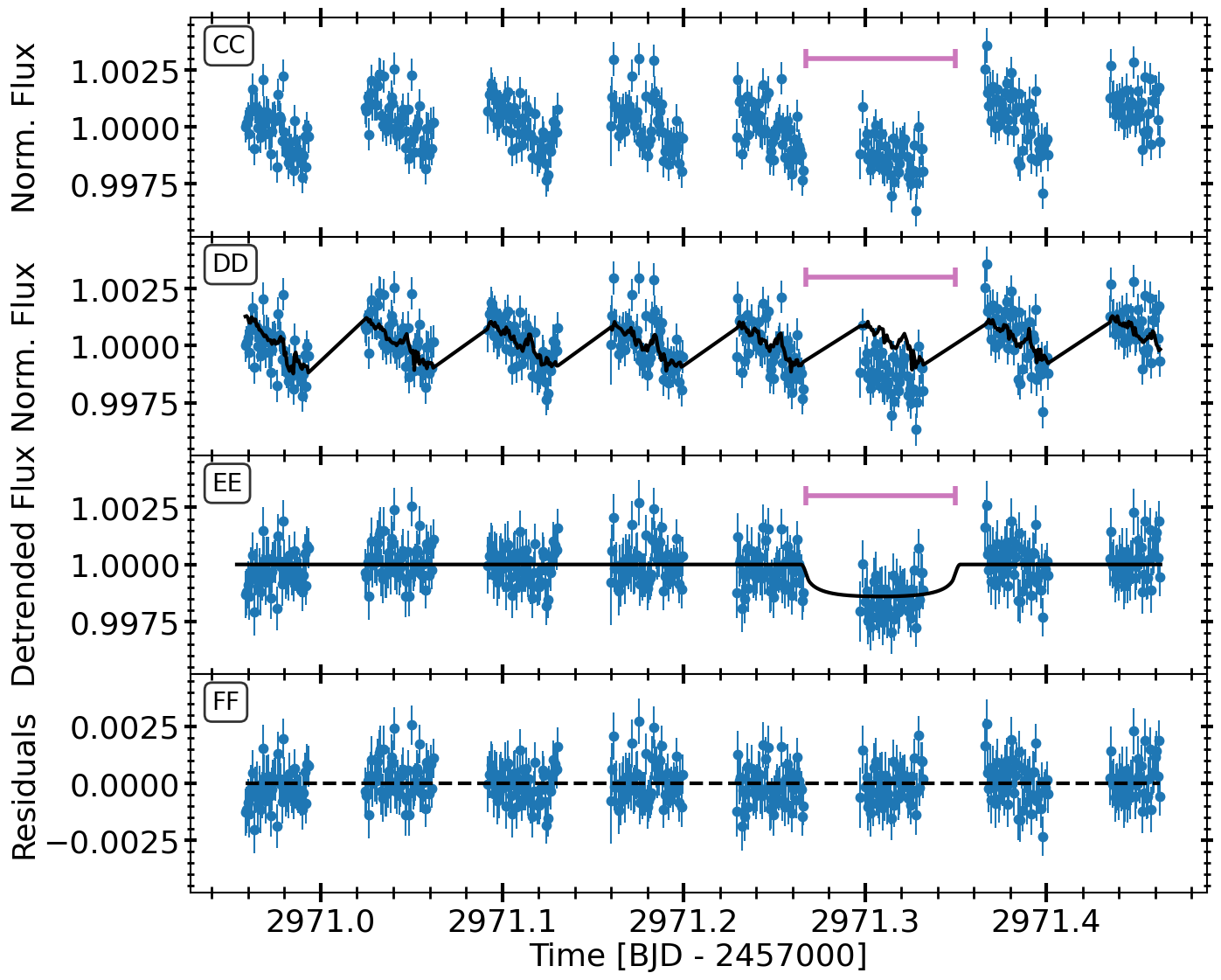} }}
    \hspace{-0.25cm}
    {{\includegraphics[width=0.329\textwidth]{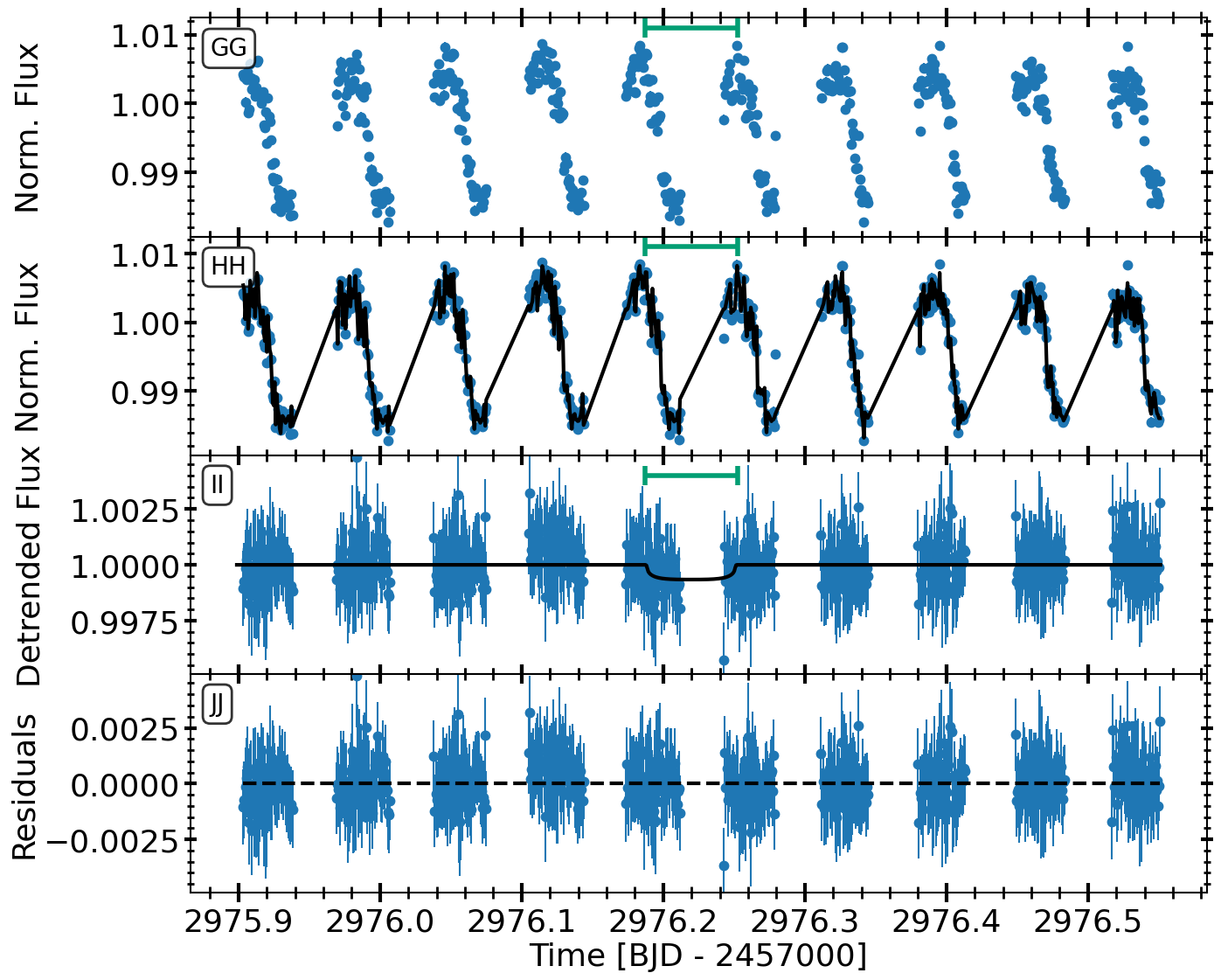} }}
    
    {{\includegraphics[width=0.329\textwidth]{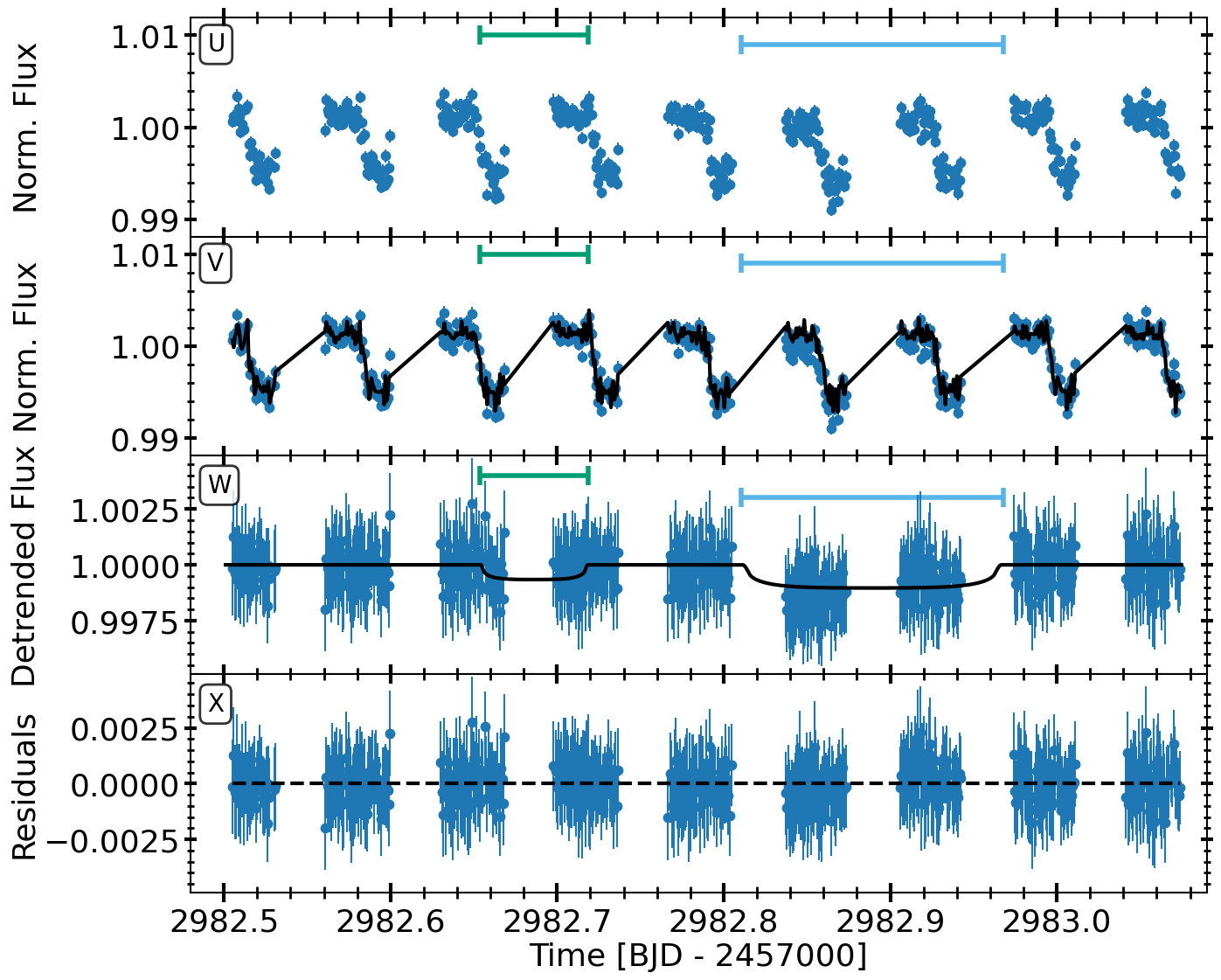} }}

\noindent
\begin{small}{\bf Fig. S3: CHEOPS transit photometric data and 1$\sigma$ uncertainties of the LHS\,1903 system}. Time-ordered normalised light curves with truncated visit file keys are listed in Table~S1. Transits of LHS\,1903\,b,\,c,\,d, and\,e are indicated by green, purple, orange, and cyan bars. ({\bf A, E, I, M, Q, U, Y, CC, GG, KK}) The flux produced by the CHEOPS DRP. ({\bf B, F, J, N, R, V, Z, DD, HH, LL}) The linear noise model from the instrument and PSF-{\sc scalpels} components fitted to the DRP fluxes. ({\bf C, G, K, O, S, W, AA, EE, II, MM}) The detrended data and transit model with the flux errors inflated by the linear model uncertainties. ({\bf D, H, L, P, T, X, BB, FF, JJ, NN}) The residuals to the fit. \end{small} 
\end{figure}

\begin{figure}[htbp]
\centerline{\includegraphics[width=0.8\columnwidth]{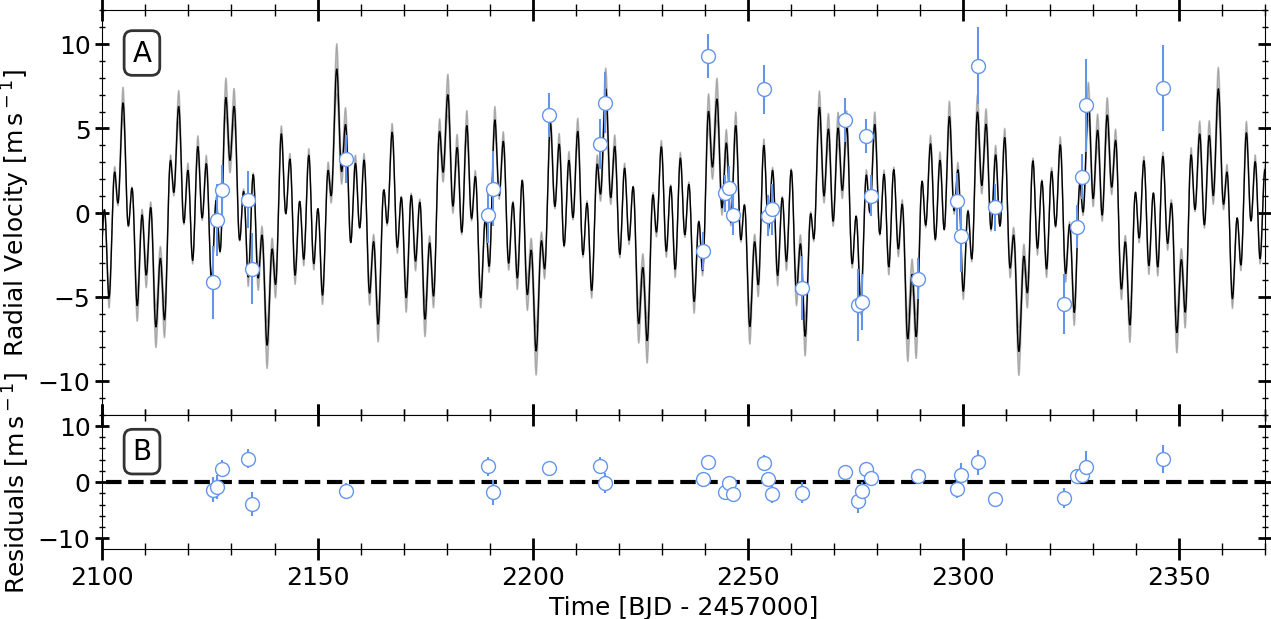}}\vspace{0.2cm}
\centerline{\includegraphics[width=0.8\columnwidth]{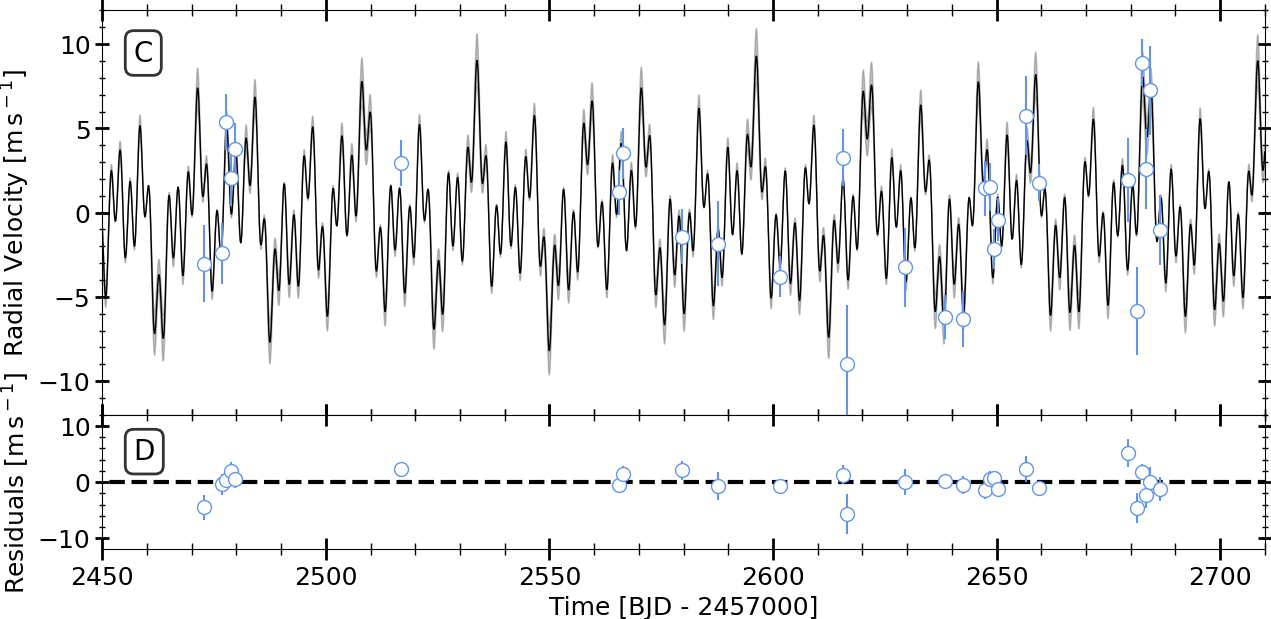}}\vspace{0.2cm}
\centerline{\includegraphics[width=0.8\columnwidth]{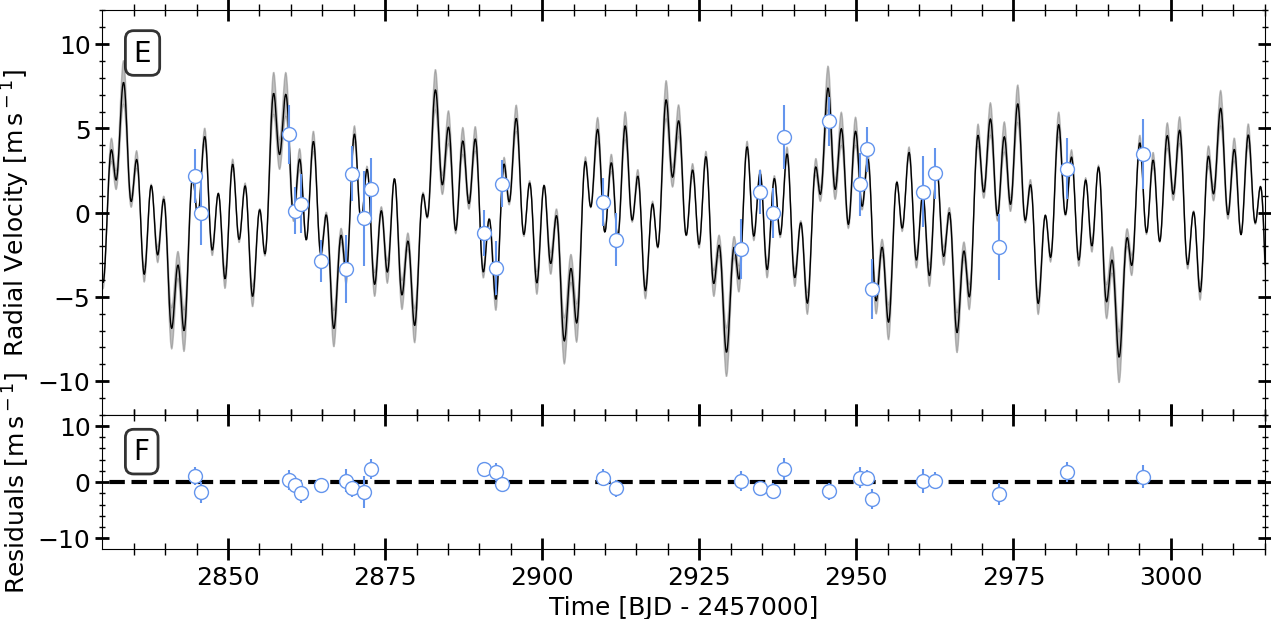}}
\noindent
\begin{small}{\bf Fig. S4: HARPS-N radial velocity data and 1$\sigma$ uncertainties of the LHS\,1903 system}. {\sc s-bart} extracted RVs detrended with {\sc scalpels} and a GP plotted as blue open circles. ({\bf A, C, E}) The best-fitting four Keplerian signal model in black with 1$\sigma$ uncertainty shown in grey from our simultaneous model fitting for highlighting the three HARPS-N seasons separately. ({\bf B, D, F}) Residuals to the model fit for highlighting the three HARPS-N seasons separately. \end{small} 
\end{figure} 

\begin{figure}[htbp]
\centerline{\includegraphics[width=\columnwidth]{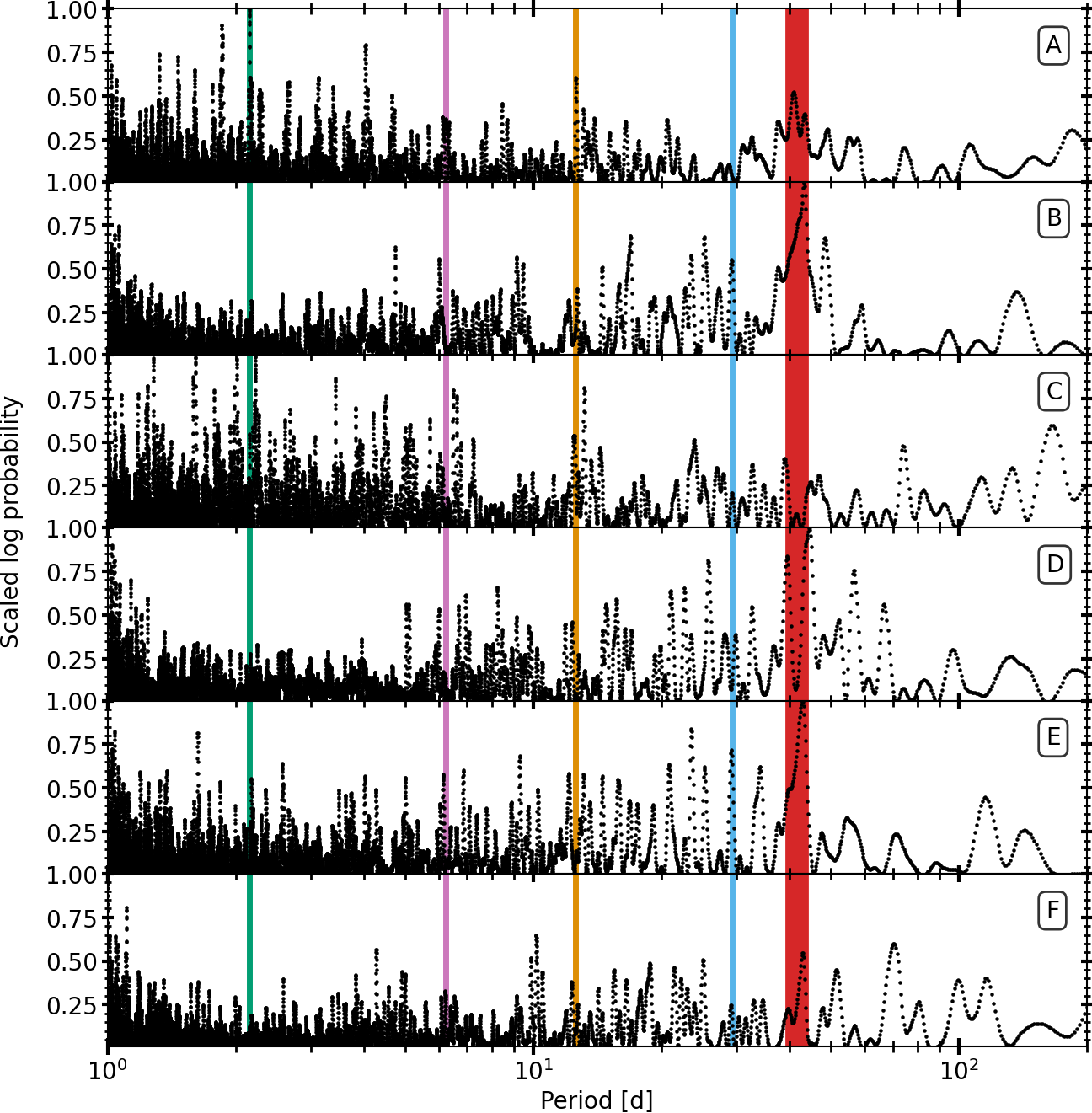}}
\noindent
\begin{small}{\bf Fig. S5: Bayesian Generalised Lomb-Scargle periodograms of the HARPS-N radial velocities and stellar activity indicators}. ({\bf A}) RVs, ({\bf B}) CCF FWHM, ({\bf C}) CCF bisector span (BIS), ({\bf D}) CCF contrast, ({\bf E}) Ca\,{\sc ii} H\,\&\,K (S-index), and ({\bf F}) H$\alpha$. The y-axis shows the scaled log probability for each dataset. The orbital periods of the LHS\,1903\,b,\,c,\,d, and\,e derived from the transit photometry are shown as green, purple, orange, and cyan vertical bars, respectively. The vertical red region indicates the stellar rotation period. The removal of this stellar signal with a GP reveals the RV signal of the transiting planet LHS\,1903\,e, see Fig.~S6. \end{small} 
\end{figure} 

\begin{figure}[htbp]
\centerline{\includegraphics[width=\columnwidth]{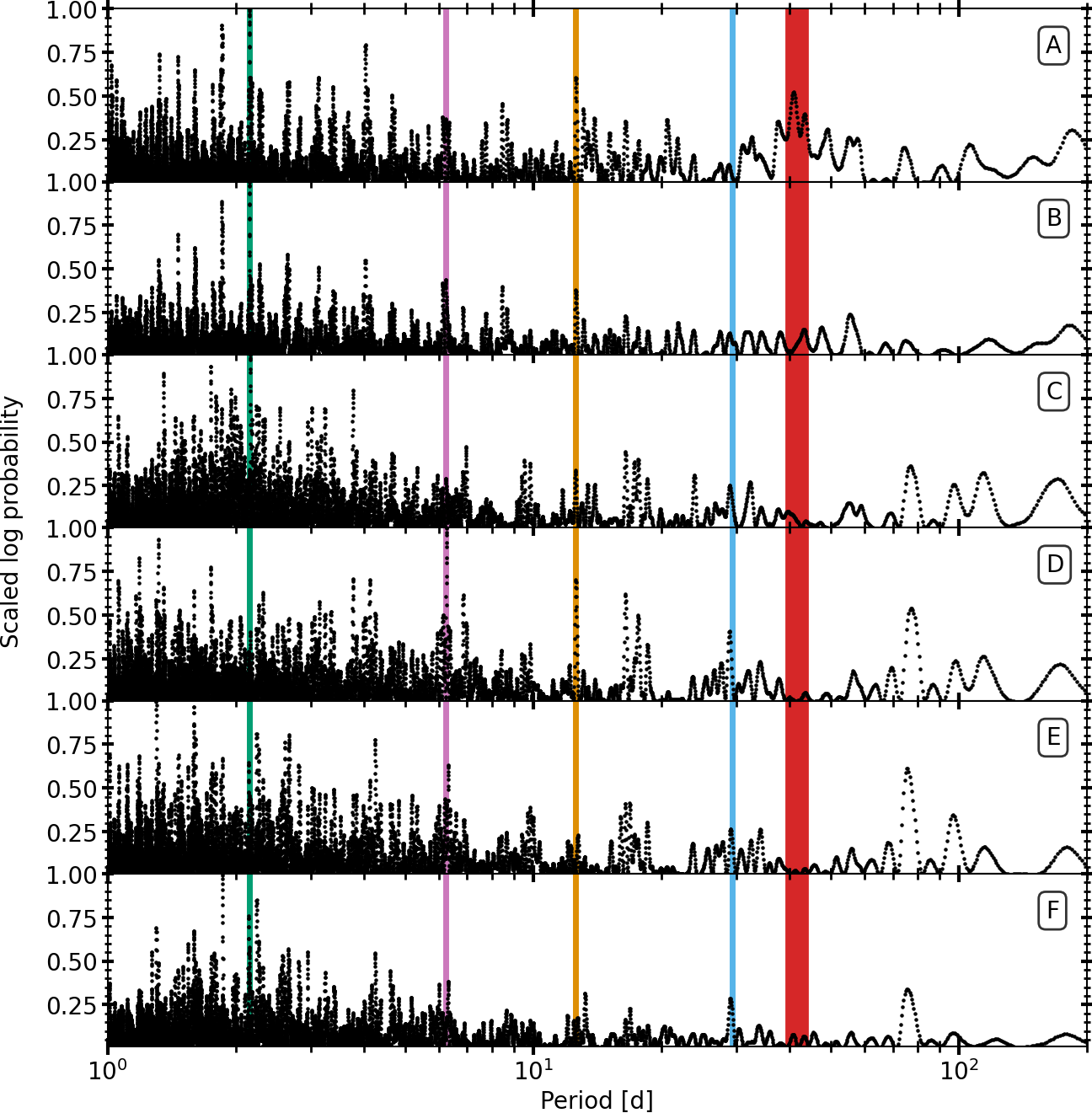}}
\noindent
\begin{small}{\bf Fig. S6: Bayesian Generalised Lomb-Scargle periodograms of the HARPS-N radial velocities after removal of the stellar rotation and planetary signals}. ({\bf A}) Same as Figure S4A, but ({\bf B}) after the stellar rotation has been removed using a GP, and ({\bf C-F}) removal of each planet (LHS\,1903\,b,\,c,\,d, and\,e) signal in sequence. \end{small} 
\end{figure}

\begin{figure}[htbp]
    \centerline{\includegraphics[width=0.85\columnwidth]{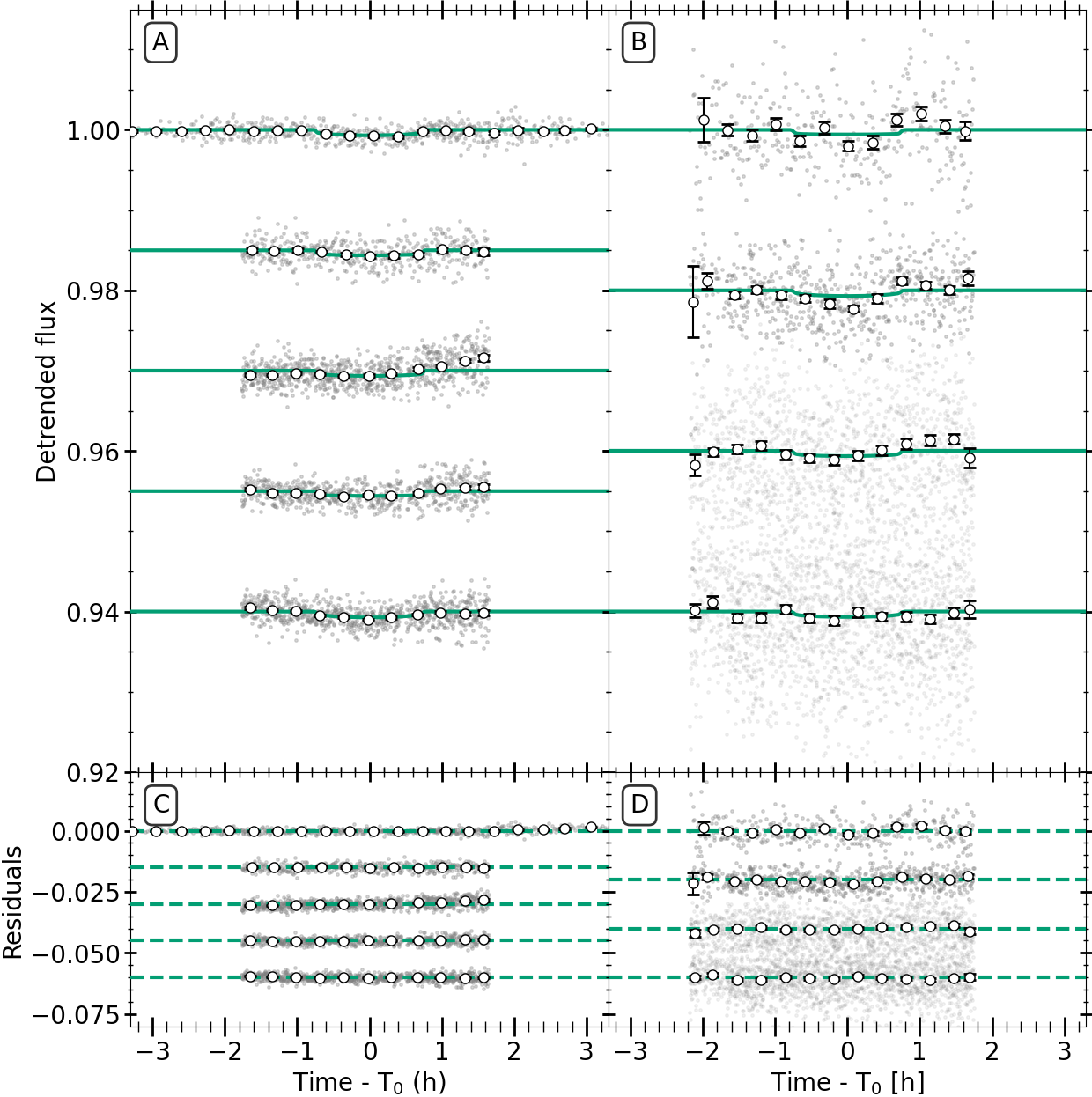}}\vspace{-0.cm}
    {{\includegraphics[width=0.312\columnwidth]{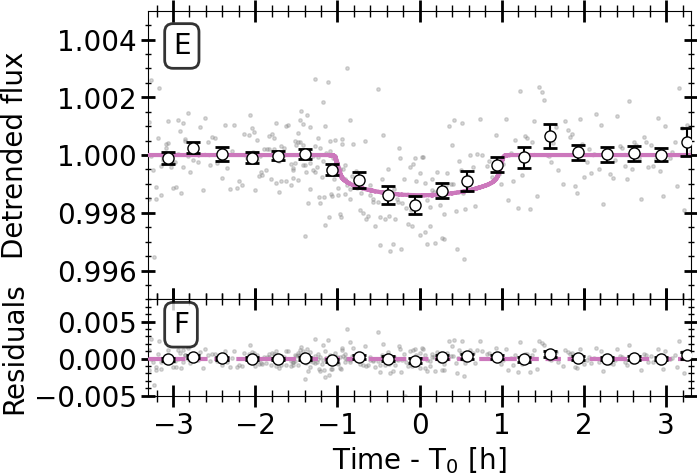} }}\hspace{-0.885cm}
    \qquad
    {{\includegraphics[width=0.312\columnwidth]{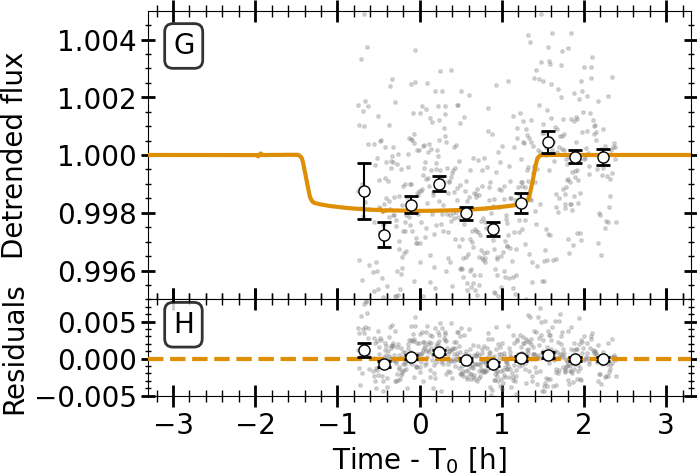} }}\hspace{-0.885cm}
    \qquad
    {{\includegraphics[width=0.312\columnwidth]{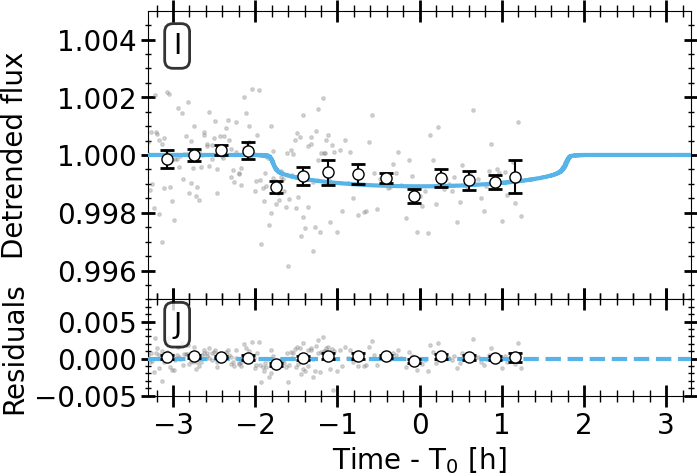} }} \\
\noindent
\begin{small}{\bf Fig. S7: Detrended ground-based phase-folded transit photometry of the LHS\,1903\, planets}. Individual observations are shown as grey points, with open circles indicating data binned every 20\,minutes. ({\bf A}) LCOGT and MuSCAT3 (in $g$, $r$, $i$, and $Z_s$ photometric bandpasses), and ({\bf B}): MuSCAT2 (in $g$, $r$, $i$, and $Z_s$ photometric bandpasses) of LHS\,1903\,b. ({\bf E}): LCOGT, ({\bf G}): SAINT-EX, and ({\bf I}):  LCOGT observations of LHS\,1903\,c,\,d, and\,e phase-folded to the orbital period found by our global analysis. Fitted transit models for LHS\,1903\,b,\,c,\,d, and\,e are shown as green, purple, orange, and cyan solid lines, respectively. ({\bf C, D, F, H, and J}): Residuals between the model and the data with average values shown as dashed lines. \end{small} 
\end{figure}

\begin{figure}[htbp]
    {\includegraphics[width=0.5\columnwidth]{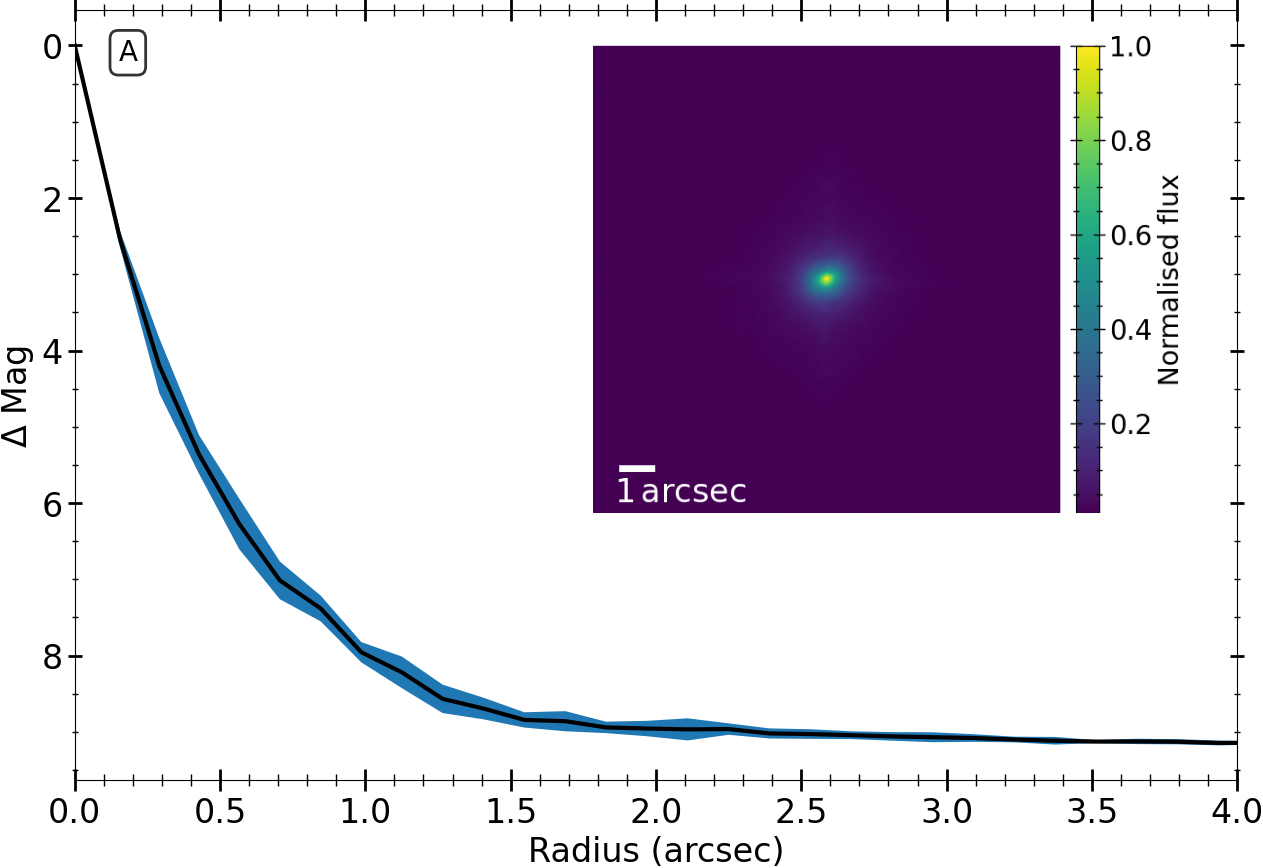}} 
    {\includegraphics[width=0.5\columnwidth]{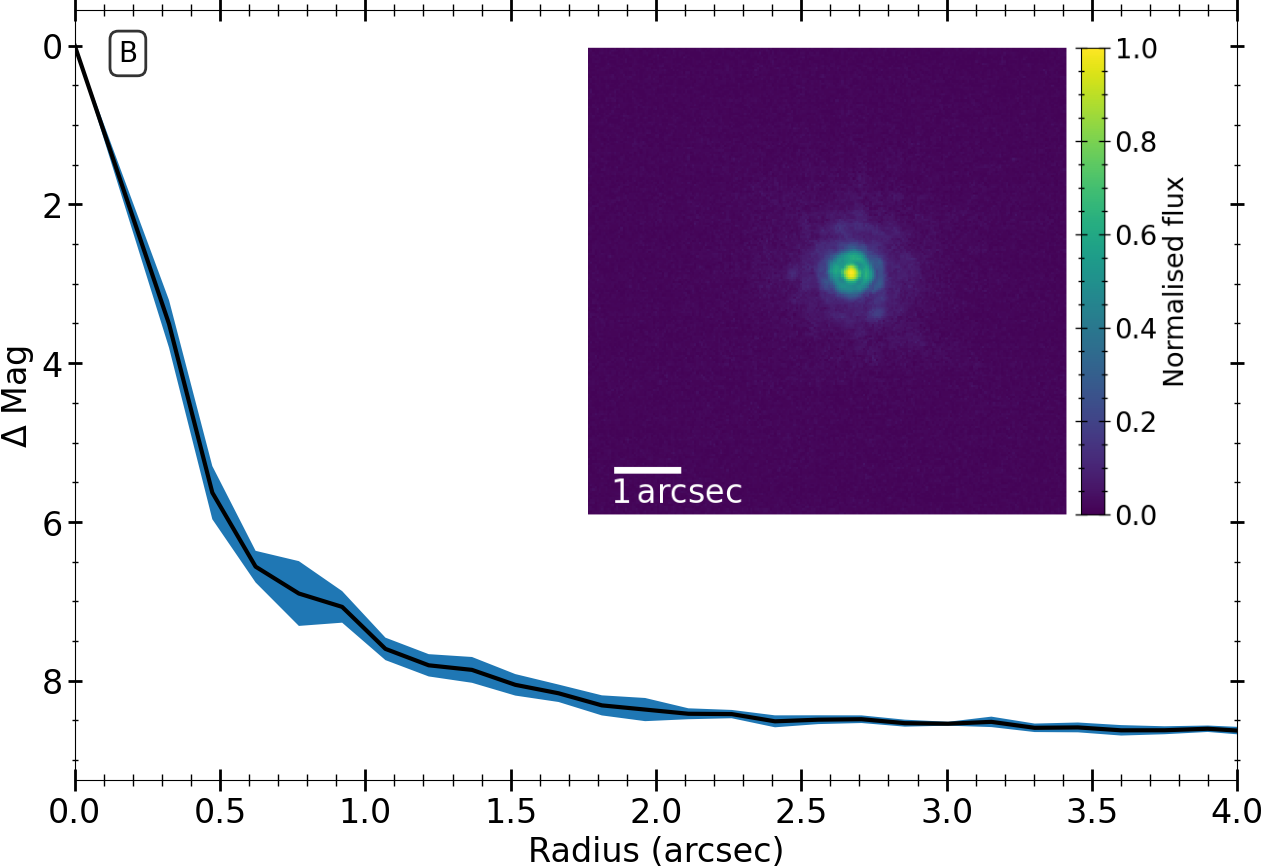}}\\
\noindent
\begin{small}{\bf Fig. S8: AO imaging and sensitivity curves for LHS\,1903}. Black lines report the observed data with 1$\sigma$ uncertainties shown as shaded regions. ({\bf A}) NIR Br$\gamma$ Palomar/PHARO and ({\bf B}) $K_{\rm s}$ Shane/ShARCS. Images of the central portion of the data centred on the star are shown in the insets. \end{small} 
\end{figure}

\begin{figure}[htbp]
\centerline{\includegraphics[width=1.25\columnwidth]{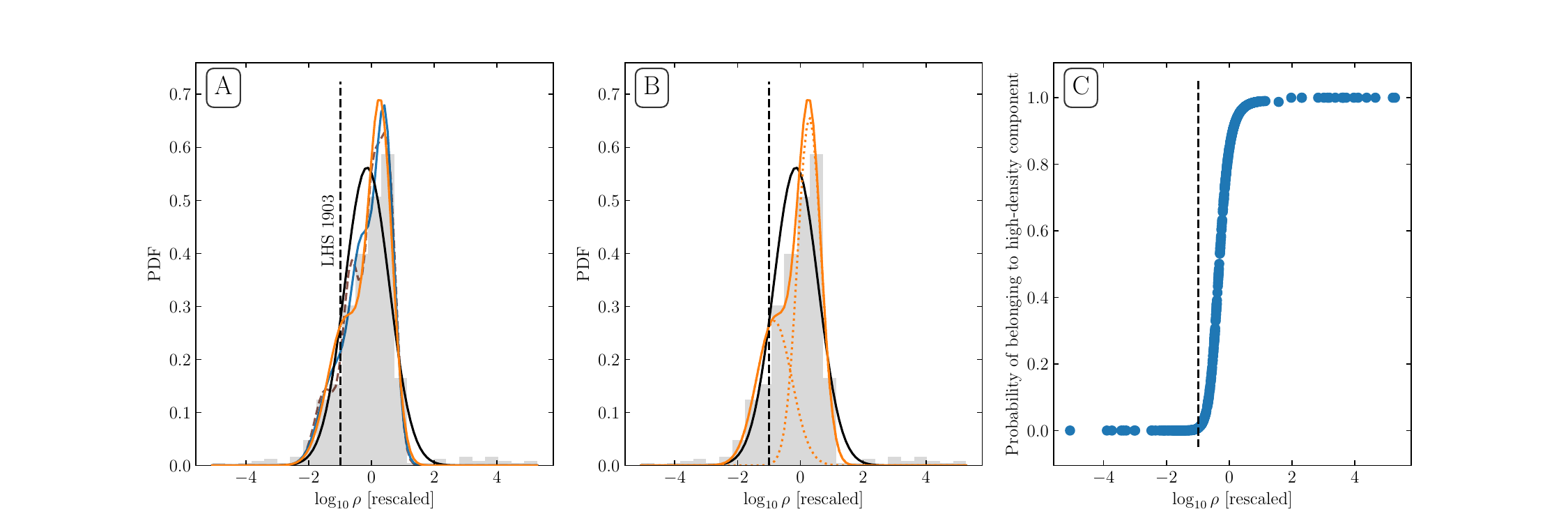}}
\noindent
\begin{small}{\bf Fig. S9: 6D position and velocity local phase space density probabilities}. ({\bf A}) The probability density function (PDF) of rescaled local phase space density, $\rho$, of LHS\,1903 (grey histogram) that is modelled by one to four Gaussian distributions (black, orange, and blue solid lines, and a brown dashed curve) with the value for LHS\,1903 shown as a vertical black dashed line. ({\bf B}) The PDF with one (solid black curve) and two (solid orange curve) Gaussian distributions, with the individual components of the two Gaussian analysis shown as dotted orange curves. ({\bf C}) The probability of being in a high-density region as a function of rescaled local phase space density. \end{small} 
\end{figure}

\begin{figure}[htbp]
    {{\includegraphics[width=0.5\columnwidth]{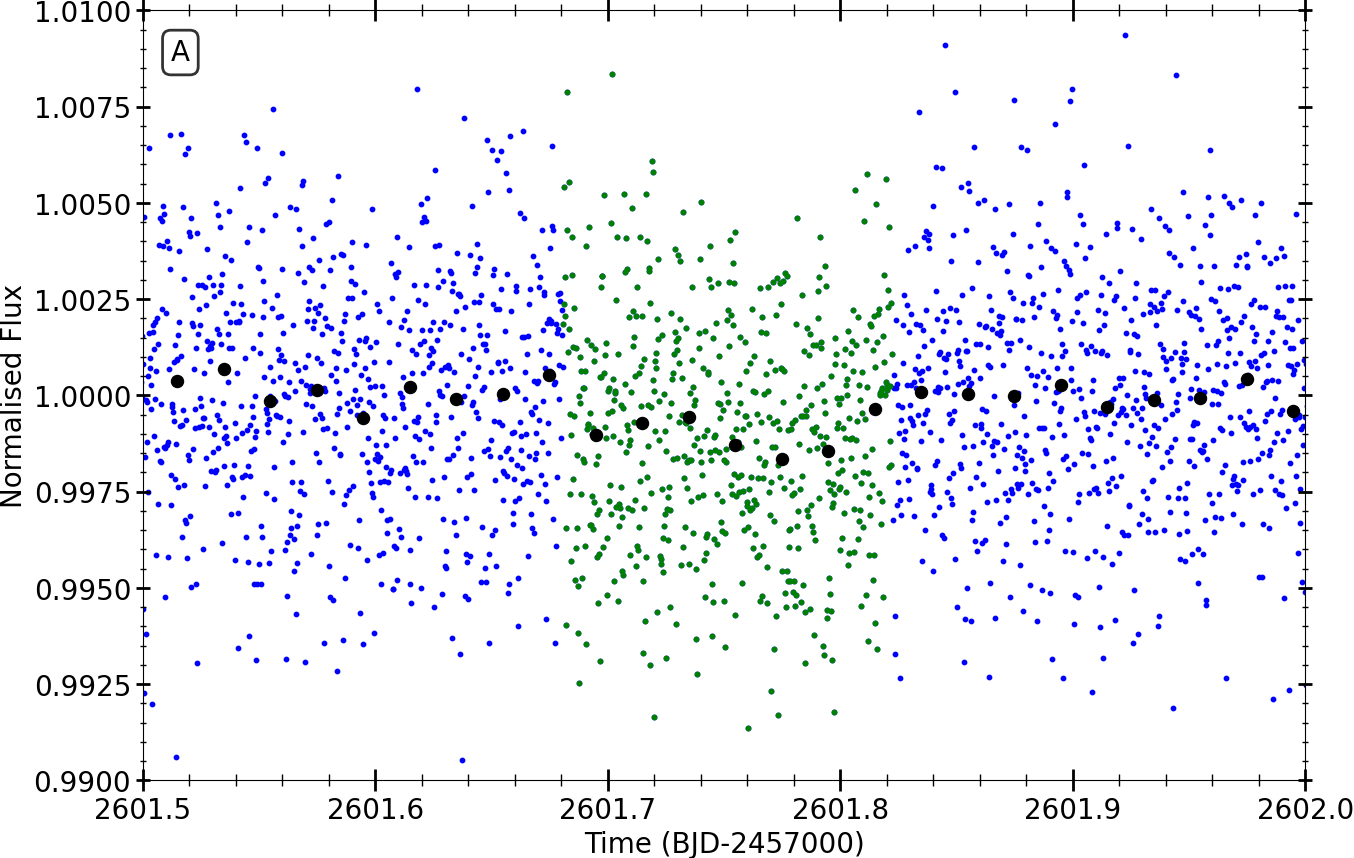} }}\hspace{-0.75cm}
    \qquad
    {{\includegraphics[width=0.5\columnwidth]{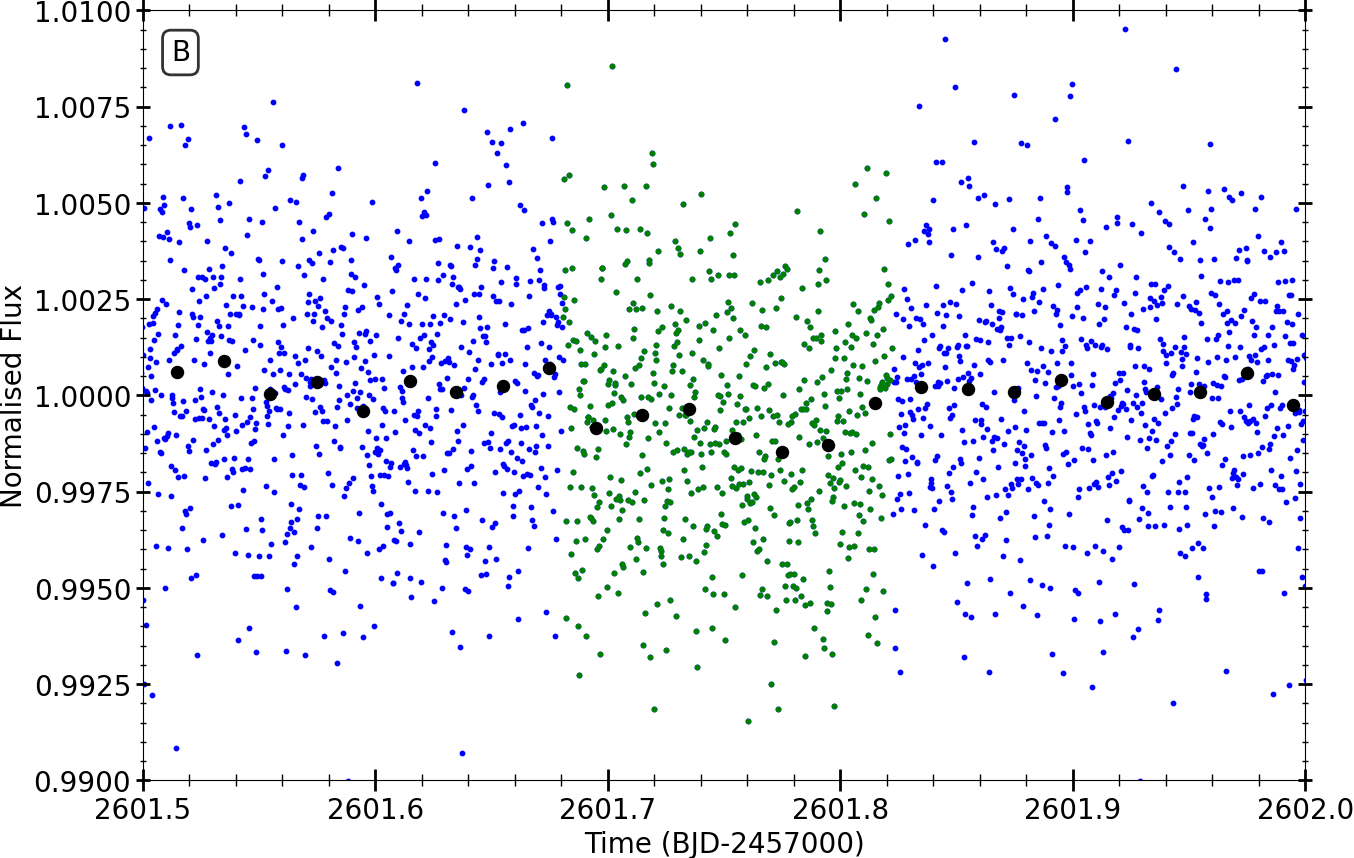} }}
    \qquad
    {{\includegraphics[width=0.5\columnwidth]{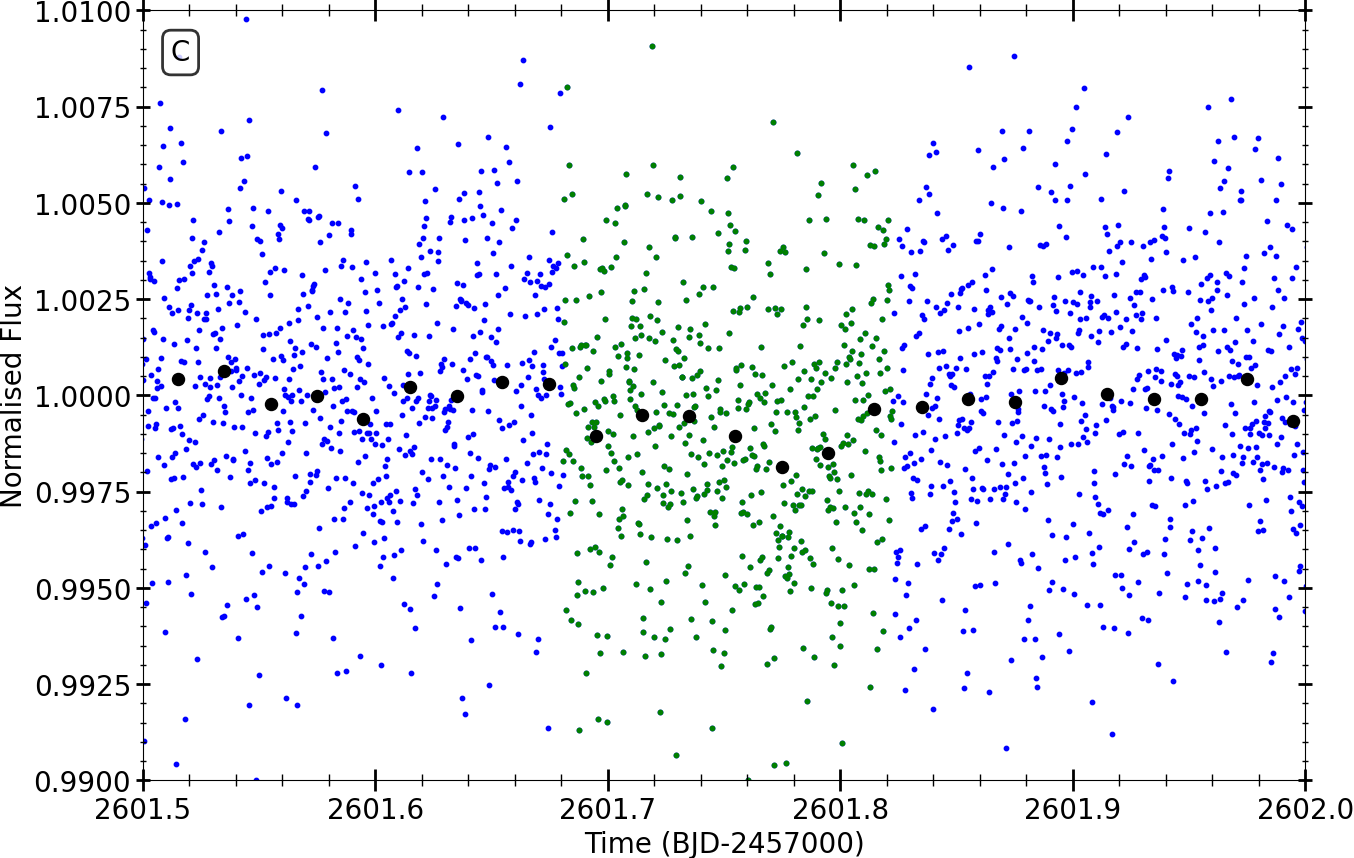} }}\hspace{-0.75cm}
    \qquad
    {{\includegraphics[width=0.5\columnwidth]{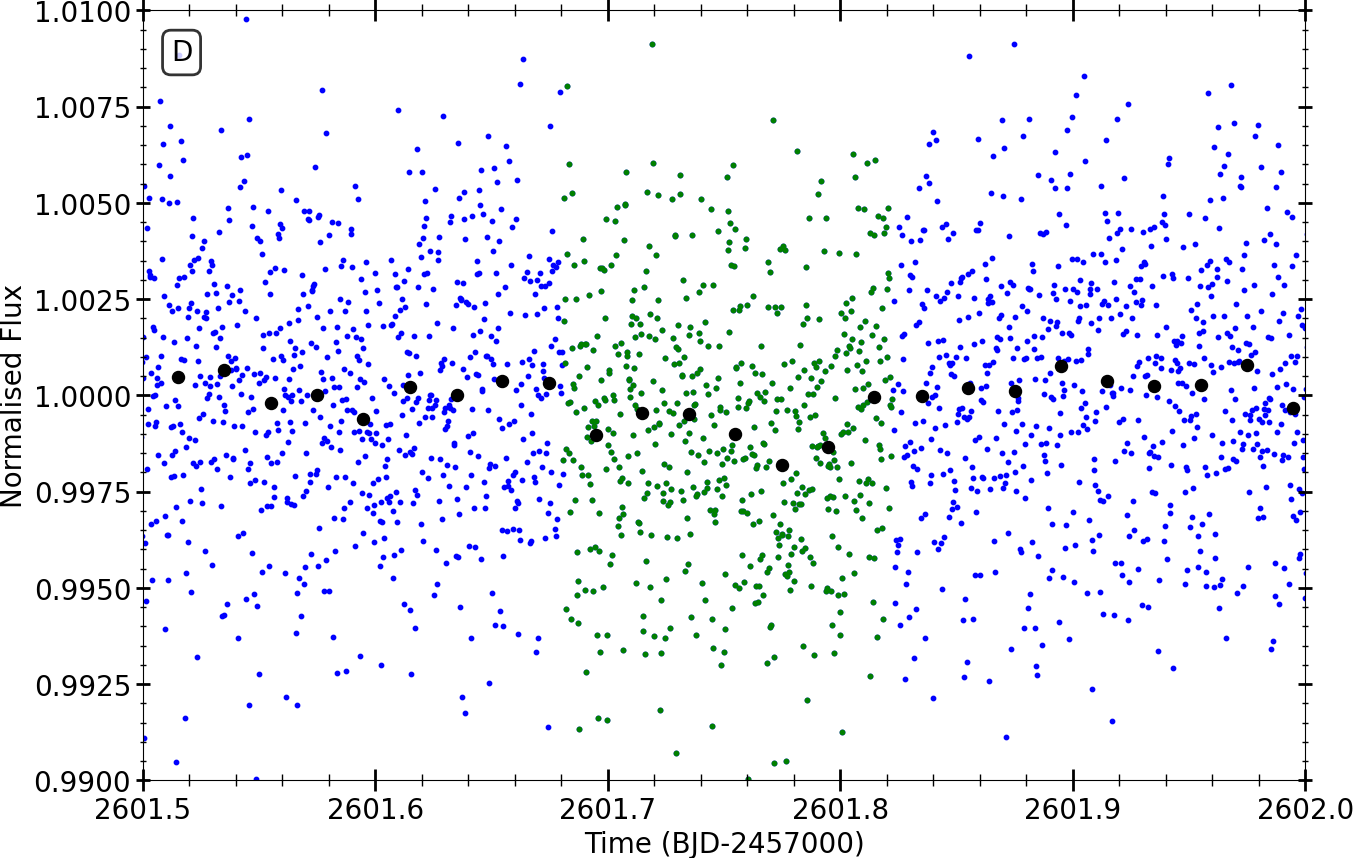} }}
    \qquad
    {{\includegraphics[width=0.5\columnwidth]{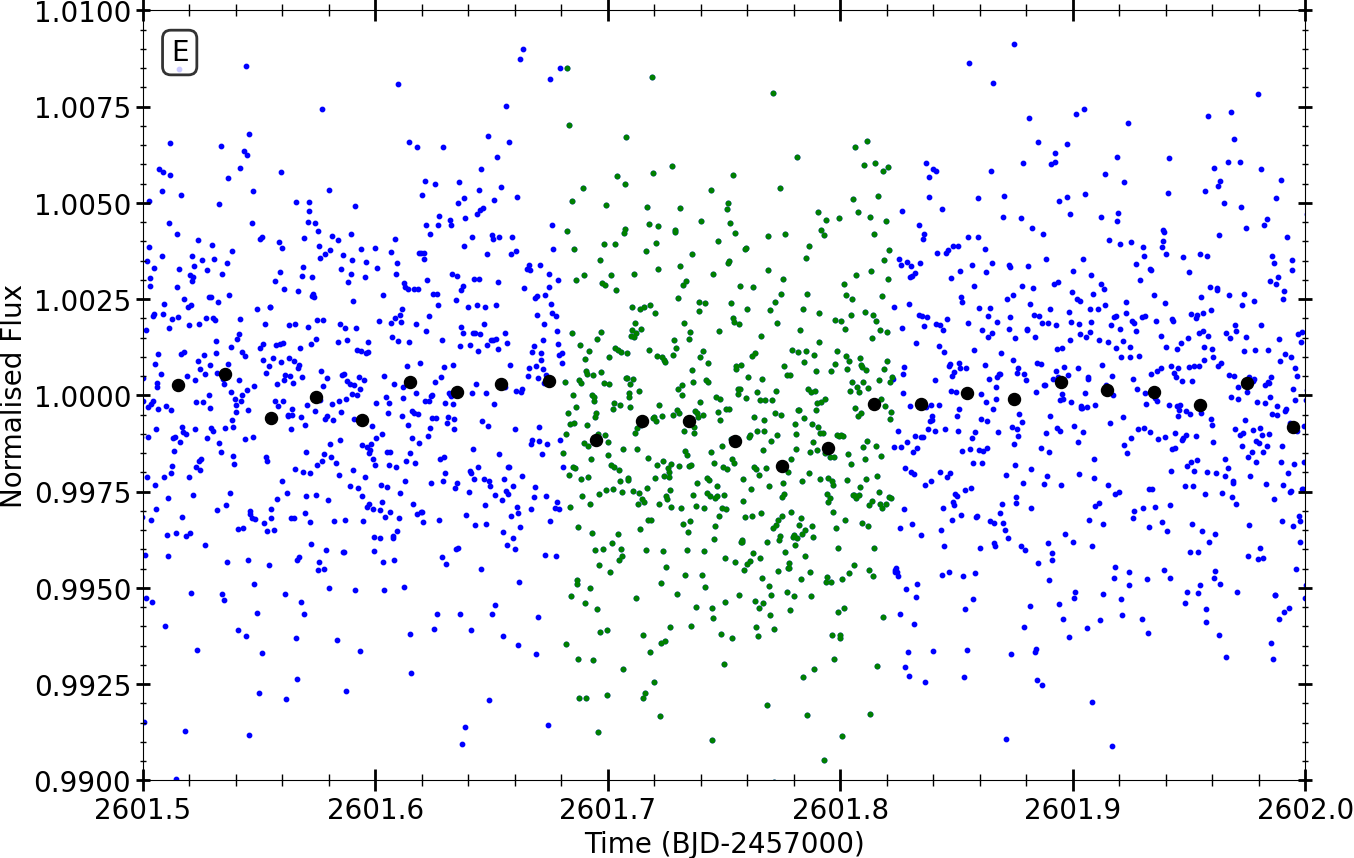} }}\hspace{-0.75cm}
    \qquad
    {{\includegraphics[width=0.5\columnwidth]{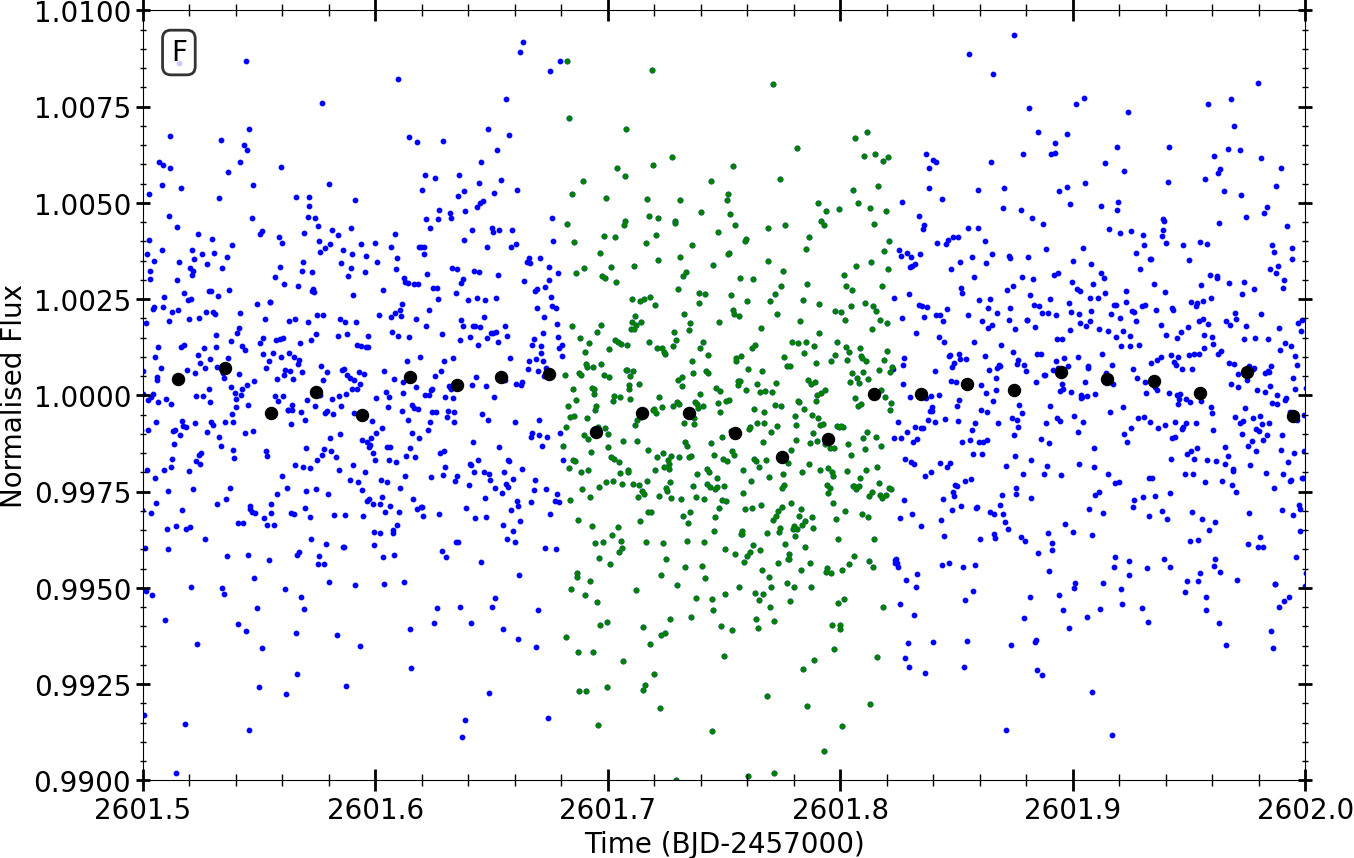} }} \\
\noindent
\begin{small}{\bf Fig. S10: Transit of LHS\,1903\,e with different extraction and detrending methods}. A segment of TESS Sector 47 photometry centred on the transit of LHS\,1903\,e, marked by green points, with out-of-transit data shown as blue points. ({\bf A}) PDCSAP extraction detrended with a GP, ({\bf B}) PDCSAP extraction detrended with {\sc wotan}, ({\bf C}) TPFED/FFIED extraction detrended with a GP, ({\bf D}) TPFED/FFIED extraction detrended with {\sc wotan}, ({\bf E}) TPFED/FFIED \& PSF-{\sc scalpels} extraction detrended with a GP, and ({\bf F}) TPFED/FFIED \& PSF-{\sc scalpels} extraction detrended with {\sc wotan}, along with the photometry binned every 30\,minutes, black points. \end{small} 
\end{figure}

\begin{figure}[htbp]
\centerline{\includegraphics[width=0.85\columnwidth]{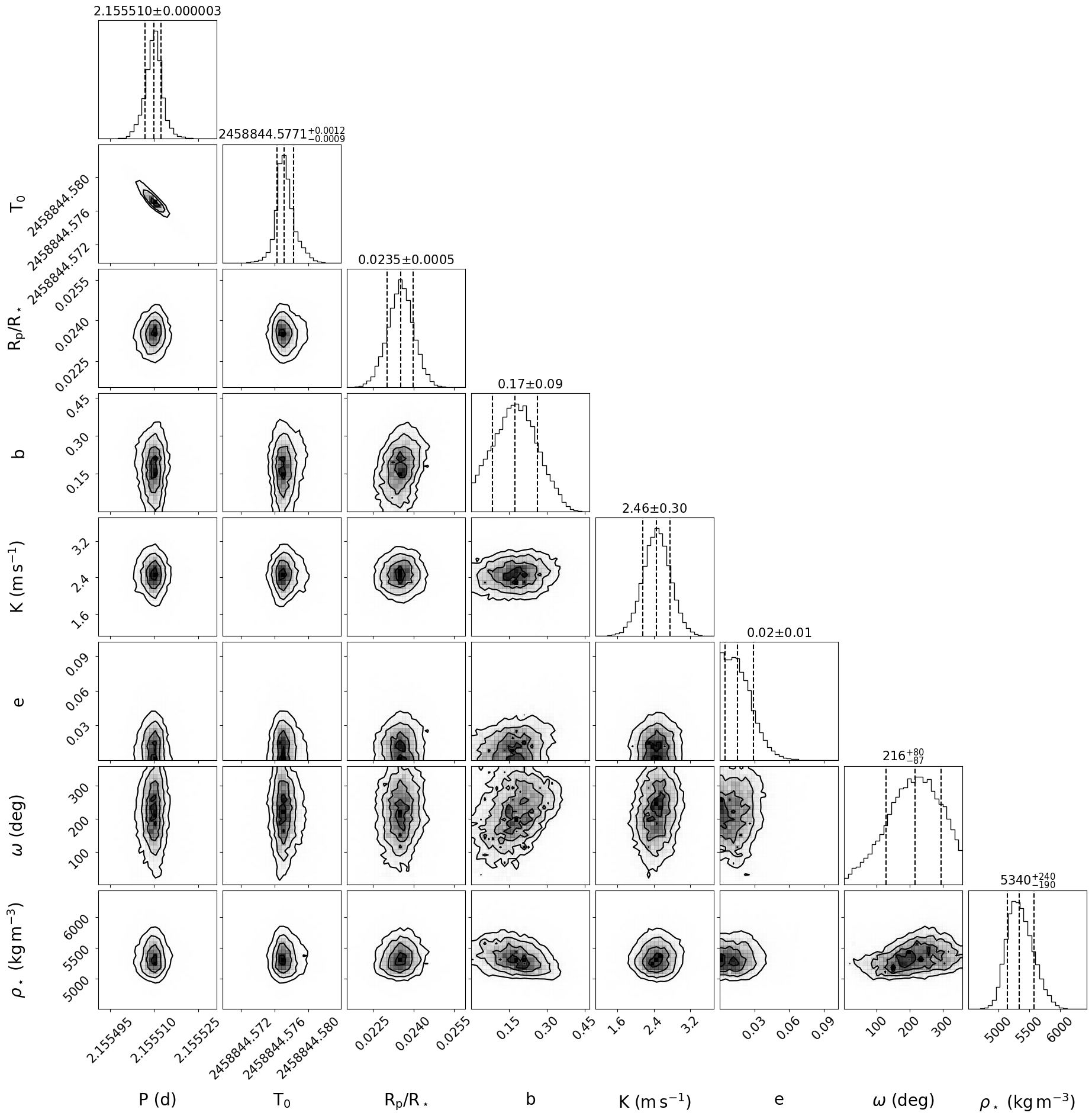} } 
\noindent
\begin{small}{\bf Fig. S11: Posterior probability distributions for the main transit photometry and RV parameters from the model fitting for LHS\,1903\,b}. The columns from left to right show planetary orbital period, $P$, transit centre time, $T_0$, planet-to-star radius ratio, $R_{\rm p}/R_\star$, transit impact parameter, $b$, RV semi-amplitude, $K$, eccentricity, $e$, argument of peristron, $\omega$, and stellar density, $\rho_*$. 1-dimensional histograms for each parameter posterior probability distribution are shown in the diagonal elements. Other panels shown correlation 2-dimensional histograms. Medians and the 16\% and 84\% percentiles for each parameter are labelled above each column. \end{small} 
\end{figure} 

\begin{figure}[htbp]
\centerline{\includegraphics[width=0.85\columnwidth]{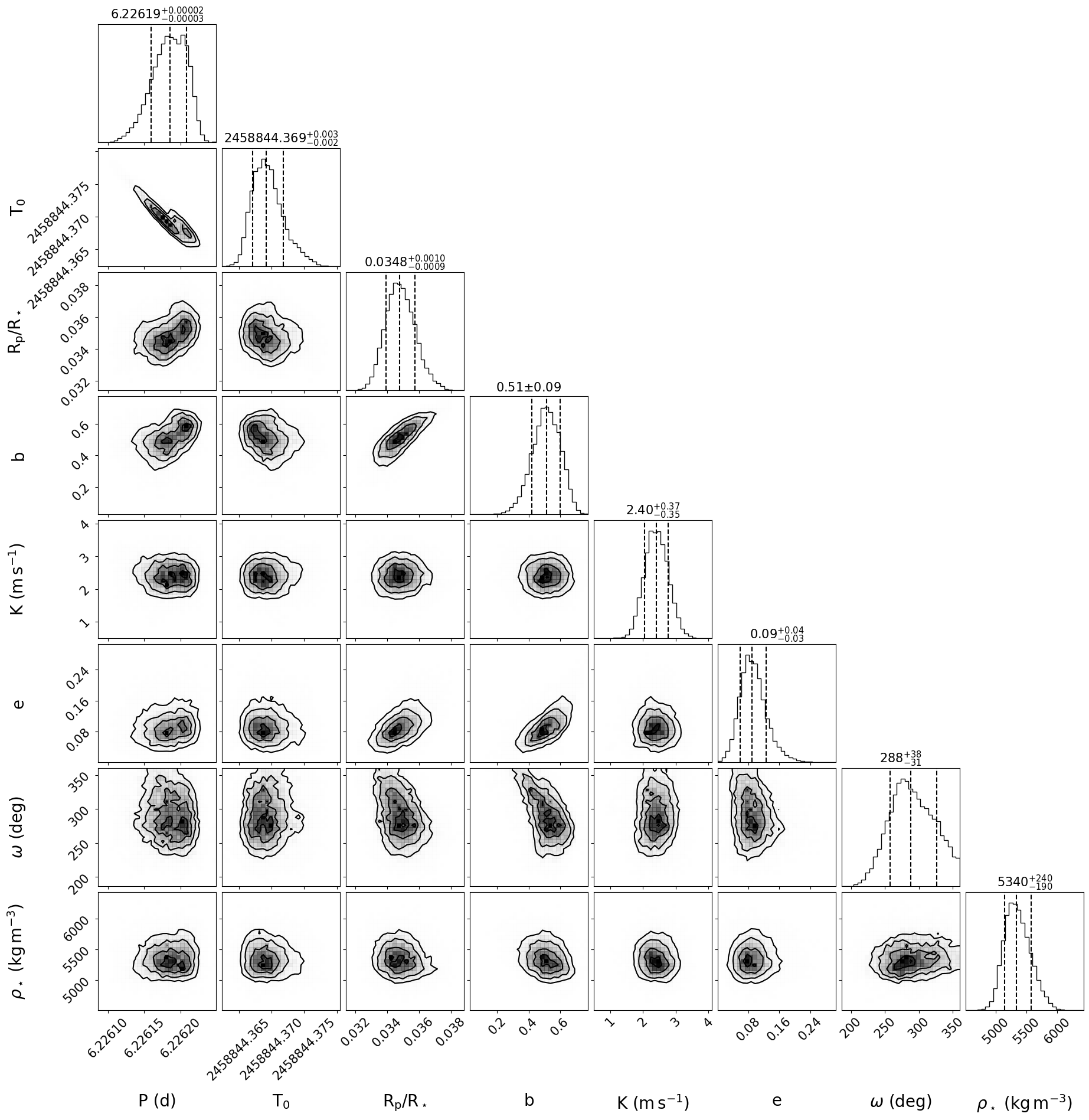} } 
\noindent
\begin{small}{\bf Fig. S12: Posterior probability distributions for the main transit photometry and RV parameters from the model fitting for LHS\,1903\,c}. The columns from left to right show planetary orbital period, $P$, transit centre time, $T_0$, planet-to-star radius ratio, $R_{\rm p}/R_\star$, transit impact parameter, $b$, RV semi-amplitude, $K$, eccentricity, $e$, argument of peristron, $\omega$, and stellar density, $\rho_*$. 1-dimensional histograms for each parameter posterior probability distribution are shown in the diagonal elements. Other panels shown correlation 2-dimensional histograms. Medians and the 16\% and 84\% percentiles for each parameter are labelled above each column. \end{small} 
\end{figure} 

\begin{figure}[htbp]
\centerline{\includegraphics[width=0.85\columnwidth]{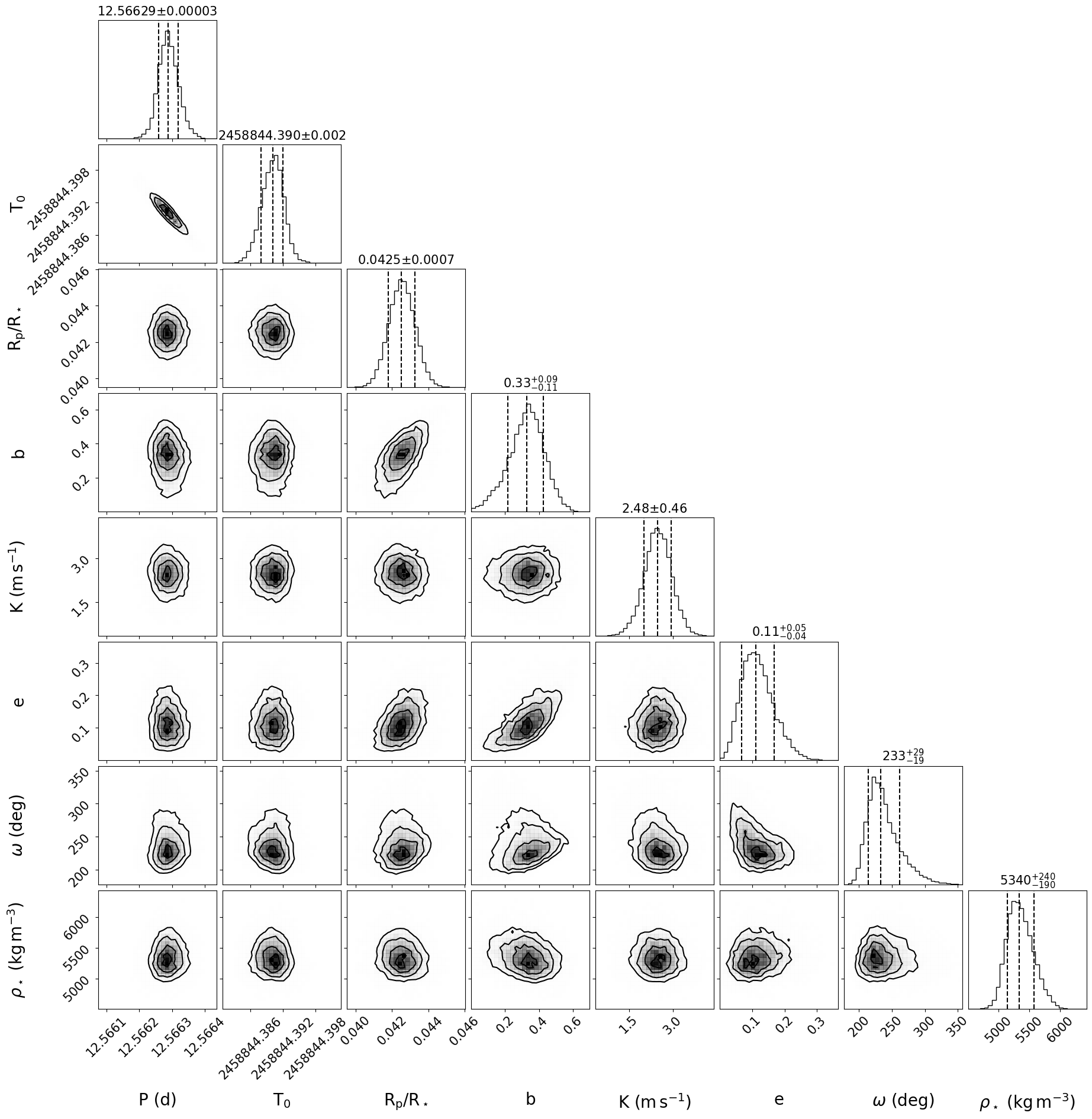} } 
\noindent
\begin{small}{\bf Fig. S13: Posterior probability distributions for the main transit photometry and RV parameters from the model fitting for LHS\,1903\,d}. The columns from left to right show planetary orbital period, $P$, transit centre time, $T_0$, planet-to-star radius ratio, $R_{\rm p}/R_\star$, transit impact parameter, $b$, RV semi-amplitude, $K$, eccentricity, $e$, argument of peristron, $\omega$, and stellar density, $\rho_*$. 1-dimensional histograms for each parameter posterior probability distribution are shown in the diagonal elements. Other panels shown correlation 2-dimensional histograms. Medians and the 16\% and 84\% percentiles for each parameter are labelled above each column. \end{small} 
\end{figure} 

\begin{figure}[htbp]
\centerline{\includegraphics[width=0.85\columnwidth]{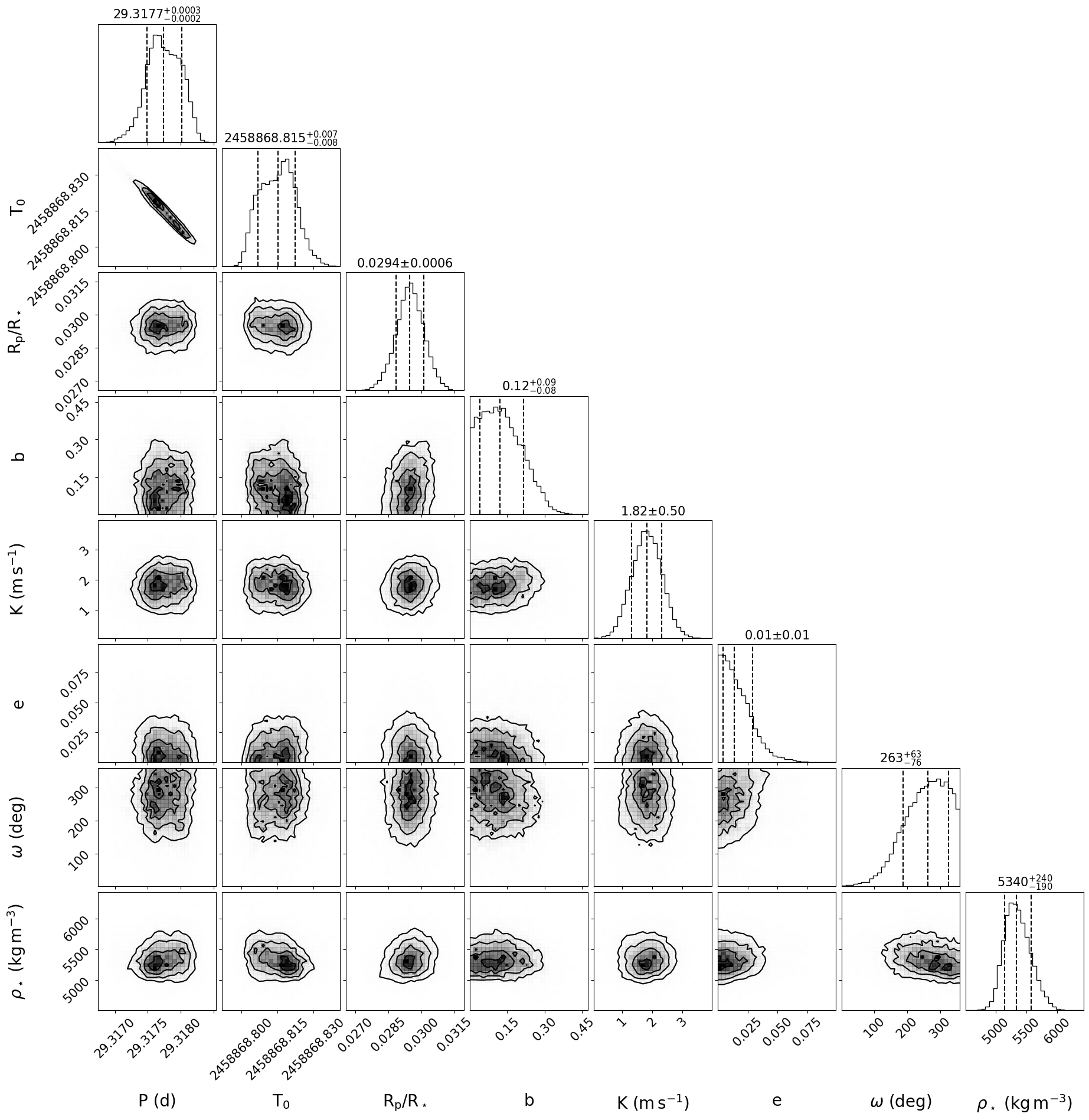} } 
\noindent
\begin{small}{\bf Fig. S14: Posterior probability distributions for the main transit photometry and RV parameters from the model fitting for LHS\,1903\,e}. The columns from left to right show planetary orbital period, $P$, transit centre time, $T_0$, planet-to-star radius ratio, $R_{\rm p}/R_\star$, transit impact parameter, $b$, RV semi-amplitude, $K$, eccentricity, $e$, argument of peristron, $\omega$, and stellar density, $\rho_*$. 1-dimensional histograms for each parameter posterior probability distribution are shown in the diagonal elements. Other panels shown correlation 2-dimensional histograms. Medians and the 16\% and 84\% percentiles for each parameter are labelled above each column. \end{small} 
\end{figure}

\begin{figure}[htbp]
\centerline{\includegraphics[width=\columnwidth]{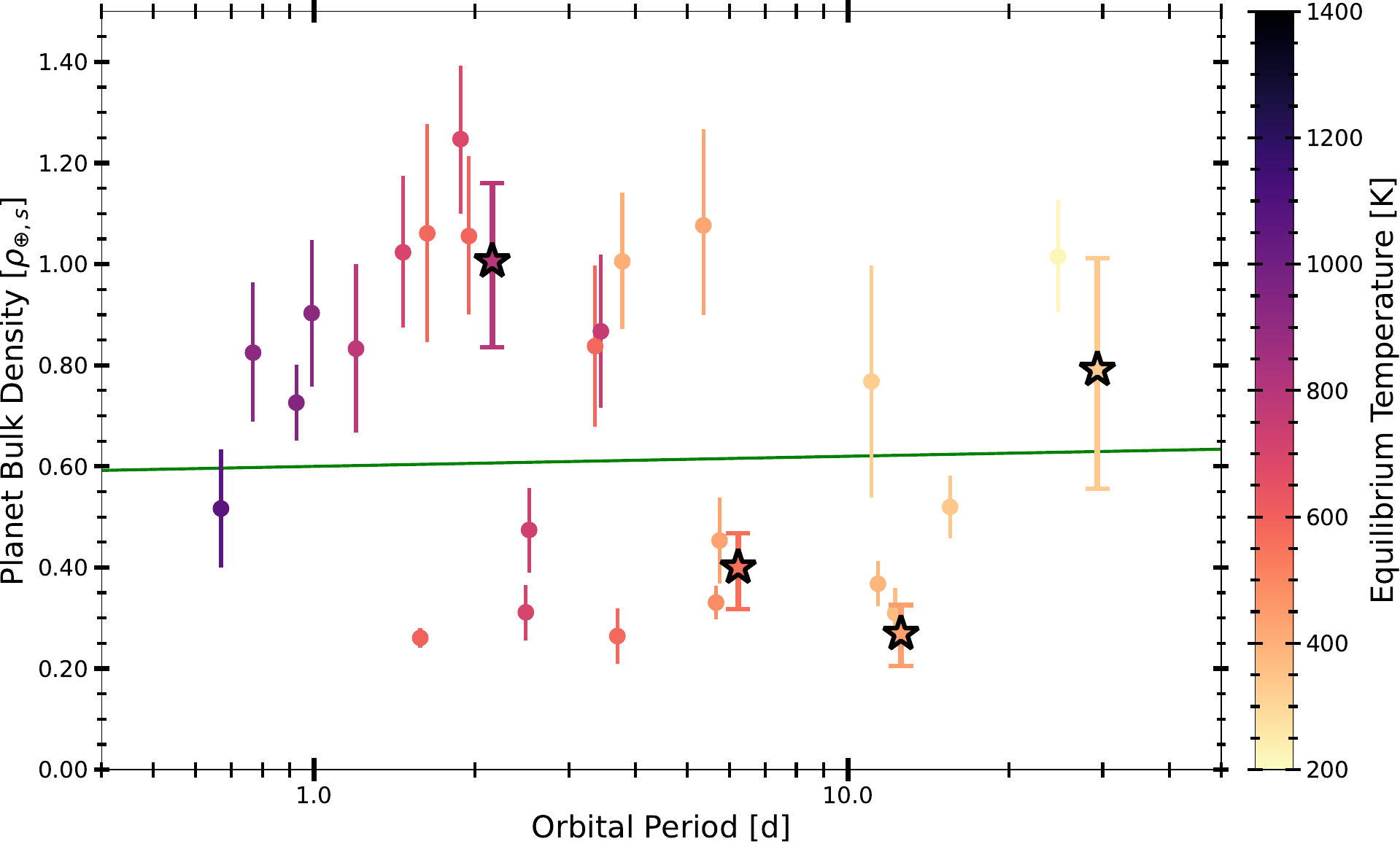}}
\noindent
\begin{small}{\bf Fig. S15: Planet densities as a function of orbital periods for selected M-dwarf planets}. Planets were selected to have precise radius and mass measurements ($\sigma\,R_{\rm p}\,<\,$5\% \& $\sigma\,M_{\rm p}\,<\,$33\%). Their bulk densities are normalised to an Earth-like density, $\rho_{\oplus,s}$,\cite{Adibekyan2021b} and coloured by zero Bond albedo equilibrium temperatures, plotted as a function of orbital period. Error bars are 1$\sigma$ uncertainties. Star symbols with black outlines indicate the LHS\,1903 planets. The green solid line is the position of the density valley\cite{Luque2022}. LHS\,1903\,e is located above this line so is probably rocky and is colder than most other planets in the sample.  \end{small}
\end{figure}

\begin{figure}[htbp]
\centerline{\includegraphics[width=0.75\columnwidth]{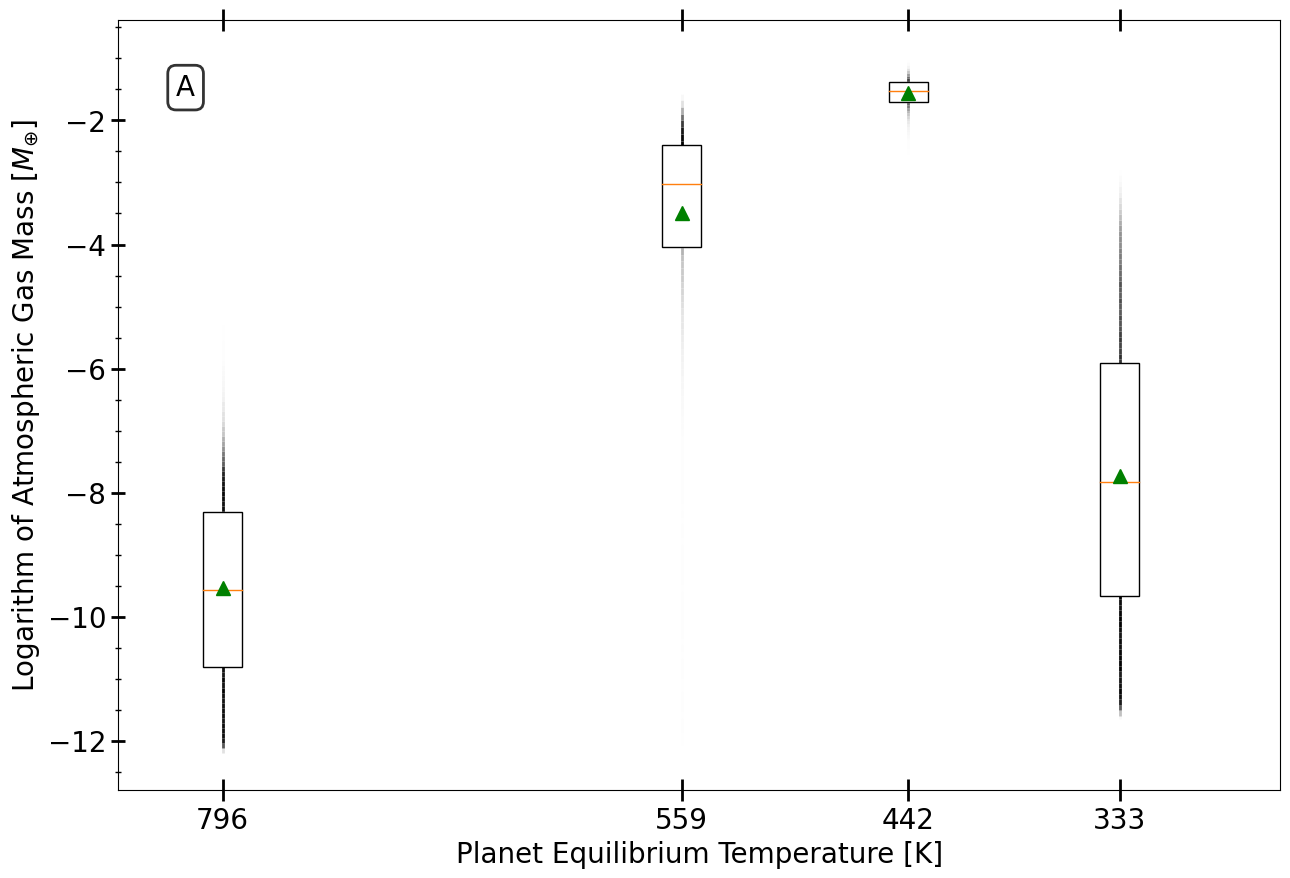}}
\centerline{\includegraphics[width=0.75\columnwidth]{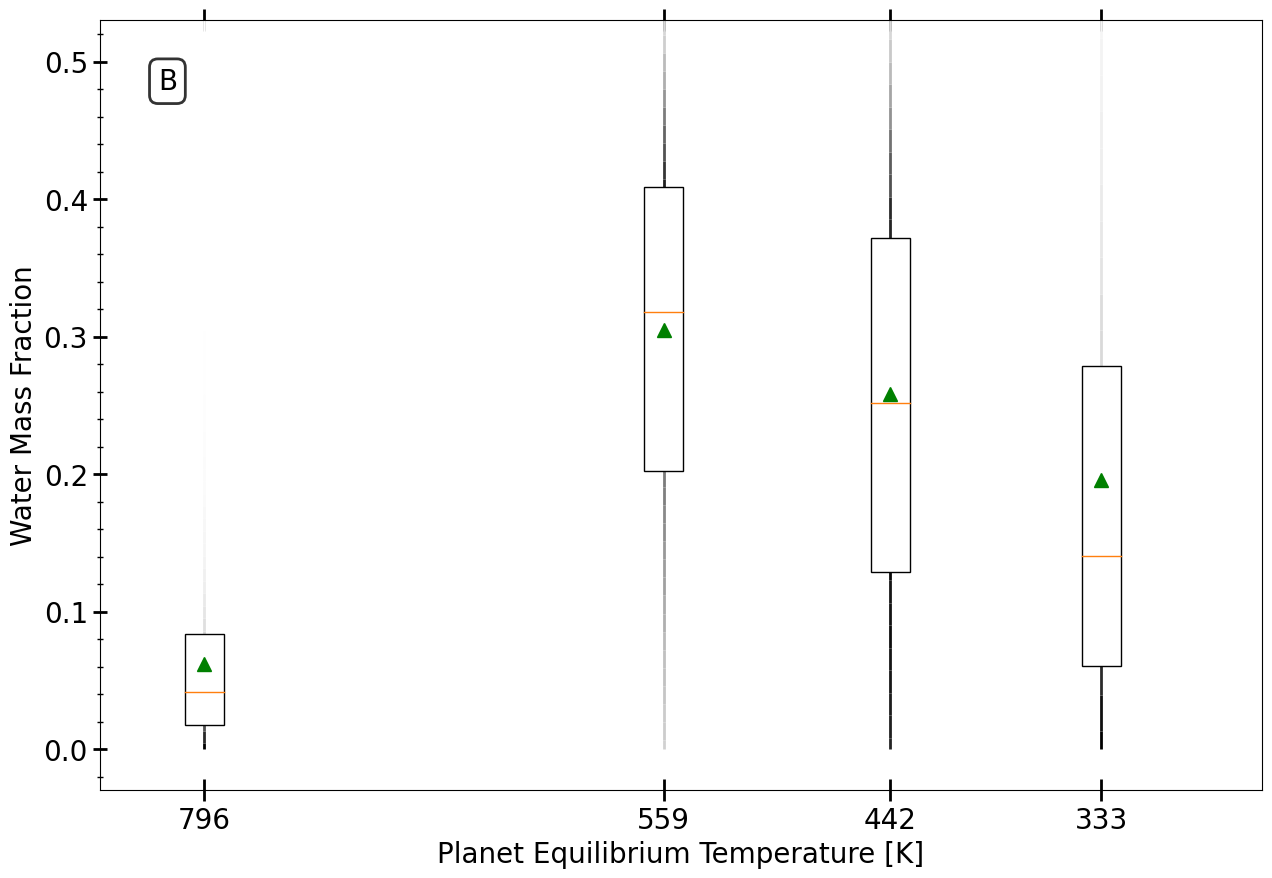}}
\noindent
\begin{small}{\bf Fig. S16: Internal structure mass fractions for the outer layers of the LHS\,1903 planets}. ({\bf A}) Logarithm of the gas mass fractions and ({\bf B}) the water fractions of {\it Left to right:} LHS\,1903\,b,\,c,\,d, \&\,e as a function of zero Bond albedo equilibrium temperature. Statistical measures of the posterior probability distributions are shown by boxes (25\% to 75\% percentiles), orange lines (medians), and green triangles (means), with the opacity of the vertical black line proportional to the posterior probability distribution of the given internal structure mass fraction. \end{small}
\end{figure}

\begin{figure}[htbp]
\centerline{\includegraphics[width=0.85\columnwidth]{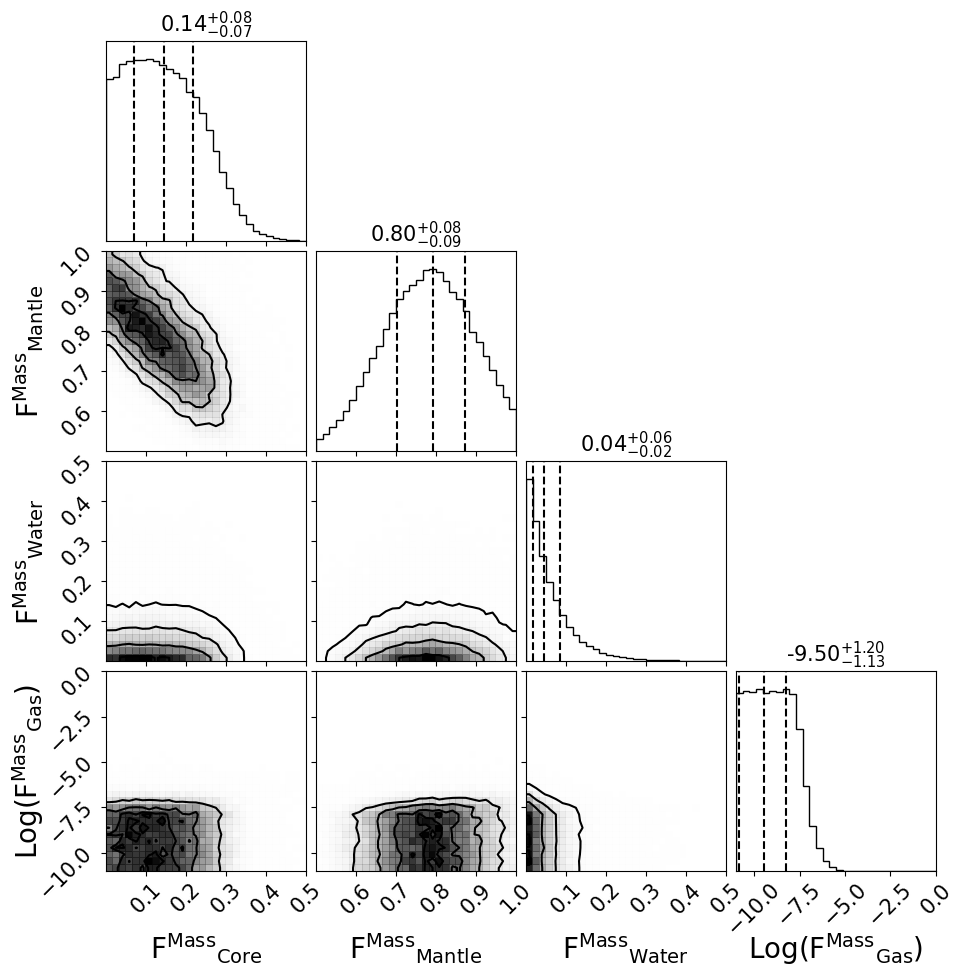} } 
\noindent
\begin{small}{\bf Fig. S17: Posterior distributions for the internal structure parameters for LHS\,1903\,b}. The columns from left to right show the core mass fraction, F$^{\rm Mass}_{\rm Core}$, mantle mass fraction, F$^{\rm Mass}_{\rm Mantle}$, water mass fraction, F$^{\rm Mass}_{\rm Water}$, and the atmospheric gas mass, Log(F$^{\rm Mass}_{\rm Gas}$), in M$_\oplus$. 1-dimensional histograms for each parameter posterior probability distribution are shown in the diagonal elements. Other panels shown correlation 2-dimensional histograms. Medians and the 16\% and 84\% percentiles for each parameter are labelled above each column.\end{small}
\end{figure} 

\begin{figure}[htbp]
\centerline{\includegraphics[width=0.85\columnwidth]{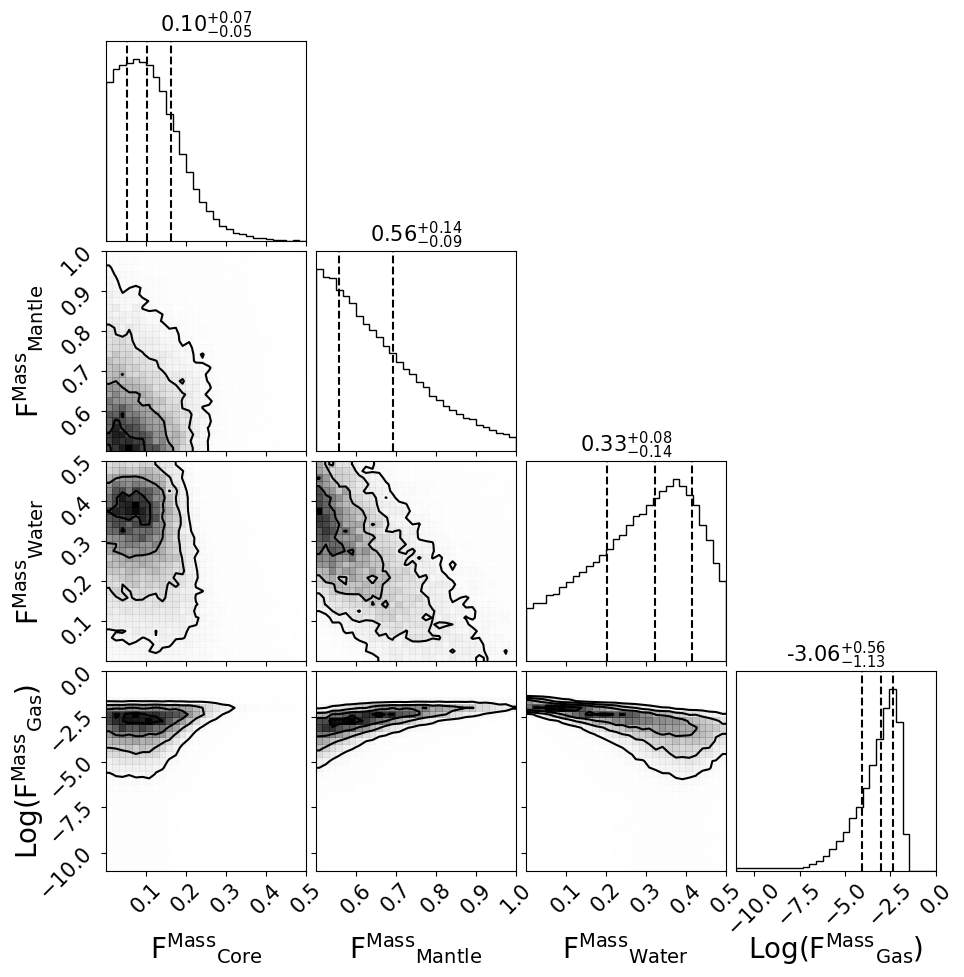} } 
\noindent
\begin{small}{\bf Fig. S18: Posterior distributions for the internal structure parameters for LHS\,1903\,c}. The columns from left to right show the core mass fraction, F$^{\rm Mass}_{\rm Core}$, mantle mass fraction, F$^{\rm Mass}_{\rm Mantle}$, water mass fraction, F$^{\rm Mass}_{\rm Water}$, and the atmospheric gas mass, Log(F$^{\rm Mass}_{\rm Gas}$), in M$_\oplus$. 1-dimensional histograms for each parameter posterior probability distribution are shown in the diagonal elements. Other panels shown correlation 2-dimensional histograms. Medians and the 16\% and 84\% percentiles for each parameter are labelled above each column.\end{small}
\end{figure} 

\begin{figure}[htbp]
\centerline{\includegraphics[width=0.85\columnwidth]{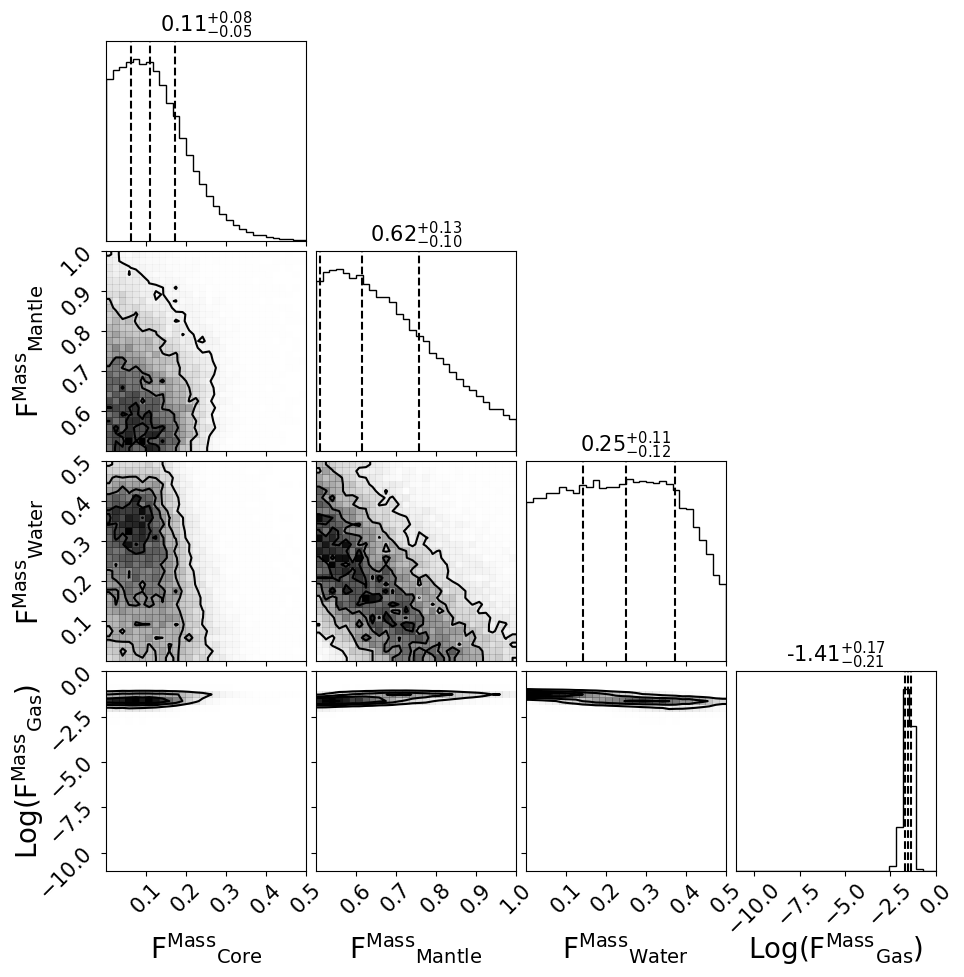} } 
\noindent
\begin{small}{\bf Fig. S19: Posterior distributions for the internal structure parameters for LHS\,1903\,d}. The columns from left to right show the core mass fraction, F$^{\rm Mass}_{\rm Core}$, mantle mass fraction, F$^{\rm Mass}_{\rm Mantle}$, water mass fraction, F$^{\rm Mass}_{\rm Water}$, and the atmospheric gas mass, Log(F$^{\rm Mass}_{\rm Gas}$), in M$_\oplus$. 1-dimensional histograms for each parameter posterior probability distribution are shown in the diagonal elements. Other panels shown correlation 2-dimensional histograms. Medians and the 16\% and 84\% percentiles for each parameter are labelled above each column.\end{small}
\end{figure} 

\begin{figure}[htbp]
\centerline{\includegraphics[width=0.85\columnwidth]{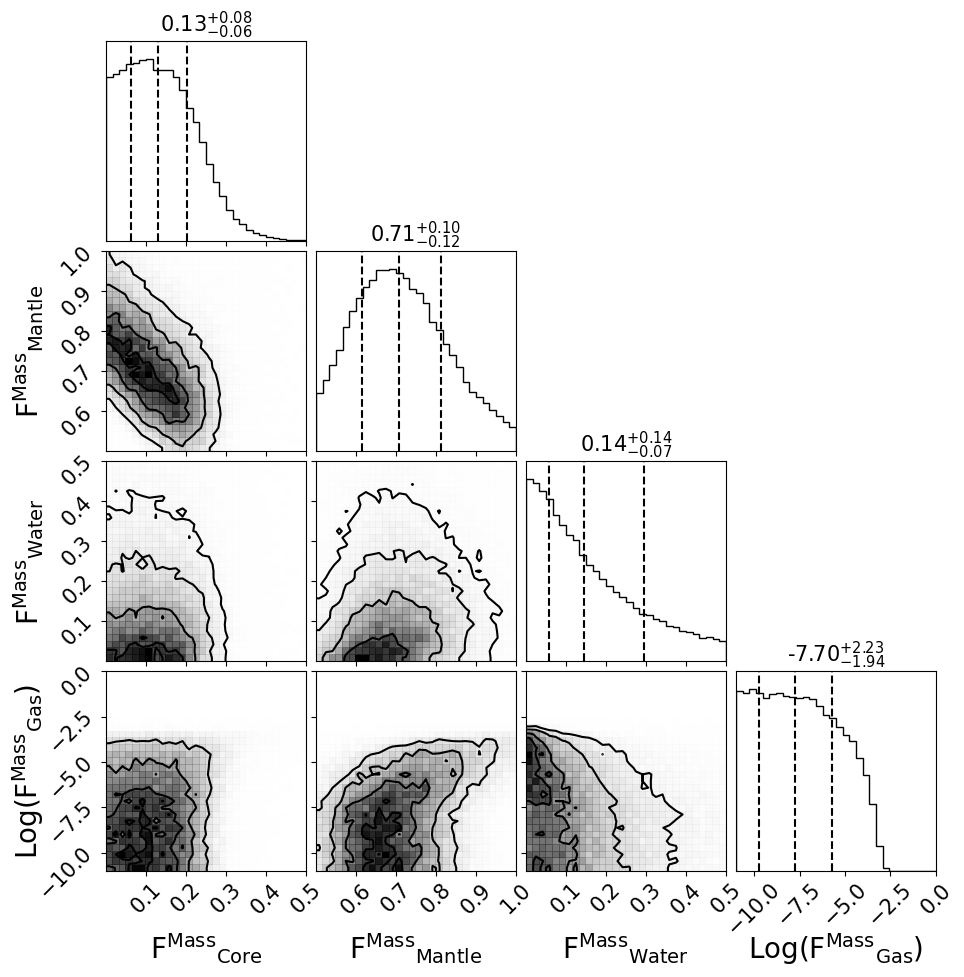} } 
\noindent
\begin{small}{\bf Fig. S20: Posterior distributions for the internal structure parameters for LHS\,1903\,e}. The columns from left to right show the core mass fraction, F$^{\rm Mass}_{\rm Core}$, mantle mass fraction, F$^{\rm Mass}_{\rm Mantle}$, water mass fraction, F$^{\rm Mass}_{\rm Water}$, and the atmospheric gas mass, Log(F$^{\rm Mass}_{\rm Gas}$), in M$_\oplus$. 1-dimensional histograms for each parameter posterior probability distribution are shown in the diagonal elements. Other panels shown correlation 2-dimensional histograms. Medians and the 16\% and 84\% percentiles for each parameter are labelled above each column.\end{small}
\end{figure}

\begin{figure}[htbp]
    \centerline{\includegraphics[width=1.2\columnwidth]{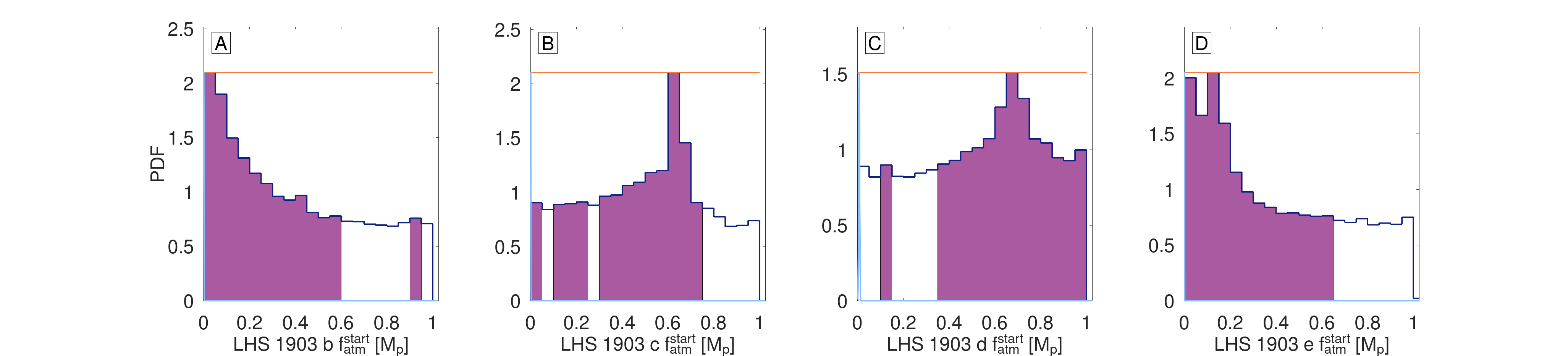}}
\noindent
\begin{small}{\bf Fig. S21: Initial atmospheric mass fraction distributions for all LHS\,1903 planets}. ({\bf A}) LHS\,1903\,b, ({\bf B}) LHS\,1903\,c, ({\bf C}) LHS\,1903\,d, and ({\bf D}) LHS\,1903\,e. The purple area shows the 68\%-highest probability density interval of the initial atmospheric mass fractions presented as dark blue histograms. The orange horizontal lines indicate the uniform prior that has been imposed on $f_{\mathrm{atm}}^{\mathrm{start}}$. The light blue curves, which are barely visible and almost coincident with the left y-axis, correspond to the present-day atmospheric mass fraction derived from our internal structure models. \end{small}
\end{figure}

\begin{figure}[htbp]
\centerline{\includegraphics[width=1.0\textwidth]{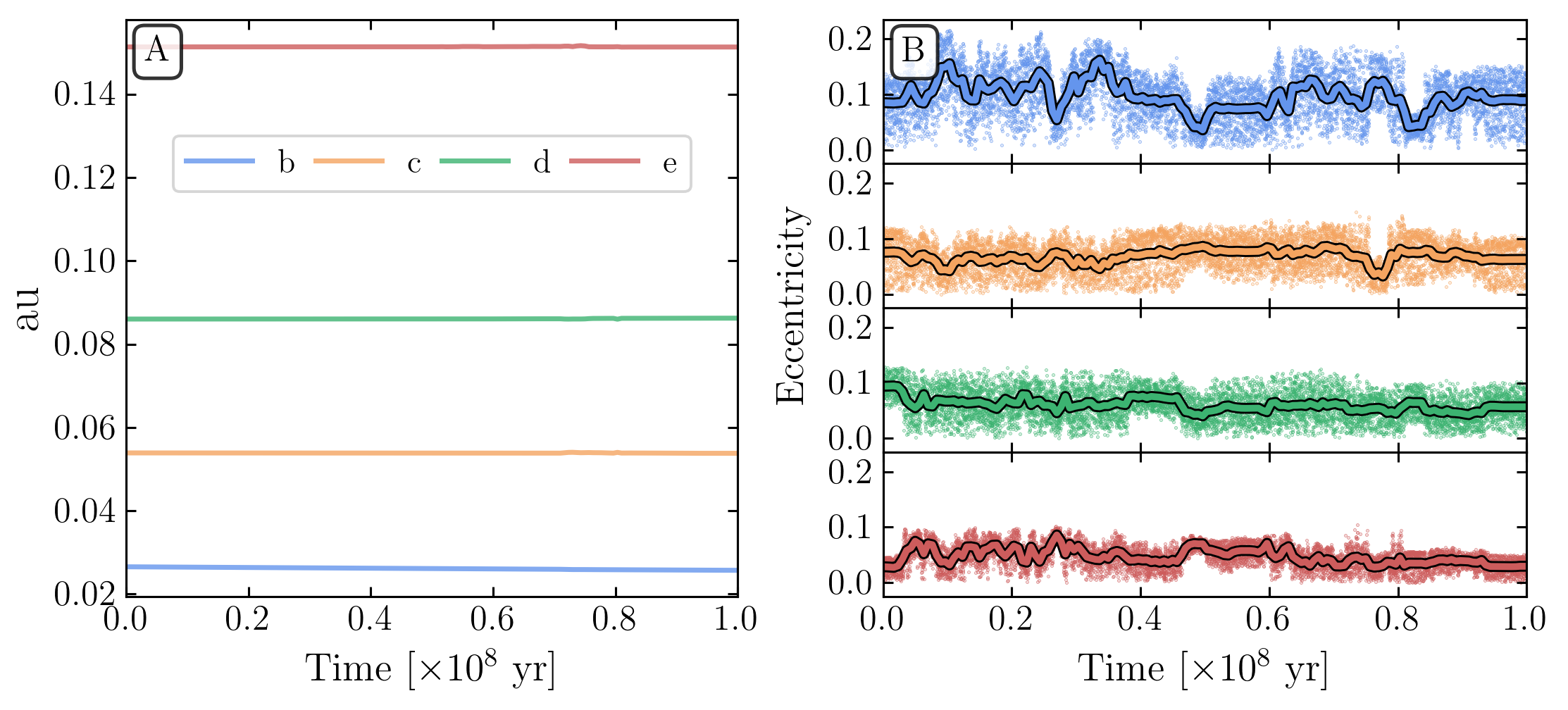}}
\noindent
\begin{small}{\bf Fig. S22: Simulated orbital evolution of the LHS\,1903 system over 100\,Myr}. ({\bf A}) Temporal evolution of the semi-major axes. ({\bf B}) Temporal evolution of the orbital eccentricity (points). The curves are the mean eccentricity variations over time. \end{small}

\end{figure}

\begin{figure}[htbp]
\centerline{\includegraphics[width=0.8\textwidth]{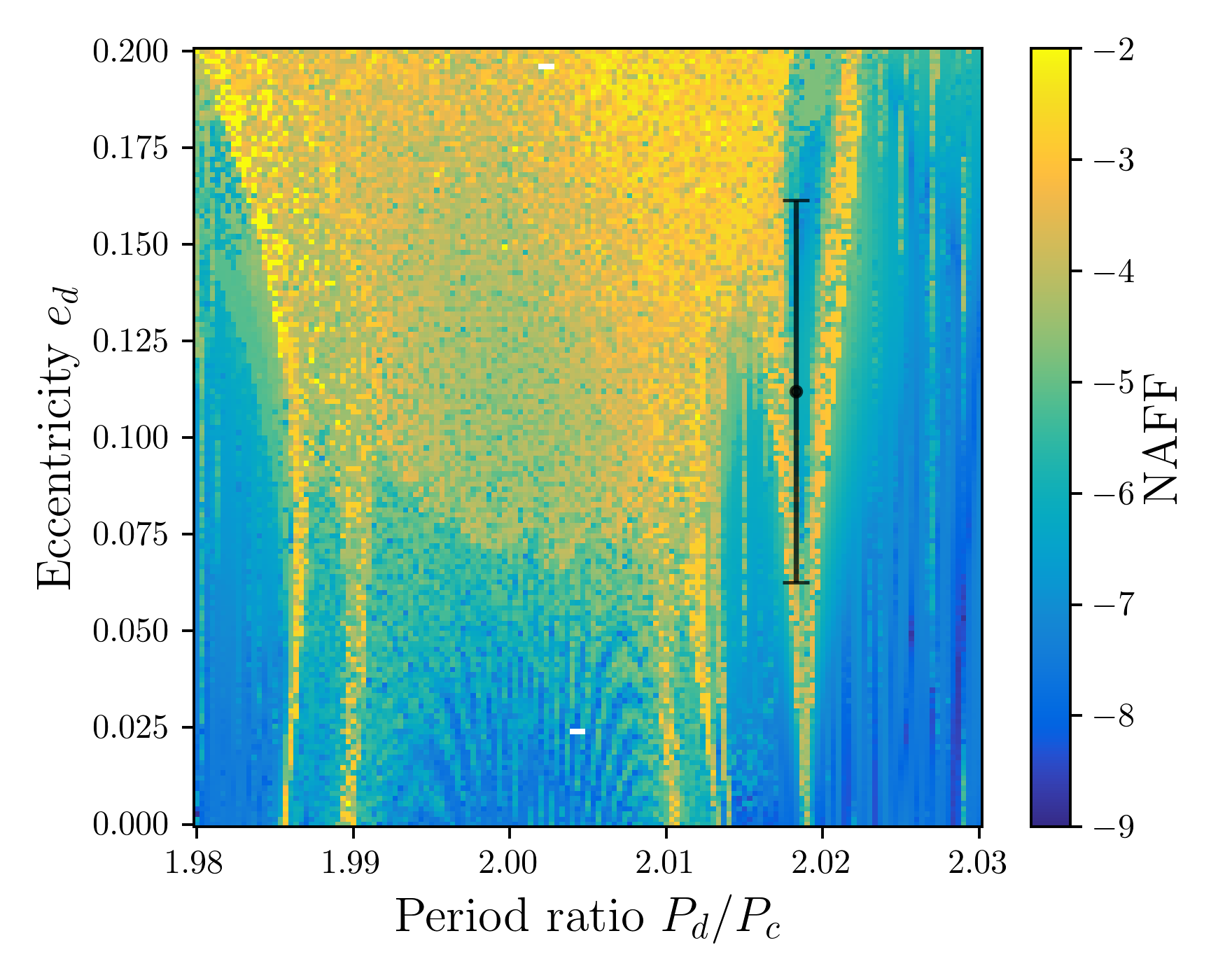}}
\noindent
\begin{small}{\bf Fig. S23: Dynamical chaos map of the LHS\,1903\,c and\,d orbits}. It shows the two dimensional sub-space ($P_d/P_c$, $e_d$) of the parameter space with colour indicating the NAFF. The black data point shows the location of the system in this space as derived from the combined global fit, with the error bar indicating its 1$\sigma$ uncertainty on $e_d$. \end{small}
\end{figure}

\begin{figure}[htbp]
\includegraphics[width=1.0\textwidth]{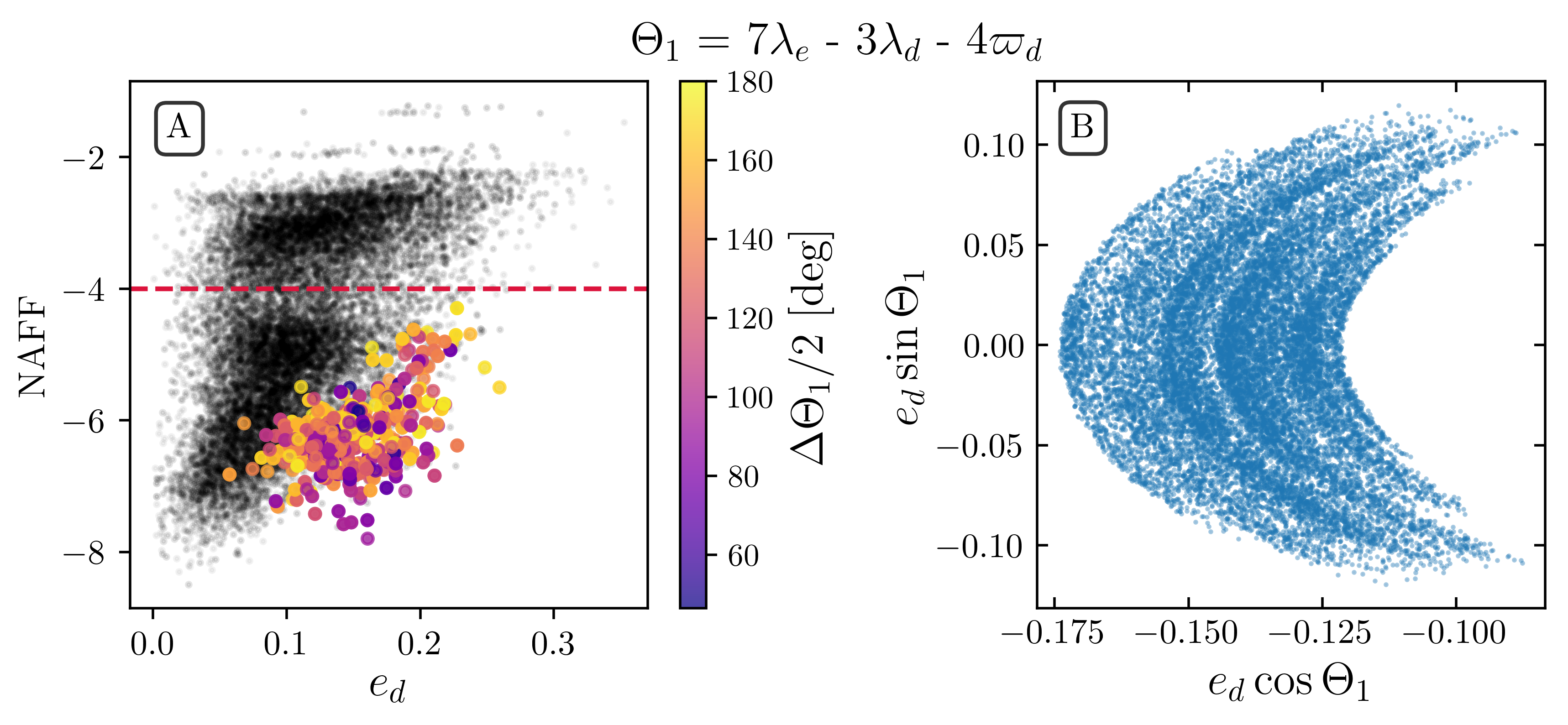}
\noindent
\begin{small}{\bf Fig. S24: Resonant configurations of the LHS\,1903\,d orbit}. ({\bf A}) Gray points are the posterior probability distribution projected onto the plane defined by the NAFF chaos indicator and the eccentricity of planet\,d, $e_d$. The highlighted solutions, in colour, have the d-e planet pair inside the 7:3 MMR, as determined by the semi-amplitude of libration of the resonant angle $\Theta_1~=~7\lambda_e-3\lambda_d-4\varpi_d$. The red dashed line indicates the defined orbital stability criteria upper bound. ({\bf B}) One possible resonant solution in the space ($e_d\cos\Theta_1$, $e_d\sin\Theta_1$) during a 10kyr simulation. \end{small}
\end{figure}

\begin{figure}[htbp]
\centerline{\includegraphics[width=\columnwidth]{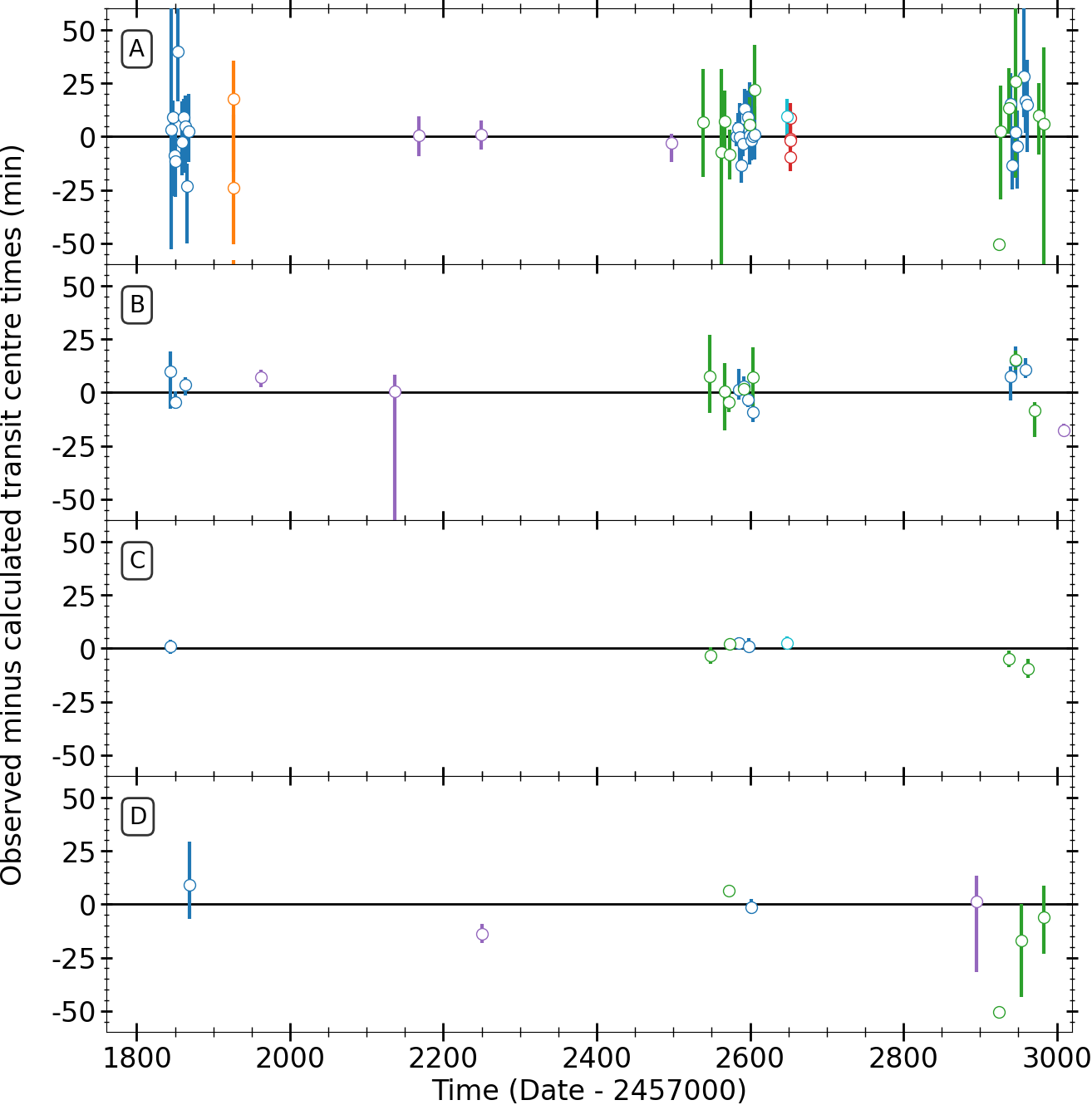}}
\noindent
\begin{small}{\bf Fig. S25: Difference and 1$\sigma$ uncertainty between observed and calculated transit centre times for the LHS\,1903 planets}. ({\bf A}) LHS\,1903\,b, ({\bf B}) LHS\,1903\,c, ({\bf C}) LHS\,1903\,d, and ({\bf D}) LHS\,1903\,e for transits observed with TESS (blue), CHEOPS (green), LCOGT (purple), MuSCAT2 (orange), MuSCAT3 (red), and SAINT-EX (cyan). \end{small}
\end{figure} 


\begin{table}
\begin{footnotesize}
\begin{small}{\bf Table S1: The CHEOPS observing log of LHS\,1903.} Visit identifier of the \textit{CHEOPS} file keys are preceded by ``CH\_PR120054\_'' (except ``TG000301'' and ``TG000401'' that are preceded by ``CH\_PR330094\_'') and followed by ``\_V0300''. See text for the detrending vector selection process; PSF = PSF-{\sc scalpels} principal components, t = time, x = x-centroid offset, y = y-centroid offset, $\phi$ = roll angle, bg = background, contam. = field contamination, T = telescope temperature. Times are in Coordinated Universal Time (UTC). \end{small}
\vspace{-0.4cm}
\begin{center}
\label{tab:obs_log}
\begin{tabular}{c c c c c c c}
\hline\hline
\vspace{-0.2cm}
 File key & Planets & Start date and time & Duration & Efficiency & Aperture & Detrending vectors  \\
\vspace{-0.1cm}
  &  & [UTC] & [h] & [\%] & [pixel] &  \\
\hline             
\vspace{-0.2cm}
 & & & & & & PSF, x$^{2}$, cos($\phi$), sin($\phi$), \\
\vspace{-0.2cm}
 TG000101 & b & 2021/11/19T19:12:21 & 13.79 & 56.0 & 27 & sin$^{2}$($\phi$), sin$^{3}$($\phi$), contam., \\%
\vspace{-0.12cm}
 & & & & & &  \& T \\
\vspace{-0.12cm} 
 TG000301 & c,d & 2021/11/28T23:50:20 & 21.86 & 55.1 & 32 & PSF, t, t$^{2}$, y$^{2}$, \& T \\%
\vspace{-0.2cm}
 TG000302 & b & 2021/12/13T05:45:21 & 23.28 & 56.3 & 34 & PSF, t$^{2}$, x$^{2}$, sin$^{2}$($\phi$), \\%
\vspace{-0.12cm}
   & & & & & & sin$^{3}$($\phi$), \& T \\
\vspace{-0.2cm}
  & & & & & & PSF, t$^{2}$, cos($\phi$), cos$^{2}$($\phi$), \\
\vspace{-0.2cm}
 TG000102 & b,c & 2021/12/17T19:59:21 & 11.57 & 56.4 & 28 & sin($\phi$), sin$^{2}$($\phi$), sin$^{3}$($\phi$), \\%
\vspace{-0.12cm}
  & & & & & & contam., \& T \\  
\vspace{-0.12cm}
 TG001501 & c,e & 2021/12/23T20:32:20 & 12.04 & 58.2 & 36 & PSF, t$^{2}$, \& T \\%
\vspace{-0.12cm}
 TG000103 & b,c,d & 2021/12/24T08:46:21 & 13.32 & 57.1 & 28 & PSF, t$^{2}$, x, bg, \& T \\%
\vspace{-0.12cm}
 TG000303 & $-$ & 2021/12/27T14:27:21 & 22.84 & 55.3 & 32 & PSF, x$^{2}$, \& T \\%
\vspace{-0.2cm}
  & & & & & & PSF, x$^{2}$, y$^{2}$, cos$^{2}$($\phi$), \\
\vspace{-0.2cm}
 TG001701 & c & 2022/01/11T16:26:21 & 12.21 & 59.3 & 23 & cos$^{3}$($\phi$), sin($\phi$), sin$^{2}$($\phi$), \\%
\vspace{-0.12cm}
  & & & & & & contam., \& T \\ 
\vspace{-0.2cm}
 TG001801 & b & 2022/01/19T04:03:21 & 11.57 & 56.5 & 33 & PSF, t$^{2}$, cos$^{3}$($\phi$), contam., \\%
\vspace{-0.12cm}
  & & & & & & \& T \\ 
\vspace{-0.2cm}
 TG001702 & c & 2022/01/24T02:58:21 & 12.79 & 55.6 & 31 & PSF, t$^{2}$, cos$^{2}$($\phi$), contam., \\%
\vspace{-0.12cm}
  & & & & & & \& T \\ 
\vspace{-0.2cm}
 TG001802 & b & 2022/01/25T17:05:21 & 11.57 & 56.2 & 33 & PSF, t, t$^{2}$, x$^{2}$, y, cos$^{2}$($\phi$), \\
\vspace{-0.12cm}
  & & & & & & sin($\phi$), sin$^{3}$($\phi$), \& T \\
\vspace{-0.12cm}
 TG002601 & b,e & 2022/12/10T15:21:22 & 11.57 & 56.8 & 36 & PSF, t, t$^{2}$, y$^{2}$, \& T \\
\vspace{-0.2cm}
  & & & & & & PSF, t$^{2}$, x, y$^{2}$, cos($\phi$), \\
\vspace{-0.2cm}
 TG002602 & b & 2022/12/12T19:55:22 & 11.57 & 56.5 & 27 & sin($\phi$), sin$^{3}$($\phi$), bg, \\
\vspace{-0.12cm}
  & & & & & & contam., \& T \\
\vspace{-0.2cm}
  & & & & & & PSF, t, t$^{2}$, x$^{2}$, y, cos($\phi$), \\
\vspace{-0.2cm}
 TG002603 & b,d & 2022/12/23T14:16:21 & 20.69 & 58.0 & 28 & cos$^{2}$($\phi$), cos$^{3}$($\phi$), sin$^{2}$($\phi$), \\
\vspace{-0.12cm}
  & & & & & &  \& T \\
\vspace{-0.12cm}
 TG000301 & $-$ & 2022/12/29T05:37:22 & 1.55 & 53.2 & 18 & PSF \\
\vspace{-0.12cm}
 TG000401 & $-$ & 2022/12/31T11:53:58 & 0.82 & 100.0 & 24 & PSF \\
\vspace{-0.2cm}
 TG002604 & b,c & 2023/01/01T05:06:22 & 19.48 & 55.3 & 31 & PSF, t$^{2}$, x$^{2}$, y$^{2}$, cos$^{3}$($\phi$), \\
\vspace{-0.12cm}
  & & & & & & smear, \& T \\
\vspace{-0.2cm}
 TG002801 & e & 2023/01/08T16:22:22 & 14.39 & 55.4 & 27 & PSF, t, t$^{2}$, sin($\phi$), sin$^{2}$($\phi$), \\
\vspace{-0.12cm}
  & & & & & & sin$^{3}$($\phi$), contam., \& T \\
\vspace{-0.12cm}
 TG002501 & d & 2023/01/18T00:05:22 & 12.16 & 58.6 & 35 & PSF, t, t$^{2}$, \& T \\
\vspace{-0.12cm}
 TG002701 & c & 2023/01/26T10:53:21 & 12.22 & 59.2 & 37 & PSF, t$^{2}$, x$^{2}$, y$^{2}$, cos($\phi$), \& T \\
\vspace{-0.2cm}
  & & & & & & PSF, t$^{2}$, x, x$^{2}$, y, cos$^{2}$($\phi$), \\
\vspace{-0.2cm}
 TG002605 & d & 2023/01/31T09:34:21 & 15.66 & 57.6 & 24 & sin($\phi$), sin$^{2}$($\phi$), sin$^{3}$($\phi$), \\
\vspace{-0.12cm}
  & & & & & & contam., \& T \\
\vspace{-0.2cm}
 TG002802 & b,e & 2023/02/07T00:01:22 & 13.77 & 58.7 & 29 & PSF, t$^{2}$, y$^{2}$, cos$^{2}$($\phi$), sin($\phi$), \\
\vspace{-0.12cm}
  & & & & & & \& T \\
\hline\hline                  
\end{tabular}
\end{center}

\vspace{-0.4cm}
\noindent
\end{footnotesize}
\end{table}

\begin{table}
\begin{footnotesize}
\begin{small}{\bf Table S2: Ground-based photometric observing log of LHS\,1903}. \end{small}
\vspace{-0.4cm}
\begin{center}
\label{tab:ground_obs_log}      
\begin{tabular}{c c c c c c}
\hline\hline
 Telescope & Planets & Start date & Duration & Filter & Exposure time \\ 
  &  & [UTC] & [h] &  & [s] \\
\hline             
 LCOGT McDonald & c & 2020/04/23T02:27:37 & 2.23 & z$_{\rm s}$ & 75 \\
 LCOGT McDonald & c & 2020/10/14T08:02:25 & 3.10 & z$_{\rm s}$ & 75 \\
 LCOGT McDonald & b & 2020/11/14T07:09:26 & 5.28 & z$_{\rm s}$ & 75 \\
 LCOGT McDonald & b & 2021/02/04T03:41:27 & 5.83 & z$_{\rm s}$ & 75 \\
 LCOGT Teide & b & 2021/10/10T02:31:04 & 3.70 & z$_{\rm s}$ & 75 \\
 LCOGT McDonald & e & 2022/11/10T06:15:17 & 5.43 & z$_{\rm s}$ & 90 \\
 LCOGT McDonald & c & 2023/03/05T01:55:14 & 5.98 & z$_{\rm s}$ & 90 \\
 MuSCAT2 & b & 2020/03/17T21:30:07 & 3.92 & g & 40 \\
 MuSCAT2 & b & 2020/03/17T21:29:58 & 3.92 & r & 20 \\
 MuSCAT2 & b & 2020/03/17T21:29:52 & 3.92 & i & 10 \\
 MuSCAT2 & b & 2020/03/17T21:29:50 & 3.92 & z$_{\rm s}$ &6 \\
 MuSCAT3 & b & 2022/03/14T07:40:26 & 3.42 & g & 23 \\
 MuSCAT3 & b & 2022/03/14T07:40:19 & 3.42 & r & 8 \\
 MuSCAT3 & b & 2022/03/14T07:40:21 & 3.42 & i & 12 \\
 MuSCAT3 & b & 2022/03/14T07:40:20 & 3.42 & z$_{\rm s}$ & 11 \\
 SAINT-Ex & d & 2022/03/10T02:25:37 & 3.13 & I+z & 10 \\
\hline\hline                  
\end{tabular}
\end{center}
\end{footnotesize}
\end{table}

\begin{table}
\begin{footnotesize}
\begin{small}{\bf Table S3: Additional stellar properties of LHS\,1903}. Right ascension ($\alpha$) and declination ($\delta$) are given for the J2000 equinox. Proper motions in right ascension ($\mu_\alpha$) and declination ($\mu_\delta$) are in milliarcsecond per year (mas~yr$^{-1}$) The parallax ($\Pi$) in milliarcsecond (mas) and distance ($d$) in parsec (pc) have been offset-corrected using known formulation\cite{Lindegren2021b}.\end{small}
\vspace{-0.4cm}
\begin{center}
\label{tab:stellarParam}      
\begin{tabular}{lll}        
\hline\hline                 
Parameter (unit) & Value & Reference \\ 
\hline
  $\alpha$ & 07$^{\rm h}$11$^{\rm m}$27$\overset{\rm s}{.}$79 & \cite{GaiaCollaboration2023} \\
  $\delta$ & 48$^{\circ}$19$^{'}$40$\overset{\rm ''}{.}$27 & \cite{GaiaCollaboration2023} \\
  $\mu_\alpha$ (mas~yr$^{-1}$) & -92.924$\pm$0.018 & \cite{GaiaCollaboration2023} \\
  $\mu_\delta$ (mas~yr$^{-1}$) & -570.393$\pm$0.015 & \cite{GaiaCollaboration2023} \\
  $\Pi$ (mas) & 28.027$\pm$0.024 & \cite{GaiaCollaboration2023} \\
  $d$ (pc) & 35.68$\pm$0.03 & This work \\
  RV (km~s$^{-1}$) & 30.83$\pm$0.59 & \cite{GaiaCollaboration2023} \\
  U (km~s$^{-1}$) & -60.54$\pm$0.53 & This work$^{\rm *}$ \\
  V (km~s$^{-1}$) & -78.68$\pm$0.13 & This work$^{\rm *}$ \\
  W (km~s$^{-1}$) & -25.49$\pm$0.23 & This work$^{\rm *}$ \\
\hline
  $V$ (mag) & 12.23$\pm$0.04 & \cite{Zacharias2013} \\
  $G_{\rm BP}$ (mag) & 12.460$\pm$0.003 & \cite{GaiaCollaboration2023} \\
  $G$ (mag) & 11.393$\pm$0.003 & \cite{GaiaCollaboration2023} \\
  $G_{\rm RP}$ (mag) & 10.365$\pm$0.004 & \cite{GaiaCollaboration2023} \\
  $J$ (mag) & 9.06$\pm$0.03 & \cite{Skrutskie2006} \\
  $H$ (mag) & 8.42$\pm$0.02 & \cite{Skrutskie2006} \\
  $K$ (mag) & 8.21$\pm$0.02 & \cite{Skrutskie2006} \\
  $W1$ (mag) & 8.10$\pm$0.02 & \cite{Wright2010} \\
  $W2$ (mag) & 8.04$\pm$0.02 & \cite{Wright2010} \\
\hline\hline                                   
\end{tabular}
\end{center}
\vspace{-0.3cm}
$^{\rm *}$ Calculated via a right-handed, heliocentric Galactic spatial velocity formulation\cite{Johnson1987} using the coordinates, proper motions, parallax, RV, and Galactic reference coordinates\cite{GaiaCollaboration2023}.
\end{footnotesize}
\end{table}

\begin{table}
\begin{footnotesize}
\begin{small}{\bf Table S4: TESS extraction and detrending BLS results}. For each pair of methods the top row indicates the first planet found, second row shows the second discovered planet, and so on. The orbital period, $P$, and transit centre times, $T_0$, listed are at the maximum $\Delta$log(L). \end{small}
\vspace{-0.4cm}
\begin{center}
\label{tab:BLS_logL}      
\begin{tabular}{c c c c c }
\hline\hline
 Extraction method & Detrending & $P$ (d) & $T_0$ (BJD-2457000) & $\Delta$log(L) \\ 
\hline             
 PDCSAP & GP Mat\'{e}rn-3/2 kernel & 12.5659 & 1844.42 & 1053 \\
  &  & 6.2263 & 1844.36 & 393 \\
  &  & 29.3182 & 1868.80 & 264 \\
  &  & 2.1555 & 1844.58 & 210 \\
 PDCSAP & {\sc wotan} & 12.5659 & 1844.42 & 944 \\
  &  & 6.2263 & 1844.36 & 361 \\
  &  & 29.3168 & 1868.84 & 246 \\
  &  & 2.1555 & 1844.58 & 215 \\
  \hline
 TPFED/FFIED & GP Mat\'{e}rn-3/2 kernel & 12.5666 & 1844.37 & 1134 \\
  &  & 6.2263 & 1844.35 & 423 \\
  &  & 29.3173 & 1868.83 & 277 \\
  &  & 2.1555 & 1844.57 & 260 \\
 TPFED/FFIED & {\sc wotan} & 12.5666 & 1844.37 & 966 \\
  &  & 6.2263 & 1844.36 & 354 \\
  &  & 29.3168 & 1868.84 & 260 \\
  &  & 2.1555 & 1844.57 & 247 \\
  \hline
 TPFED/FFIED \& PSF-{\sc scalpels} & GP Mat\'{e}rn-3/2 kernel & 12.5659 & 1844.42 & 1125 \\
  &  & 6.2263 & 1844.35 & 391 \\
  &  & 29.3173 & 1868.83 & 310 \\
  &  & 2.1555 & 1844.57 & 246 \\
 TPFED/FFIED \& PSF-{\sc scalpels} & {\sc wotan} & 12.5659 & 1844.42 & 1058 \\
  &  & 6.2263 & 1844.36 & 351 \\
  &  & 29.3173 & 1868.83 & 269 \\
  &  & 2.1555 & 1844.58 & 240 \\
\hline\hline                  
\end{tabular}
\end{center}
\end{footnotesize}
\end{table}

\begin{table}
\begin{footnotesize}
\begin{small}{\bf Table S5: Priors adopted in the joint fitting of the transit photometry and RV data}. Subscripts (b, c, d, and e) refer to previously defined parameters for the LHS\,1903\,b,\,c,\,d, and\,e planets, respectively. Uniform and log-uniform priors are represented by $\mathcal{U}(a, b)$ and $\mathcal{L}(a, b)$, with lower and upper bounds of $a$ and $b$, $\mathcal{N}(\mu,\sigma)$ indicates a Normal (Gaussian) prior with mean, $\mu$, and standard deviation, $\sigma$, and $\mathcal{T}(\mu,\sigma,a,b)$ is a Truncated Gaussian with mean, $\mu$, standard deviation, $\sigma$, and lower and upper bounds of $a$ and $b$. \end{small}
\vspace{-0.4cm}
\begin{center}
\label{tab:trrv_priors}
\begin{tabular}{c c}
\hline\hline                
 Parameter (unit) & Prior \\ 
\hline 
\vspace{-0.1cm}
$P_{\rm b}$ (d) & $\mathcal{U}(2.13,2.18)$ \\ 
\vspace{-0.1cm}
$T_{\rm 0,b}$ (BJD-2457000) & $\mathcal{U}(1844.4,1844.6)$  \\ 
\vspace{-0.1cm}
$R_\mathrm{p,b}/R_\star$ & $\mathcal{U}(0,0.05)$  \\ 
\vspace{-0.1cm}
$b_{\rm b}$ & $\mathcal{U}(0,1)$ \\ 
\vspace{-0.1cm}
$K_{\rm b}$ (${\rm m\,s}^{-1}$) & $\mathcal{U}(0,1)$ \& $\mathcal{L}(1,100)$ \\ 
\vspace{-0.1cm}
$\omega_{\rm b}$ (deg) & $\mathcal{U}(0,360)$ \\ 
\vspace{-0.1cm}
$e_{\rm b}$ & $\mathcal{T}(0.0,0.098,0.0,1.0)$  \\ 
\vspace{-0.1cm}
$P_{\rm c}$ (d) & $\mathcal{U}(6.20,6.25)$ \\ 
\vspace{-0.1cm}
$T_{\rm 0,c}$ (BJD-2457000) & $\mathcal{U}(1844.1,1844.4)$  \\ 
\vspace{-0.1cm}
$R_\mathrm{p,c}/R_\star$ & $\mathcal{U}(0,0.05)$  \\ 
\vspace{-0.1cm}
$b_{\rm c}$ & $\mathcal{U}(0,1)$ \\ 
\vspace{-0.1cm}
$K_{\rm c}$ (${\rm m\,s}^{-1}$) & $\mathcal{U}(0,1)$ \& $\mathcal{L}(1,100)$ \\ 
\vspace{-0.1cm}
$\omega_{\rm c}$ (deg) & $\mathcal{U}(0,360)$ \\ 
\vspace{-0.1cm}
$e_{\rm c}$ & $\mathcal{T}(0.0,0.098,0.0,1.0)$  \\ 
\vspace{-0.1cm}
$P_{\rm d}$ (d) & $\mathcal{U}(12.54,12.59)$ \\ 
\vspace{-0.1cm}
$T_{\rm 0,d}$ (BJD-2457000) & $\mathcal{U}(1844.3,1844.5)$  \\ 
\vspace{-0.1cm}
$R_\mathrm{p,d}/R_\star$ & $\mathcal{U}(0,0.05)$  \\ 
\vspace{-0.1cm}
$b_{\rm d}$ & $\mathcal{U}(0,1)$ \\ 
\vspace{-0.1cm}
$K_{\rm d}$ (${\rm m\,s}^{-1}$) & $\mathcal{U}(0,1)$ \& $\mathcal{L}(1,100)$ \\ 
\vspace{-0.1cm}
$\omega_{\rm d}$ (deg) & $\mathcal{U}(0,360)$ \\ 
\vspace{-0.1cm}
$e_{\rm d}$ & $\mathcal{T}(0.0,0.098,0.0,1.0)$  \\ 
\vspace{-0.1cm}
$P_{\rm e}$ (d) & $\mathcal{U}(29.29,29.34)$ \\ 
\vspace{-0.1cm}
$T_{\rm 0,e}$ (BJD-2457000) & $\mathcal{U}(1868.7,1886.9)$  \\ 
\vspace{-0.1cm}
$R_\mathrm{p,e}/R_\star$ & $\mathcal{U}(0,0.05)$  \\ 
\vspace{-0.1cm}
$b_{\rm e}$ & $\mathcal{U}(0,1)$ \\ 
\vspace{-0.1cm}
$K_{\rm e}$ (${\rm m\,s}^{-1}$) & $\mathcal{U}(0,1)$ \& $\mathcal{L}(1,100)$ \\ 
\vspace{-0.1cm}
$\omega_{\rm e}$ (deg) & $\mathcal{U}(0,360)$ \\ 
\vspace{-0.1cm}
$e_{\rm e}$ & $\mathcal{T}(0.0,0.098,0.0,1.0)$  \\ 
\hline 
\vspace{-0.1cm}
$\rho_\star$ (${\rm kg\,m^{-3}}$) &  $\mathcal{N}(4840,490)$ \\ 
\hline\hline
\end{tabular}
\end{center}
\end{footnotesize}
\end{table}

\begin{table}
\begin{footnotesize}
\begin{small}{\bf Table S6: Priors and posteriors for the limb-darkening coefficients in the joint fitting}. Subscripts (TESS, CHEOPS, LCOGT, MuSCAT2$_{\rm g}$, MuSCAT2$_{\rm i}$, MuSCAT2$_{\rm Zs}$, MuSCAT3$_{\rm g}$, MuSCAT3$_{\rm i}$, MuSCAT3$_{\rm Zs}$, and SAINT-EX$_{\rm I+z}$) refer to previously defined parameters for the denoted telescopes and filters, respectively. Notation for priors is the same as in Table S5. Fitted values and uncertainties are the medians and 16th/84th percentiles of the posterior probability distributions. \end{small}
\vspace{-0.4cm}
\begin{center}
\label{tab:noise_priors}
\begin{tabular}{c c c}
\hline\hline                
 Parameter (unit) & Prior & Fitted value \\ 
\hline 
$q_{1,{\rm TESS}}$ & $\mathcal{U}(0,1)$ & 0.45$^{+0.15}_{-0.13}$ \\
$q_{2,{\rm TESS}}$ & $\mathcal{U}(0,1)$ & 0.64$^{+0.19}_{-0.22}$  \\
$q_{1,{\rm CHEOPS}}$ & $\mathcal{U}(0,1)$ & 0.78$^{+0.13}_{-0.14}$  \\
$q_{2,{\rm CHEOPS}}$ & $\mathcal{U}(0,1)$ & 0.057$^{+0.076}_{-0.040}$  \\
$q_{1,{\rm LCOGT}}$ & $\mathcal{U}(0,1)$ & 0.62$^{+0.13}_{-0.15}$  \\
$q_{1,{\rm MuSCAT2_{\rm g}}}$ & $\mathcal{U}(0,1)$ & 0.21$^{+0.16}_{-0.13}$  \\
$q_{1,{\rm MuSCAT2_{\rm r}}}$ & $\mathcal{U}(0,1)$ & 0.70$^{+0.18}_{-0.22}$  \\
$q_{1,{\rm MuSCAT2_{\rm i}}}$ & $\mathcal{U}(0,1)$ & 0.48$^{+0.24}_{-0.26}$  \\
$q_{1,{\rm MuSCAT2_{\rm Zs}}}$ & $\mathcal{U}(0,1)$ & 0.59$^{+0.20}_{-0.22}$  \\
$q_{1,{\rm MuSCAT3_{\rm g}}}$ & $\mathcal{U}(0,1)$ & 0.54$^{+0.23}_{-0.22}$  \\
$q_{1,{\rm MuSCAT3_{\rm r}}}$ & $\mathcal{U}(0,1)$ & 0.49$^{+0.22}_{-0.23}$  \\
$q_{1,{\rm MuSCAT3_{\rm i}}}$ & $\mathcal{U}(0,1)$ & 0.24$^{+0.18}_{-0.15}$  \\
$q_{1,{\rm MuSCAT3_{\rm Zs}}}$ & $\mathcal{U}(0,1)$ & 0.78$^{+0.14}_{-0.19}$  \\
$q_{1,{\rm SAINT-EX_{\rm I+z}}}$ & $\mathcal{U}(0,1)$ & 0.22$^{+0.15}_{-0.12}$  \\

\hline\hline
\end{tabular}
\end{center}
\end{footnotesize}
\end{table}

\begin{table}
\begin{footnotesize}
\begin{small}{\bf Table S7: Priors and posteriors for the noise parameters in the joint fitting}. Notation for parameters is the same as in Tables~S5 and~S6. Fitted values and uncertainties are the medians and 16th/84th percentiles of the posterior probability distributions. \end{small}
\vspace{-0.4cm}
\begin{center}
\label{tab:noise_priors}
\begin{tabular}{c c c}
\hline\hline       
\vspace{-0.15cm}
 Parameter (unit) & Prior & Fitted value \\ 
\hline 
\vspace{-0.15cm}
$\sigma_{{\mathrm{jitter,TESS}}}$ (ppm) & $\mathcal{L}(0.1,1000.0)$ & 2.1$^{+11.7}_{-1.8}$ \\
\vspace{-0.15cm}
$\sigma_{{\mathrm{jitter,CHEOPS}}}$ (ppm) & $\mathcal{L}(0.1,1000.0)$ & 11.2$^{+41.6}_{-9.8}$ \\
\vspace{-0.15cm}
$\sigma_{{\mathrm{jitter,LCOGT}}}$ (ppm) & $\mathcal{L}(0.1,10000.0)$) & 794$^{+30}_{-31}$ \\
\vspace{-0.15cm}
$\sigma_{{\mathrm{jitter,MuSCAT2_{\rm g}}}}$ (ppm) & $\mathcal{L}(0.1,10000.0)$ & 88$^{+365}_{-79}$ \\
\vspace{-0.15cm}
$\sigma_{{\mathrm{jitter,MuSCAT2_{\rm r}}}}$ (ppm) & $\mathcal{L}(0.1,10000.0)$ & 1.3$^{+11.1}_{-1.0}$ \\
\vspace{-0.15cm}
$\sigma_{{\mathrm{jitter,MuSCAT2_{\rm i}}}}$ (ppm) & $\mathcal{L}(0.1,10000.0)$ & 16$^{+126}_{-15}$ \\
\vspace{-0.15cm}
$\sigma_{{\mathrm{jitter,MuSCAT2_{\rm Zs}}}}$ (ppm) & $\mathcal{L}(0.1,10000.0)$ & 41$^{+201}_{-38}$ \\
\vspace{-0.15cm}
$\sigma_{{\mathrm{jitter,MuSCAT3_{\rm g}}}}$ (ppm) & $\mathcal{L}(0.1,10000.0)$ & 2.3$^{+29.5}_{-2.0}$ \\
\vspace{-0.15cm}
$\sigma_{{\mathrm{jitter,MuSCAT3_{\rm r}}}}$ (ppm) & $\mathcal{L}(0.1,10000.0)$ & 44$^{+112}_{-37}$ \\
\vspace{-0.15cm}
$\sigma_{{\mathrm{jitter,MuSCAT3_{\rm i}}}}$ (ppm) & $\mathcal{L}(0.1,10000.0)$ & 9.2$^{+73.9}_{-8.5}$ \\
\vspace{-0.15cm}
$\sigma_{{\mathrm{jitter,MuSCAT3_{\rm Zs}}}}$ (ppm) & $\mathcal{L}(0.1,10000.0)$ & 5.4$^{+50.2}_{-4.9}$ \\
\vspace{-0.15cm}
$\sigma_{{\mathrm{jitter,SAINT-EX}}}$ (ppm) & $\mathcal{L}(0.1,10000.0)$ & 2371$^{+82}_{-77}$ \\
\vspace{-0.05cm}
$\sigma_{{\mathrm{jitter,HARPS-N}}}$ (${\rm m\,s}^{-1}$) & $\mathcal{L}(0.001,100.0)$ & 1.08$^{+0.48}_{-0.43}$ \\
\hline 
\vspace{-0.15cm}
$\mu_{\rm TESS}$ & $\mathcal{N}(0.0,0.1)$ & -0.000034$^{+0.000081}_{-0.000083}$ \\ 
\vspace{-0.15cm}
$\mu_{\rm CHEOPS}$ & $\mathcal{N}(0.0,0.1)$ & 0.000022$^{+0.000108}_{-0.000092}$ \\ 
\vspace{-0.15cm}
$\mu_{\rm LCOGT}$ & $\mathcal{N}(0.0,0.1)$ & 0.000009$^{+0.000035}_{-0.000036}$ \\ 
\vspace{-0.15cm}
$\mu_{\rm MuSCAT2_{\rm g}}$ & $\mathcal{N}(0.0,0.1)$ & -0.00001$^{+0.00064}_{-0.00059}$ \\ 
\vspace{-0.15cm}
$\mu_{\rm MuSCAT2_{\rm r}}$ & $\mathcal{N}(0.0,0.1)$ & -0.00002$^{+0.00036}_{-0.00035}$ \\ 
\vspace{-0.15cm}
$\mu_{\rm MuSCAT2_{\rm i}}$ & $\mathcal{N}(0.0,0.1)$ & -0.00002$^{+0.00025}_{-0.00024}$ \\ 
\vspace{-0.15cm}
$\mu_{\rm MuSCAT2_{\rm Zs}}$ & $\mathcal{N}(0.0,0.1)$ & -0.00002$^{+0.00017}_{-0.00016}$ \\ 
\vspace{-0.15cm}
$\mu_{\rm MuSCAT3_{\rm g}}$ & $\mathcal{N}(0.0,0.1)$ & -0.000216$^{+0.000060}_{-0.000061}$ \\ 
\vspace{-0.15cm}
$\mu_{\rm MuSCAT3_{\rm r}}$ & $\mathcal{N}(0.0,0.1)$ & -0.000007$^{+0.000051}_{-0.000052}$ \\ 
\vspace{-0.15cm}
$\mu_{\rm MuSCAT3_{\rm i}}$ & $\mathcal{N}(0.0,0.1)$ & 0.000024$^{+0.000046}_{-0.000045}$ \\ 
\vspace{-0.15cm}
$\mu_{\rm MuSCAT3_{\rm Zs}}$ & $\mathcal{N}(0.0,0.1)$ & -0.000005$^{+0.000043}_{-0.000042}$ \\ 
\vspace{-0.15cm}
$\mu_{\rm SAINT-EX}$ & $\mathcal{N}(0.0,0.1)$ & -0.000080$\pm$0.000120 \\ 
\vspace{-0.15cm}
$\mu_{\rm HARPS-N}$ (${\rm m\,s}^{-1}$) & $\mathcal{N}(-25.5,26.6)$ & -0.19$\pm$1.02 \\ 
\hline 
$\sigma_{{\mathrm{GP,TESS}}}$ (ppm) & $\mathcal{L}(1\times10^{-6},1000.0)$ & 0.000495$^{+0.000049}_{-0.000041}$ \\
\vspace{-0.15cm}
$\rho_{{\mathrm{GP,TESS}}}$ (d) & $\mathcal{L}(0.001,1000.0)$ & 0.566$^{+0.071}_{-0.063}$ \\
\vspace{-0.15cm}
$\sigma_{{\mathrm{GP,CHEOPS}}}$ (ppm) & $\mathcal{L}(1\times10^{-6},1000.0)$ & 0.000082$^{+0.000217}_{-0.000066}$ \\
\vspace{-0.15cm}
$\rho_{{\mathrm{GP,CHEOPS}}}$ (d) & $\mathcal{L}(0.001,1000.0)$ & 300$^{+380}_{-200}$ \\
\vspace{-0.15cm}
$\sigma_{{\mathrm{GP,HARPS-N}}}$ (${\rm m\,s}^{-1}$) & $\mathcal{L}(1,100)$ & 3.67$^{+0.69}_{-0.54}$ \\
\vspace{-0.15cm}
$\alpha_{{\mathrm{GP,HARPS-N}}}$ (d$^{-2}$) & $\mathcal{L}(5\times10^{-5},0.0003)$ & 0.000100$^{+0.000058}_{-0.000034}$ \\
\vspace{-0.15cm}
$\Gamma_{{\mathrm{GP,HARPS-N}}}$ & $\mathcal{L}(2,2000)$ & 5.10$^{+3.60}_{-2.00}$ \\
\vspace{-0.05cm}
${\rm Prot}_{{\mathrm{GP,HARPS-N}}}$ (d) & $\mathcal{U}(35,50)$ & 40.88$^{+1.01}_{-0.75}$ \\
\hline 
\vspace{-0.15cm}
$\theta_{{\mathrm{0,HARPS-N}}}$ & $\mathcal{N}(0,1)$ & 0.68$^{+0.53}_{-0.49}$ \\
\vspace{-0.15cm}
$\theta_{{\mathrm{1,HARPS-N}}}$ & $\mathcal{N}(0,1)$ & -0.15$^{+0.53}_{-0.52}$ \\
\vspace{-0.15cm}
$\theta_{{\mathrm{2,HARPS-N}}}$ & $\mathcal{N}(0,1)$ & -0.22$^{+0.80}_{-0.85}$ \\
\vspace{-0.15cm}
$\theta_{{\mathrm{3,HARPS-N}}}$ & $\mathcal{N}(0,1)$ & -0.41$^{+0.76}_{-0.81}$ \\
\vspace{-0.15cm}
$\theta_{{\mathrm{4,HARPS-N}}}$ & $\mathcal{N}(0,1)$ & 0.33$\pm$0.37 \\
\vspace{-0.15cm}
$\theta_{{\mathrm{5,HARPS-N}}}$ & $\mathcal{N}(0,1)$ & 0.65$^{+0.35}_{-0.33}$ \\
\vspace{-0.15cm}
$\theta_{{\mathrm{6,HARPS-N}}}$ & $\mathcal{N}(0,1)$ & 0.73$^{+0.72}_{-0.64}$ \\
\vspace{-0.05cm}
$\theta_{{\mathrm{7,HARPS-N}}}$ & $\mathcal{N}(0,1)$ & 0.37$^{+0.25}_{-0.26}$ \\

\hline\hline
\end{tabular}
\end{center}
\end{footnotesize}
\end{table}

\begin{table}
\begin{small}{\bf Table S8: Results of the internal structure Bayesian analysis modelling}. The columns list the core, mantle and water mass fraction in the non-gaseous part of the planet, as well as the gas mass (in Earth masses, log scale) are presented. Listed values are the medians with 1$\sigma$ uncertainties.
\end{small}
\vspace{-0.3cm}
\begin{center}
\label{tab:intstruc}   
\begin{tabular}{l c c c c}
\hline\hline
 Planet & $f_{\rm core}$ & $f_{\rm mantle}$ & $f_{\rm water}$ & Log $(M_{\rm gas})$ \\  
\hline             
 b & $0.14^{+0.08}_{-0.07}$ & $0.80^{+0.08}_{-0.09}$ & $0.04^{+0.06}_{-0.02}$ & $-9.50^{+1.20}_{-1.13}$ \\
 c & $0.10^{+0.07}_{-0.05}$ & $0.56^{+0.14}_{-0.09}$ & $0.33^{+0.08}_{-0.14}$ & $-3.06^{+0.56}_{-1.66}$ \\
 d & $0.11^{+0.08}_{-0.05}$ & $0.62^{+0.13}_{-0.10}$ & $0.25^{+0.11}_{-0.12}$ & $-1.41^{+0.17}_{-0.21}$ \\
 e & $0.13^{+0.08}_{-0.06}$ & $0.71^{+0.10}_{-0.12}$ & $0.14^{+0.14}_{-0.07}$ & $-7.70^{+2.23}_{-1.94}$ \\
 \hline\hline                  
\end{tabular}
\end{center}
\end{table}


\clearpage 



\end{document}